\documentclass[11pt,prd,groupedaddress,showpacs,doublecolumn,nofootinbib]{revtex4-1}

 \pdfoutput=1 

\usepackage{graphicx}
\usepackage{dcolumn}
\usepackage{bm}
\usepackage{amssymb}
\usepackage{amsmath}
\usepackage{epsfig}
\usepackage[T1]{fontenc}
\usepackage[utf8]{inputenc} 
\usepackage{tgpagella}

\usepackage{hyperref}
\usepackage{hhline}
\usepackage{color}
\usepackage{multirow}
\usepackage[normalem]{ulem}
\usepackage{slashed}
\usepackage{slashed}
\usepackage{tikz}
\usetikzlibrary{snakes}
\usepackage{blindtext}
\usepackage{hyperref}

\usepackage[section]{placeins}

\definecolor{Zsug}{RGB}{0, 145, 33} 
\definecolor{Zcor}{RGB}{210, 0, 210}
\definecolor{Zque}{RGB}{0, 180, 190} 
 
\definecolor{jd}{rgb}{0.858, 0.188, 0.478}

\def\lapp{\mathrel{\rlap{\raise.5ex\hbox{$<$}}
                    {\lower.5ex\hbox{$\sim$}}}}
\def\gapp{\mathrel{\rlap{\raise.5ex\hbox{$>$}}
                    {\lower.5ex\hbox{$\sim$}}}}

{\newcommand{\lsim}{\mbox{\raisebox{-.6ex}{~$\stackrel{<}{\sim}$~}}}
{\newcommand{\gsim}{\mbox{\raisebox{-.6ex}{~$\stackrel{>}{\sim}$~}}}

\newcommand{\bmt}{\begin{pmatrix}}
\newcommand{\emt}{\end{pmatrix}}
\newcommand{\ba}{\begin{array}{c}}
\newcommand{\ea}{\end{array}}
\newcommand{\be}{\begin{equation}}
\newcommand{\ee}{\end{equation}}
\newcommand{\bea}{\begin{eqnarray}}
\newcommand{\eea}{\end{eqnarray}}

\newcommand{\bi}{\begin{itemize}}
\newcommand{\ei}{\end{itemize}}

\newcommand{\baz}{\begin{array}{cc}}

\newcommand{\mathsym}[1]{{}}

\newcommand{\bt}{\begin{tabular}}
\newcommand{\et}{\end{tabular}}

\newcommand{\benu}{\begin{enumerate}}
\newcommand{\eenu}{\end{enumerate}}

\newcommand{\bav}{\begin{array}{cccc}}

\begin{document}

\renewcommand*{\thefootnote}{\fnsymbol{footnote}}

\begin{center}
{\Large\bf Singlet-Doublet Fermionic Dark Matter and Gravitational Wave in Two Higgs Doublet Extension of the Standard Model}
 \\
 \vskip .5cm
 {
Basabendu Barman$^{a,}$\footnote{bb1988@iitg.ac.in},
Amit Dutta Banik$^{b,}$\footnote{amitdbanik@mail.ccnu.edu.cn },
Avik Paul$^{c,}$\footnote{avik.paul@saha.ac.in},
 }\\[3mm]
 {\it{
 $^a$ Department of Physics, Indian Institute of Technology Guwahati, Assam 781039, India \\
 $^b$Key Laboratory of Quark and Lepton Physics (MoE) and Institute of Particle Physics, Central China Normal University, Wuhan 430079, China}\\
 $^c$ Astroparticle Physics and Cosmology Division, Saha Institute of Nuclear Physics, HBNI, 1/AF Bidhannagar, Kolkata 700064, India}
 \end{center}
\vspace{1cm}

\begin{center}
 {\bf{Abstract}}
\end{center}

\begin{abstract}

We present a study of singlet-doublet vector-like leptonic dark matter (DM) in the framework of two Higgs doublet model (2HDM), where the dark sector is comprised of one doublet and one singlet vectorlike fermions (VLFs). The DM, that arises as an admixture of the neutral components of the VLFs, is stabilized by an imposed discrete symmetry $\mathcal{Z}_2^{'}$ . We test the viability of the DM candidate in the light of observations from PLANCK and recent limits on spin-independent direct detection experiments, and search for its possible collider signals. In addition, we also look for the stochastic gravitational wave (GW) signatures resulting from strong first order phase transition due to the presence of the second Higgs doublet. The model thus offers a viable parameter space for a stable DM candidate that can be probed from direct search, collider and GW experiments.
\end{abstract}

\def\a{\alpha}
\def\b{\beta}
\def\l{\lambda}

\maketitle
\flushbottom


\setcounter{footnote}{0}
\renewcommand*{\thefootnote}{\arabic{footnote}}

\section{Introduction}
\label{sec:intro}

 Despite strong evidence of the existence of the dark matter (DM) from several astrophysical and cosmological observations like rotation curves of spiral galaxies~\cite{Zwicky:1933gu,Rubin:1970zza}, the bullet cluster~\cite{Clowe:2006eq}, gravitational lensing~\cite{Massey:2010hh} etc., particle nature of dark matter (DM) is unknown till date. While Planck~\cite{Aghanim:2018eyx} results claim that nearly 26.5\% of matter in the Universe is indeed DM, the Standard Model (SM) of particle physics is incapable of accounting for a viable DM candidate. Interestingly, if DM interactions with the SM particles are similar to those of electroweak interactions, and the particle DM has a mass around the electroweak scale, then such a DM can be thermally produced in the early universe, followed by its freeze-out, leaving a thermal relic very close to the observed DM abundance ($\Omega h^2\sim$ 0.12). This remarkable coincidence is often referred to as the weakly interacting massive particle (WIMP) miracle~\cite{Kolb:1990vq}. Although WIMP remains as an elusive DM candidate (for a recent review on the status of WIMP see~\cite{Arcadi:2017kky}) as it indicates new physics signature around TeV-scale, but the non-observation of any excess both at the colliders and at the DM scattering experiments such as LUX~\cite{Akerib:2017kat}, PandaX-II~ \cite{Tan:2016zwf,Cui:2017nnn} and XENON1T~\cite{Aprile:2015uzo,Aprile:2018dbl} etc. compels us to strive for either an alternative to WIMP-paradigm~\cite{Hall:2009bx,Hochberg:2014dra,Kuflik:2017iqs,Davoudiasl:2019xeb} or to come up with some search strategies for DM detection other than the usual scattering experiments (for an overview see~\cite{Lin:2019uvt}).  
 
 

Motivated from these, in this work we propose a particle DM model by extending the SM with a second Higgs doublet, and adding one vector-like lepton doublet and one vector-like lepton singlet. We consider a Type-I two Higgs doublet model (2HDM) where one of the Higgs doublets is odd under a discrete $\mathcal{Z}_2$ symmetry. Such an imposition of a discrete symmetry is very commonplace in context with 2HDM as this necessarily prevents the appearance of tree-level flavour changing neutral current (FCNC) via Higgs~\cite{Branco:2011iw,Mader:2012pm,Chen:2013qda,Bhattacharyya:2015nca,Basler:2017nzu,Gori:2017qwg}. The newly added fermions are assumed to be odd under a second $\mathcal{Z}_2^{'}$ symmetry. All the SM particles are even under both of the discrete symmetries $\mathcal{Z}_2$ and $\mathcal{Z}_2^{'}$. The imposition of two different discrete symmetries ensures a first order phase transition (FOPT) without hampering the stability of the DM. Under these circumstances the DM emerges as the lightest particle odd under $\mathcal{Z}_2^{'}$ due to an admixture of the neutral component of the doublet and the singlet. Studies of minimal singlet-doublet vectorlike DM in SM has been exhaustively performed in the literature~\cite{Mahbubani:2005pt,DEramo:2007anh,Enberg:2007rp,Cohen:2011ec,Cheung:2013dua,Restrepo:2015ura,Calibbi:2015nha,Cynolter:2015sua, Bhattacharya:2015qpa,Banik:2015aya,Bhattacharya:2016rqj,Bhattacharya:2016lts,Bhattacharya:2017sml,Bhattacharya:2018fus,Bhattacharya:2018cgx,DuttaBanik:2018emv,Arcadi:2018pfo,Barman:2019tuo,Barman:2019aku}. As it is understandable, a purely singlet vector-like fermion (VLF) DM does not have any renormalizable portal interaction with the SM to obtain the observed thermal relic abundance. The purely doublet VLF, on the other hand, annihilates too much to the SM due to its electroweak gauge interactions, thus making it under abundant unless the mass is $\gsim$ TeV. A purely doublet VLF also faces stringent constraints from DM direct detection experiments because of large scattering cross-section mediated by $Z$ boson. In the present model, as we shall see, due to the presence of the second Higgs doublet the bound from direct search is less stringent. This is possible due to some destructive interference~\footnote{The presence of ``bind spot'' in direct search cross-section for a 2HDM model has been studied in~\cite{Cheung:2012qy,Huang:2014xua,Berlin:2015wwa,Arcadi:2018pfo,Cabrera:2019gaq}} between the scalar mediated direct search diagrams. Even in the absence of such destructive interference, a small direct search cross-section is still conceivable due to the suppression coming from the heavy Higgs mass and small scalar mixing. This provides some freedom of choosing a moderate $\sin\theta$. However, as the $Z$-mediated DM-nucleon scattering is still present, hence the constraint is not completely alleviated, and that confines the singlet-doublet mixing to some extent. This, in turn, affects the collider signature for this model. Here we would like to mention that our model is different from the one present in~\cite{Arcadi:2018pfo} where the DM particles do not couple to SM $Z$ boson due to its  Majorana nature.

The origin of the baryon asymmetry of the Universe (BAU) is another long-standing puzzle of particle physics. The electroweak baryogenesis (EWBG) is a possible way to account for the BAU exploiting the three Sakharov conditions~\cite{Sakharov:1967dj}. However, it is not possible to have a successful EWBG within the SM paradigm as the SM neither provides sufficient CP-violation or strong first-order phase transition (SFOPT)~\cite{Gavela:1993ts,Huet:1994jb,Gavela:1994dt,Morrissey:2012db}. Therefore, a successful EWBG invokes new physics at the electroweak scale that can be obtained via an extended scalar sector. The two Higgs doublet model (2HDM) is a very well motivated non-supersymmetric extension of SM where the scalar sector of the SM is augmented with additional Higgs doublet giving rise to a plethora of different phenomenological  implications~\cite{HABER1979493,HALL1981397,PhysRevD.19.945,PhysRevD.41.3421,Aoki:2009ha,Branco:2011iw,Mader:2012pm,Chen:2013qda,Bhattacharyya:2015nca,Basler:2017nzu,Gori:2017qwg}. In context with electroweak phase transition (EWPT), 2HDM has been extensively studied both in the CP-conserving case~\cite{Dorsch:2013wja,Dorsch:2014qja,Basler:2016obg} and CP-violating scenario~\cite{Cline:1996mga,Fromme:2006cm,Haarr:2016qzq,Basler:2017uxn}. It was also shown that 2HDM framework is capable of generating a SFOPT~\cite{Bernon:2017jgv,Wang:2019pet}. 


The production  of GW spectrum happens mainly via three processes: bubble collisions \cite{Kosowsky:1991ua,Paul:2019pgt,Kosowsky:1992vn,Huber:2008hg,Kosowsky:1992rz,Kamionkowski:1993fg,Caprini:2007xq}, sound wave \cite{Hindmarsh:2013xza,Giblin:2013kea,Giblin:2014qia,Hindmarsh:2015qta} and turbulence in the plasma \cite{Caprini:2006jb,Kahniashvili:2008pf,Kahniashvili:2008pe,Kahniashvili:2009mf,Caprini:2009yp}. The signal thus produced can be detected in different GW detectors, for example, space-based detectors like Advanced Laser Interferometer Antenna (ALIA) \cite{Gong:2014mca}, Big Bang Observer(BBO)~\cite{Harry:2006fi}, Deci-hertz Interferometer Gravitational wave Observatory (DECIGO)~\cite{Seto:2001qf}, Laser Interferometer Space Antenna(LISA)~\cite{Caprini:2015zlo}, ground-based detector advanced Laser Interferometer Gravitational-Wave Observatory (aLIGO)~\cite{Harry_2010} etc. The GW signature as a complimentary search strategy in context with DM models has already been studied both in case of freeze-out and freeze-in~\cite{Schwaller:2015tja,Beniwal:2017eik,Cai:2017cbj,Buckley:2017ijx,Alves:2018jsw,Shajiee:2018jdq,Bian:2018mkl,Paul:2019pgt,Mohamadnejad:2019vzg}(for a review on GW probes of DM see~\cite{Bertone:2019irm}). We have shown, within a consistent framework, our model is also capable of providing a detectable GW signal by satisfying all stringent DM, collider and other theoretical constraints.






The paper is organised as follows: in Sec.~\ref{sec:model} we have introduced the particle content of the model along with the necessary interaction terms, in Sec.~\ref{sec:constr} we have discussed the constraints on the model parameters arising due to tree-level unitarity, precision observables and collider bounds, we next move on to Sec.~\ref{sec:dmpheno} where we illustrate the parameter space satisfying relic abundance and direct detection bounds from which we choose a couple of benchmark points to perform the collider analysis in Sec.~\ref{sec:colpheno}, then in Sec.~\ref{sec:gw} we detail the generation of gravitational wave due to SFOPT and show the detector reach for this model. Finally we conclude in Sec.~\ref{sec:concl}.  

\section{Model}
\label{sec:model}

\subsection{Fields and interactions}
\label{sec:fields}

We extend the Standard Model (SM) with the addition of a second Higgs doublet ($\Phi_2$), along with two vector-like fermions (VLF): one doublet $\psi$ and one singlet $\chi$. In order to have a stable DM candidate, we need to impose a $\mathcal{Z}_2^{'}$ symmetry on the dark sector fermions, different from the existing $\mathcal{Z}_2$ symmetry of 2HDM, which is anyway required to forbid tree-level Higgs-mediated FCNC. A second discrete symmetry ($\mathcal{Z}_2^{'}$) is needed in this framework because of the presence of soft $\mathcal{Z}_2$-breaking term in the 2HDM scalar potential {\footnote{The presence of these soft breaking terms has implications in ensuring decoupling behavior of 2HDM.}}, which can potentially lead to the decay of the DM to SM fermions if the DM is also stabilized under the same $\mathcal{Z}_2$. The $\mathcal{Z}_2^{'}$, on the other hand, is exact. All SM fermions are even under both $\mathcal{Z}_2$ and $\mathcal{Z}_2^{'}$, which forbids the Yukawa interactions of the dark sector with the SM sector. Different charge assignments of new particles are listed in Table~\ref{tab:charges}.

\begin{table}[htb]
 \begin{center}
 \begin{tabular}{|c| c| c| c|c|c|} 
 \hline
 Particles & $SU(3)_c$ & $SU(2)$ & $U(1)_Y$ & $\mathcal{Z}_2$ & $\mathcal{Z}_2^{'}$\\ [0.5ex] 
 \hline\hline
 $\psi^T:\left(\psi^0,\psi^{-}\right)$ & 1 & 2 & 1 & + &-\\ 
 \hline
 $\chi^0$ & 1 & 1 & 0 & + &-\\
 \hline
 $\Phi_2$ & 1 & 2 & 1 & + &+\\
 \hline
 $\Phi_1$ & 1 & 2 & 1 & - &+\\
 \hline
 \hline
 \end{tabular}
\end{center}
\caption{New particle content of the model and their charge assignments.}
\label{tab:charges}
\end{table}

In this set-up the Lagrangian for the model can be written as:

\bea
\mathcal{L} = \mathcal{L}_{SM}+\mathcal{L}_{f}+\mathcal{L}_{s}+\mathcal{L}_{yuk},
\label{eq:lagrang}
\eea

where $\mathcal{L}_f$ is the Lagrangian for the VLFs, $\mathcal{L}_s$ involves the SM doublet and the additional Higgs doublet, and $\mathcal{L}_{yuk}$ contains the Yukawa interaction terms. 

The interaction Lagrangian for the VLFs reads:

\bea
\mathcal{L}_f =  \bar{\psi}\slashed{D}\psi + \bar{\chi^0}\slashed{\partial}\chi^0 - M_{\psi}\bar{\psi}\psi - M_{\chi}\bar{\chi^0}\chi^0~,
\label{eq:lfermi}
\eea

where $D_{\mu}$ is the covariant derivative under $SU(2)\times U(1)$:

\bea
\begin{split}
D_{\mu}\psi &= \partial_{\mu}\psi-i g \frac{\sigma^{a}}{2} W_{\mu}^a\psi + i \frac{g^{'}}{2} B_{\mu}\psi~,
\label{eq:covderiv}
\end{split}
\eea

where $g$ and $g^{'}$ are the gauge couplings corresponding to $SU(2)$ and $U(1)_Y$ and $a=1,2,3$ are the indices for the generators of $SU(2)$. $W_{\mu}$ and $B_{\mu}$ are the gauge bosons corresponding to SM $SU(2)$ and $U(1)_Y$ gauge groups respectively. 

Lagrangian of the scalar sector involving SM Higgs doublet ($H$) and the new Higgs doublet ($\Phi_2$) can be written as:

\bea
\mathcal{L}_s = \left(D^{\mu}\Phi_1\right)^{\dagger}\left(D_{\mu}\Phi_1\right) + \left(D^{\mu}\Phi_2\right)^{\dagger}\left(D_{\mu}\Phi_2\right) -V(\Phi_1,\Phi_2).
\label{eq:pot}
\eea

The model with such a modified scalar sector thus resembles with the standard two Higgs doublet model (2HDM) of Type-I~\cite{Branco:2011iw,Chen:2013qda,Bhattacharyya:2015nca,Basler:2017nzu}, where all SM fermions have Yukawa interactions with only one of the doublets {\it e.g.,} $\Phi_2$. The most general renormalizable scalar potential can then be written as:


\bea
V(\Phi_1,\Phi_2) &=&
m^2_{11}\, \Phi_1^\dagger \Phi_1
+ m^2_{22}\, \Phi_2^\dagger \Phi_2 -
 m^2_{12}\, \left(\Phi_1^\dagger \Phi_2 + \Phi_2^\dagger \Phi_1\right)
+ \frac{\lambda_1}{2} \left( \Phi_1^\dagger \Phi_1 \right)^2
+ \frac{\lambda_2}{2} \left( \Phi_2^\dagger \Phi_2 \right)^2
+ \lambda_3\, \Phi_1^\dagger \Phi_1\, \Phi_2^\dagger \Phi_2 \nonumber \\  &&
+ \lambda_4\, \Phi_1^\dagger \Phi_2\, \Phi_2^\dagger \Phi_1
+ \frac{\lambda_5}{2} \left[
\left( \Phi_1^\dagger\Phi_2 \right)^2
+ \left( \Phi_2^\dagger\Phi_1 \right)^2 \right].
\label{treepot}
\eea

As we shall see, the coefficient $m_{12}$ of the softly $\mathcal{Z}_2$-breaking term plays the pivotal role in deciding the nature of the phase transition. Finally, the charge assignment allows to write a Yukawa interaction~\cite{Barman:2019tuo,Bhattacharya:2015qpa,Bhattacharya:2018fus}:

\bea
-\mathcal{L}_{yuk} = Y\left(\overline{\psi}\widetilde{\Phi_2}\chi^0+H.c.\right),
\label{eq:lyuk}
\eea

where $Y$ is the Yukawa coupling between the VLFs and SM Higgs and $\widetilde{\Phi_2}=i\sigma_2\Phi_2^{*}$.


\subsection{Mixing in the scalar sector}
\label{sec:sclrmix}

We parametrize the scalar doublets as:

\begin{equation}
\Phi_i = 
\begin{pmatrix} 
G_i^{+}  \\
\frac{h_i+v_i+i z_i}{\sqrt{2}}
\end{pmatrix},
\end{equation}

for $i=1,2$. After spontaneous symmetry breaking (SSB), doublet Higgs fields acquires vacuum expectation values (VEVs) $\langle \Phi_1\rangle = v_1$ and $\langle \Phi_2 \rangle =v_2$ such that $\sqrt{v_1^2+v_2^2}=v=246$ GeV. The ratio of  VEVs is given as $\tan\beta=\frac{v_2}{v_1}$. The physical states are obtained by diagonalizing the charged and neutral scalar mass matrices. There are then altogether eight mass eigenstates, three of which become the longitudinal components of the $W^\pm$ and $Z$ gauge bosons. Of the remaining five, there is one charged scalars $H^\pm$, two neutral CP even scalars $h,H$ and one neutral pseudoscalar $A$. The mixing between CP even scalars is denoted by an angle $\alpha$ . For our analysis we shall follow the {\it alignment limit} : $(\beta-\alpha)=\frac{\pi}{2}$, under which $h$ is recognized as the SM Higgs boson of mass $125.09$ GeV~\cite{PhysRevD.98.030001} with exactly the same gauge, Yukawa and self couplings at tree level as those of the SM Higgs bosons, while $H$ is the beyond SM (heavy) Higgs. 


Different couplings occurring in the 2HDM scalar potential can be expressed in terms of physical masses \{  $m_h,m_H,m_{H^{\pm}},m_{A}\}$, mixings $\{\alpha,\beta\}$, VEV $v$ and $m_{12}$: 

\bea
\label{e:l1}
\l_1 &=& \frac{1}{v^2 c^2_\b}~\Big(c^2_\a m^2_H + s^2_\a m^2_h - m^2_{12}\frac{s_\b}{c_\b}  \Big),\\
\label{e:l2}
\l_2 &=& \frac{1}{v^2 s^2_\b}~\Big(s^2_\a m^2_H + c^2_\a m^2_h - m^2_{12}\frac{c_\b}{s_\b}\Big),\\
\label{e:l4}
\l_4 &=& \frac{1}{v^2}~(m^2_A - 2 m^2_{H^+}) + \frac{m^2_{12}}{v^2 s_\b c_\b},\\
\label{e:l5}
\l_5 &=& \frac{m^2_{12}}{v^2 s_\b c_\b} - \frac{m^2_A}{v^2}, \\
\label{e:l3}
\l_3 &=& \frac{1}{v^2 s_\b c_\b}((m^2_H - m^2_h)s_\a c_\a + m^2_A s_\b c_\b ) - 
\l_4\, ,
\label{coup}
\eea
where we denote $s_{\alpha}=\sin{\alpha},~c_{\alpha}=\cos\alpha$ and similarly $s_{\beta}=\sin\beta$ and $c_{\beta}=\cos\beta$. 

%
%

\subsection{Mixing in the VLF sector}
\label{sec:vlfmix}

After EWSB, the neutral components of the doublet ($\psi^0$) and singlet ($\chi^0$) mix via the Yukawa interaction (Eq.~\ref{eq:lyuk}). The mass matrix can be diagonalized in the usual way using a $2\times 2$ orthogonal rotation matrix to find the masses in the physical basis $(\psi_1,\psi_2)^T$:

\bea
\quad
\begin{pmatrix} 
m_{\psi_1} & 0 \\
0 & m_{\psi_2} 
\end{pmatrix}
\quad = ~\mathcal{R^T}\quad
\begin{pmatrix} 
M_{\psi} & m \\
m & M_{\chi} 
\end{pmatrix}
\quad \mathcal{R},
\label{eq:massmatrix}
\eea

where the non-diagonal terms are present due to  Eq.~\ref{eq:lyuk}, and the rotation matrix is given by

$\mathcal{R}=\begin{pmatrix} 
\cos\theta & \sin\theta \\
-\sin\theta  & \cos\theta 
\end{pmatrix}
\quad$. The mixing angle is related to the masses in the weak (flavour) basis:

\bea
\tan 2\theta=\frac{2 m}{M_{\psi}-M_{\chi}}.
\label{eq:mixangle}
\eea

The physical eigenstates (in mass basis) are, therefore, a linear superposition of the neutral weak eigenstates. These can be expressed in terms of the mixing angles as:

\bea
\psi_1 &= \cos\theta\chi^0+\sin\theta\psi^0,~
\psi_2 = -\sin\theta\chi^0+\cos\theta\psi^0.
\label{eq:vlfmix}
\eea

The lightest electromagnetic charge neutral $\mathcal{Z}_2$ odd particle is a viable DM candidate of this model. From now on we shall refer $\psi_1$ as the lightest stable particle (LSP) of the model. In the small mixing limit, the charged component of the VLF doublet $\psi^{\pm}$ acquires a mass as:

\bea
m_{\psi^{\pm}} = m_{\psi_1}\sin^2\theta+ m_{\psi_2}\cos^2\theta\approx m_{\psi_2}.
\eea

From Eq.~\ref{eq:mixangle}, we see that the VLF Yukawa is related to the mass difference between two physical eigenstates and is no more an independent parameter:
\bea
Y = \frac{(m_{\psi_2}-m_{\psi_1})\sin 2\theta\cot\beta}{\sqrt{2} v_1} = \frac{\Delta m\sin 2\theta\cot\beta}{\sqrt{2} v_1}.
\label{eq:vlfyuk}
\eea

Therefore one can have three new parameters:  $\{m_{\psi_1},\Delta m,\sin\theta\}$ from DM phenomenology apart from 2HDM parameters. These three parameters will play the key role in determining the relic abundance of the DM, also deciding the fate of the model in direct and collider searches.

\section{Constraints on the model parameters}
\label{sec:constr}

In this section we would like to summarize constraints on the masses, mixings and couplings arising in the model due to theoretical and experimental bounds. We are particularly interested in the choice of $\tan\beta$ and heavy scalar masses in the 2HDM sector in order to have a SFOPT, while the free parameters appearing in the VLF sector is mostly constrained by oblique parameters and later from DM phenomenology. 


\subsection*{Vacuum Stability}
\label{sec:stability}

Stability of the 2HDM potential is ensured by the following conditions~\cite{Bhattacharyya:2015nca,Chakrabarty:2016smc,Basler:2017nzu},
 \begin{gather}
   \begin{gathered}
     \lambda_1,\,\lambda_2 > 0\, ;
     \lambda_3 + 2\sqrt{\lambda_1\lambda_2} > 0\, ;
     \\
     \lambda_3 + \lambda_4 - |\lambda_5| + 2\sqrt{\lambda_1\lambda_2} > 0\, .
   \end{gathered}
   \label{4}
 \end{gather}
 
 These conditions have been shown to be necessary and sufficient~\cite{Ivanov:2006yq} to ensure that the scalar potential is bounded from below~\cite{Maniatis:2006fs}.
 
\subsection*{Perturbativity}
\label{sec:perturb}

Tree-level unitarity imposes bounds on the size of the quartic couplings $\lambda_i$ or various combinations of them. The quartic couplings and the Yukawa couplings appearing in the theory need to satisfy~\cite{Bhattacharyya:2015nca,Chakrabarty:2016smc,Basler:2017nzu}:

\bea
|\lambda_i|<4\pi,~|Y|<\sqrt{4\pi},
\eea

in order to remain within the perturbative limit. Here $\lambda_i=\lambda,\lambda_{1,2,3,4,5}$. Here we would like to mention that absolute stability of the vacuum at tree-level puts a bound on $\tan\beta$ depending on the choice of $m_H$ as derived in~\cite{Xu:2017vpq}. For $m_H\simeq 300-400~\rm GeV$ this typically allows $1\lsim\tan\beta\lsim 30$. However, as it has been discussed in~\cite{Arhrib:2000is,Ginzburg:2003fe,Ginzburg:2005dt,Dorsch:2016nrg}, for $1\leq\tan\beta\leq5$, the quartic couplings are within the perturbativity bound and tree-level unitarity is also satisfied.

\subsection*{Constraints from phase transition}
\label{sec:conpt}

In~\cite{Dorsch:2013wja,Dorsch:2016nrg,Bernon:2017jgv,Wang:2019pet}, a detailed study of phase transition in context with 2HDM has been done and hence we do not repeat it here, rather our aim is to see whether our DM parameter space is in agreement with the choice of the parameters of the scalar potential that can trigger a strong first order phase transition (SFOPT) giving rise to a measurable GW signal. It is possible to have a SFOPT in 2HDM for $m_{H^\pm}\simeq m_A\approx 600~\rm GeV$ and a large positive mass difference between $m_{H^\pm}$ and $m_H$: $m_{H^\pm}-m_H\gsim 300~\rm GeV$~\cite{Bernon:2017jgv,Wang:2019pet} with $\tan\beta\sim 1$~\cite{Dorsch:2016nrg}. In our entire analysis we thus keep $m_{H^\pm}=m_A=650~\rm GeV$ and $m_H=300~\rm GeV$ for two different choices of $\tan\beta=1.3$ and $\tan\beta=5$. 
Such choices of the $\tan\beta$ is in agreement with the the DM phenomenology. As discussed in~\cite{Dorsch:2016nrg,Bernon:2017jgv}, SFOPT can take place in Type-I 2HDM even if the masses of the three extra Higgs bosons are degenerate $\sim 350~\rm GeV$. Such scenario leads to potentially testable premise through the $A\to H h$ decay channel at colliders. 


\subsection*{Electroweak precision observables (EWPO)}
\label{sec:ewpo}

The splitting between the heavy scalar masses is constrained by the oblique electroweak $T$-parameter whose expression in the alignment limit is given by~\cite{Bhattacharyya:2015nca,Chakrabarty:2016smc,He:2001tp,Grimus:2007if,Grimus:2008nb}:

\bea
\Delta T = \frac{g^2}{64\pi^2 m_W^2}\left(\xi\left(m_{H^\pm}^2,m_A^2\right)+\xi\left(m_{H^\pm}^2,m_H^2\right)-\xi\left(m_A^2,m_H^2\right)\right),
\eea

with,

\bea
\xi\left(x,y\right) = \begin{cases}                       
\frac{x+y}{2}-\frac{xy}{x-y}\ln\left(\frac{x}{y}\right), & \text{if $x\neq y$}.\\
0, & \text{if $x=y$}.\\
\end{cases}
\label{eq:t-param}
\eea

As Eq.~\ref{eq:t-param} suggests, this new physics contribution to the $T$-parameter vanishes in the limit $m_{H^\pm}=m_A$ or $m_{H^\pm}=m_H$.  Since we are working in the exact alignment limit and a SFOPT demands $m_{H^\pm}=m_A$~\cite{Bernon:2017jgv}, hence $T$-parameter puts no bound on the scalar sector in our set-up. The presence of the new VLFs shall also contribute to the $T$-parameter~\cite{Cynolter:2008ea}:

\bea
\begin{split}
T^{\rm VLF}&=\frac{g^2}{16\pi m_W^2}\left(-2 \sin^2\theta~\Pi (M_{\psi},m_{\psi_1})\right)\\&-\frac{g^2}{16\pi m_W^2}\left(2 \cos^2\theta~\Pi (M_{\psi},m_{\psi_2})\right)\\&+\frac{g^2}{16\pi m_W^2}\left(2 \cos^2\theta \sin^2\theta~\Pi (m_{\psi_1},m_{\psi_2})\right),
\end{split}
\label{eq:tvlf}
\eea

where 

\bea
\begin{split}
\Pi(m_i,m_j)&=-\frac{1}{2} \left(m_i^2+m_j^2\right) \left(\text{div}+\log \left(\frac{\mu_{EW}^2}{m_i m_j}\right)\right)\\&+m_i m_j \left(\text{div}+\frac{\left(m_i^2+m_j^2\right) \log \left(\frac{m_j^2}{m_i^2}\right)}{2 \left(m_i^2-m_j^2\right)}+\log \left(\frac{\mu_{EW}^2}{m_i m_j}\right)+1\right)\\&-\frac{1}{4} \left(m_i^2+m_j^2\right)-\frac{\left(m_i^4+m_j^4\right) \log \left(\frac{m_j^2}{m_i^2}\right)}{4 \left(m_i^2-m_j^2\right)},
\end{split}
\label{eq:pi}
\eea

The bound on $\hat S$ comes from a global fit: $10^3\hat S=0.0\pm 1.3$~\cite{Barbieri:2004qk}. For $S$-parameter, we consider contribution only due to the VLFs as given by~\cite{Barman:2019aku,Cynolter:2008ea,Bhattacharya:2018fus}: 

\bea
\begin{split}
\hat S &=\frac{g^2}{16\pi^2}\left(\tilde\Pi^{'}\left(m_{\psi^{\pm}},m_{\psi^{\pm}},0\right)-\cos^4\theta\tilde\Pi^{'}\left(m_{\psi_1},m_{\psi_1},0\right)-\sin^4\theta\tilde\Pi^{'}\left(m_{\psi_2},m_{\psi_2},0\right)\right)\\&-\frac{g^2}{16\pi^2}\left(2\sin^2\theta\cos^2\theta\tilde\Pi^{'}\left(m_{\psi_2},m_{\psi_1},0\right)\right), 
\end{split}
\label{eq:s-param}
\eea

where $g_2$ is the $SU(2)_L$ gauge coupling. The expression for vacuum polarization for identical masses (at $q^2=0$)~\cite{Barman:2019aku,Cynolter:2008ea,Bhattacharya:2018fus}:

\bea
\tilde\Pi^{'}\left(m_i,m_i,0\right)=\frac{1}{3}{\text{div}}+\frac{1}{3}\ln\left(\frac{\mu_{EW}^2}{m_i^2}\right).
\label{eq:vp1}
\eea

\begin{figure}[htb!]
$$
\includegraphics[scale=0.32]{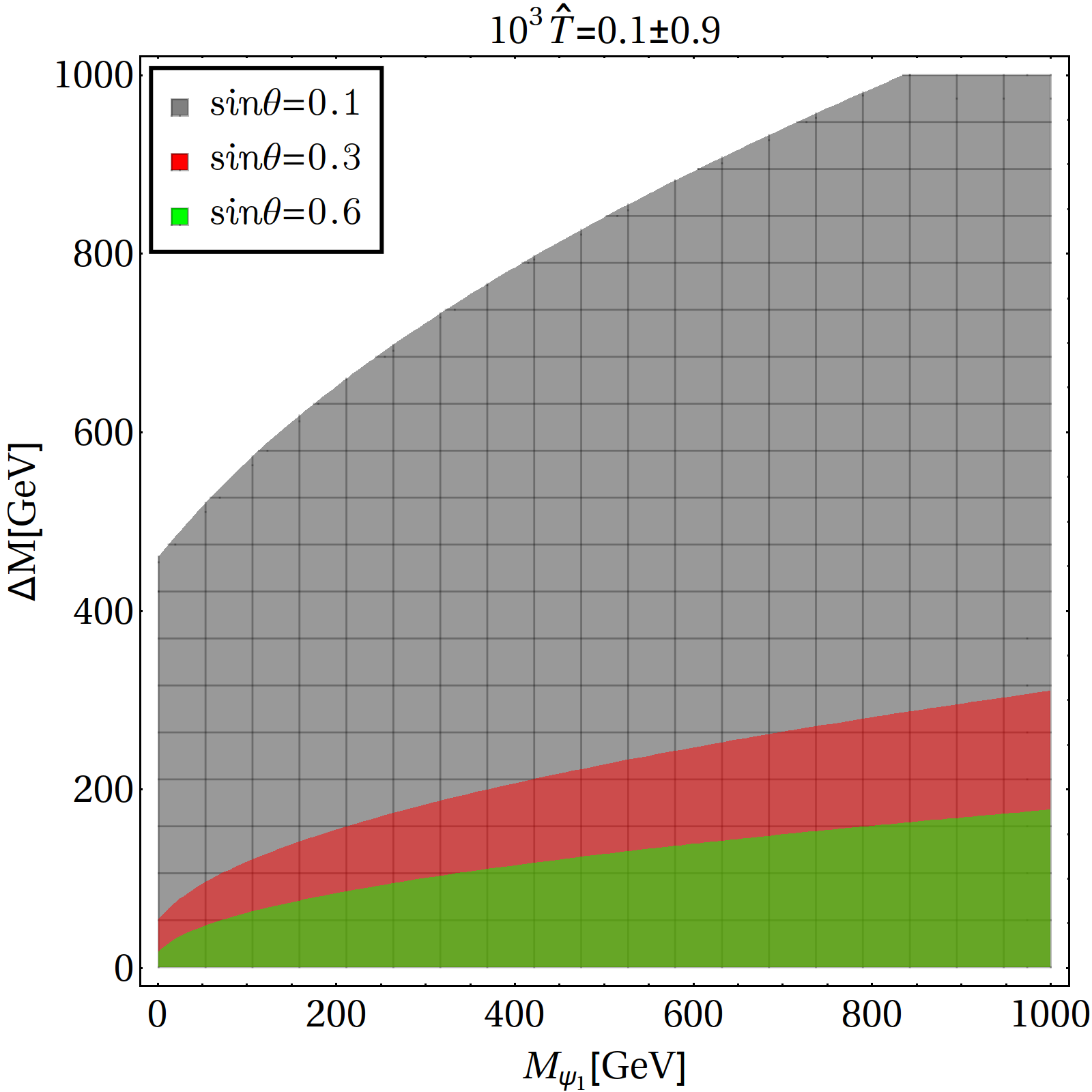}
\includegraphics[scale=0.32]{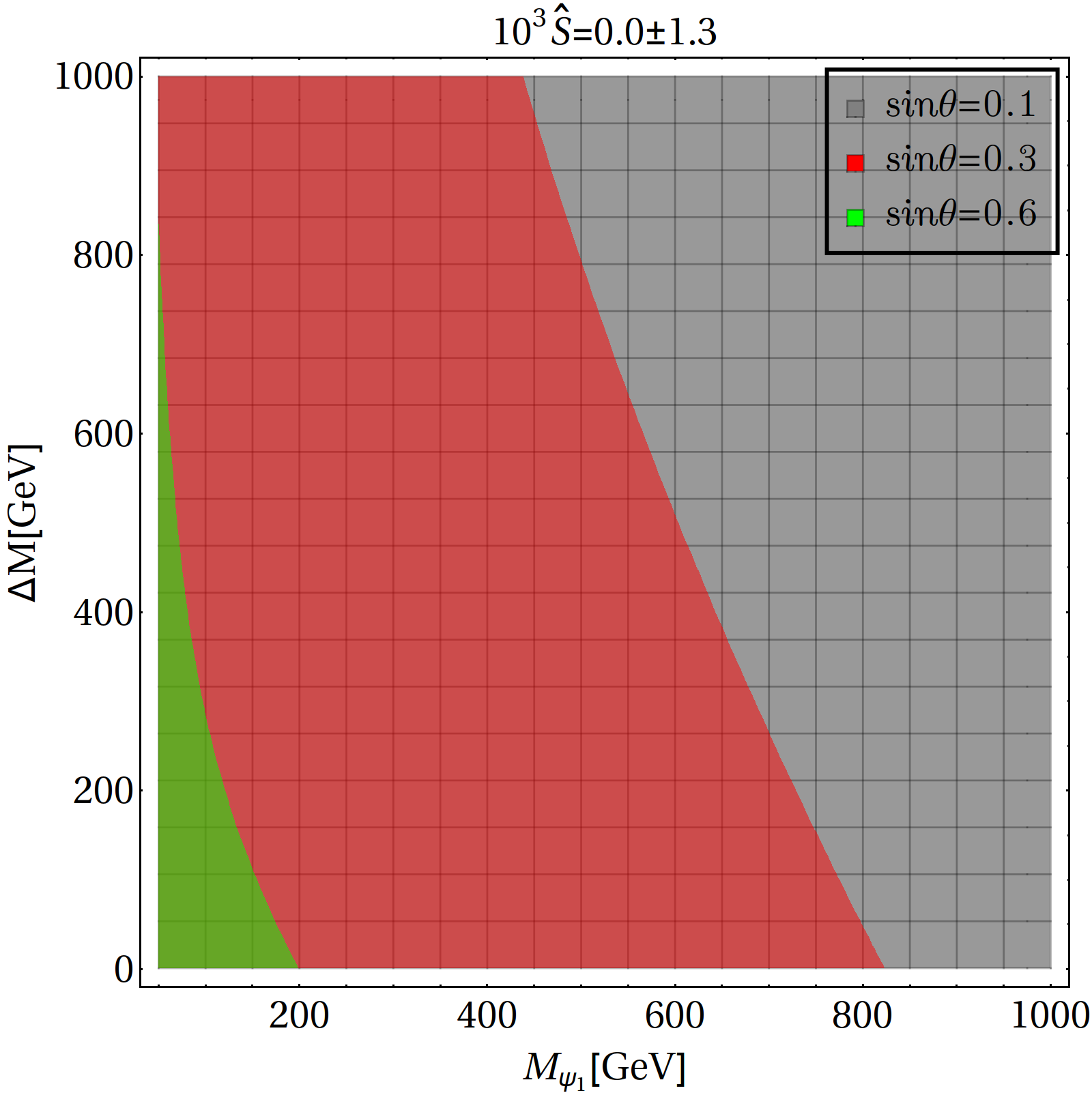}
$$
\caption{Left: Limit from $\hat T$ on DM mass $m_{\psi_1}$ and $\Delta m$ for different choices of $\sin\theta:\{0.1,0.3,0.6\}$ shown respectively in gray, red and green. Right: Limit from $\hat S$ on DM mass $m_{\psi_1}$ and $\Delta m$ for different choices of $\sin\theta:\{0.1,0.3,0.6\}$ shown respectively in gray, red and green.}
\label{fig:st}
\end{figure}

For two different masses ($m_i\neq m_j$) the expression for vacuum polarization reads~\cite{Barman:2019aku,Cynolter:2008ea,Bhattacharya:2018fus}:

\bea
\begin{split}
\tilde\Pi^{'}\left(m_i,m_j,0\right)&=\left(\frac{1}{3}{\text{div}}+\frac{1}{3}\ln\left(\frac{\mu_{EW}^2}{m_i m_j}\right)\right)+\frac{m_i^4-8 m_i^2 m_j^2+m_j^4}{9\left(m_i^2-m_j^2\right)^2}\\&    +\frac{\left(m_i^2+m_j^2\right)\left(m_i^4-4 m_i^2 m_j^2+m_j^4\right)}{6\left(m_i^2-m_j^2\right)^3}\ln\left(\frac{m_j^2}{m_i^2}\right)\\&+m_i m_j \left(\frac{1}{2}\frac{m_i^2+m_j^2}{\left(m_i^2-m_j^2\right)^2}+\frac{m_i^2 m_j^2}{\left(m_i^2-m_j^2\right)^3}\ln\left(\frac{m_j^2}{m_i^2}\right)\right). 
\end{split}
\label{eq:vp2}
\eea

Note that all the divergences appearing in Eq.~\ref{eq:vp1} and \ref{eq:vp2} along with the renormalization scale $\mu_{EW}$, are cancelled on substitution in Eq.~\ref{eq:s-param}. In the LHS of Fig.~\ref{fig:st} we have shown the region allowed by the bound on $T$-parameter in the bi-dimensional plane of $m_{\psi_1}-\Delta m$. We see that large $\Delta m$ is compatible with $T$-parameter for small $\sin\theta$, while large $\sin\theta$ heavily constraints $\Delta m\lsim 300~\rm GeV$ irrespective of the DM mass. The bound is rather complementary in case of $S$-parameter as depicted in the RHS of Fig.~\ref{fig:st}. Here we see large $\sin\theta$ constraints the DM mass $\lsim 200~\rm GeV$, but $\Delta m$ is allowed upto 1 TeV for any choice of the VLF mixing.  

\subsection*{Collider bounds}
\label{sec:colbound}

The LEP experiments have performed direct searches for charged Higgs. A combination of LEP data from searches in the $\tau\nu$ and $cs$ final states put a limit of $m_{H^\pm}\gsim 80~\rm Gev$~\cite{Abbiendi:2013hk,Arbey:2017gmh} under the assumption that the decay $H^\pm\to W^\pm h_1$ is absent. If the aforementioned decay channel is open, then DELPHI and OPAL provide complementary constraints that slightly weakens the charged Higgs mass: $m_{H^\pm}\gsim 72.5~\rm GeV$~\cite{Arbey:2017gmh,Abbiendi:2013hk} for Type-I 2HDM (in context of LEP searches~\cite{Akeroyd:1998dt} is also relevant). Moving on to LHC constraints which come from $t\to H^\pm b$ search with $H^\pm\to\tau\nu$ or $cs$ final states~\cite{Khachatryan:2015qxa,Aad:2013hla,Aad:2014kga,Khachatryan:2015uua}. Values of the charged Higgs masses for which such decay is kinematically allowed are excluded for $\tan\beta\lsim 10$~\cite{Arcadi:2018pfo} in Type-I 2HDM. Here we would like to mention that in Type-I 2HDM $\tan\beta$ is unconstrained from Higgs signal strength in the strict alignment limit, that otherwise puts a strong limit~\cite{Aad:2015gba,Khachatryan:2014jba,Bauer:2017fsw,Arcadi:2018pfo}. Flavour physics observable provide very strong constraints on charged Higgs mass. Inclusive $b\to s\gamma$ and more general $b\to s$ transitions lead to a robust exclusion of $m_{H^\pm}<570~\rm GeV$~\cite{Amhis:2016xyh} for Type-II 2HDM, while for Type-I it is excluded for $\tan\beta\gsim 2$~\cite{Arbey:2017gmh}. For Type-I 2HDM constraint from meson decay is rather weak, allowing $m_{H^\pm}\gsim 200~\rm GeV$ for $\tan\beta\sim 1.5$~\cite{Misiak:2017bgg,Karmakar:2019vnq}. 

Finally, We would like to highlight that LEP has set a lower limit on pair-produced charged heavy vector-like leptons: $m_{\psi} \gsim 101.2~\rm  GeV$ at 95 \% C.L. for $\psi^{\pm}\to\nu W^{\pm}$ final states~\cite{Achard:2001qw}. For a $SU(2)_L$ singlet charged vector-like lepton the CMS search does not improve on the LEP limits. The limits for a heavy lepton doublet decaying to $\ell\in\{e,\mu\}$ flavours are $m_L \gsim 450~\rm  GeV$~\cite{Falkowski:2013jya}. In the case of decays to the $\tau$ flavour the limits are less stringent: $m_L\gsim  270~\rm  GeV$~\cite{Falkowski:2013jya}. However, in our case the charged VLF $\psi^\pm$ has dominant decay to the DM $\psi_1$. As a result, the limits are less stringent and we only follow the LEP lmit in choosing our benchmark points for all the analyses.


\subsection*{Invisible decay constraints}
\label{sec:invdecay}

When the DM mass $m_{\psi_1}<m_{h_1}/2$ or $m_{\psi_1}<m_{Z}/2$, they can decay to a pair of the VLF DM ($\psi_1$). Higgs and $Z$ invisible decays are precisely measured at the LHC~\cite{PhysRevD.98.030001}. Our model thus can be constrained from these measurements in the low mass range of the DM. Both the Higgs and $Z$ invisible decays to DM are proportional to VLF mixing angle $\sin\theta$. In Appendix.~\ref{sec:invdecay} we have computed the invisible decay width of Higgs and $Z$ boson.


\section{Dark Matter Phenomenology}
\label{sec:dmpheno}

In this section we would like to elaborate on the DM phenomenology, where we show in detail the parameter space satisfying the PLANCK observed relic density by scanning over the free parameters of the model. Then we investigate how much of the relic density allowed parameter space also satisfies current direct detection bounds {\it e.g,} from XENON1T~\cite{Aprile:2015uzo,Aprile:2018dbl}. Finally, from the resulting parameter space satisfying relic abundance, direct search and existing theoretical and experimental bounds discussed earlier, we choose a few benchmark points for further analysis. All the relevant Feynman diagrams that contribute to the DM freeze-out are listed in Appendix.~\ref{sec:feyn-diag}.

\subsection{Relic abundance of the dark matter}
\label{sec:relic}

As we have already mentioned earlier, $\psi_1$ is the lightest VLF physical eigenstate which is odd under $\mathcal{Z}_2$, and hence a potential DM candidate in this model. The relic abundance of $\psi_1$ is mainly governed by the DM number changing annihilation and co-annihilation processes mediated by the SM Higgs $h_1$, the non-standard Higgses $h_2,h_3$ and the SM gauge bosons $Z,\gamma$ to various SM final states. The DM number density, thus, can be determined by solving the Boltzmann equation~\cite{Kolb:1990vq,Gondolo:1990dk} for single-component DM, which in our case reads:

\bea
 \frac{dn}{dt} + 3 {\rm H} n = -{\langle \sigma v\rangle}_{eff} \Big(n^2-n_{eq}^2\Big),
 \eea

where 

\bea
{\langle \sigma v\rangle}_{eff}&&= \frac{g_1^2}{g_{eff}^2} {\langle \sigma v \rangle}_{\overline{\psi_1}\psi_1}+\frac{2 g_1 g_2}{g_{eff}^2} {\langle \sigma v \rangle}_{\overline{\psi_1}\psi_2}\Big(1+\frac{\Delta m}{m_{\psi_1}}\Big)^\frac{3}{2} e^{-x \frac{\Delta m}{m_{\psi_1}}} \nonumber \\
&&+\frac{2 g_1 g_3}{g_{eff}^2} {\langle \sigma v \rangle}_{\overline{\psi_1}\psi^-}\Big(1+\frac{\Delta m}{m_{\psi_1}}\Big)^\frac{3}{2} e^{-x \frac{\Delta m}{m_{\psi_1}}} \nonumber \\
&& +\frac{2 g_2 g_3}{g_{eff}^2} {\langle \sigma v \rangle}_{\overline{\psi_2}\psi^-}\Big(1+\frac{\Delta m}{m_{\psi_1}}\Big)^3 e^{- 2 x \frac{\Delta m}{m_{\psi_1}}} \nonumber \\
&& +\frac{g_2^2}{g_{eff}^2} {\langle \sigma v \rangle}_{\overline{\psi_2}\psi_2}\Big(1+\frac{\Delta m}{m_{\psi_1}}\Big)^3 e^{- 2 x \frac{\Delta m}{m_{\psi_1}}} \nonumber \\
&& +\frac{g_3^2}{g_{eff}^2} {\langle \sigma v \rangle}_{{\psi^+}\psi^-}\Big(1+\frac{\Delta m}{m_{\psi_1}}\Big)^3 e^{- 2 x \frac{\Delta m}{m_{\psi_1}}}, 
\label{eq:vf-ann}
\eea

with $n=n_{\psi_1}+n_{\psi_2}+n_{\psi^\pm}$ and ${\rm H}$ is the Hubble parameter. In above equation, $g_{eff}$ is defined as effective degrees of freedom, given by:

\bea
g_{eff}=g_1 + g_2 \Big(1+\frac{\Delta m}{m_{\psi_1}}\Big)^\frac{3}{2} e^{-x \frac{\Delta m}{m_{\psi_1}}} + g_3\Big(1+\frac{\Delta m}{m_{\psi_1}}\Big)^\frac{3}{2} e^{-x \frac{\Delta m}{m_{\psi_1}}} ,
\eea
where $g_1 ,~ g_2~{\text and}~ g_3$ are the degrees of freedom of $\psi_1, ~\psi_2 \rm ~and~ \psi^\pm$ respectively and $x=x_f=\frac{m_{\psi_1}}{T_f}$, where $T_f$ is the freeze-out temperature. We have implemented the model in {\tt LanHEP-3.3.2}~\cite{Semenov:2008jy} and the model files are then fed into {\tt micrOMEGAs-4.3.5}~\cite{Belanger:2001fz} for determining the relic abundance and direct detection cross-section for the DM. Before delving into the detailed parameter scan, we first start by looking into the variation of relic density with DM mass $m_{\psi_1}$ for some fixed choices of two of the other free parameters: $\{\Delta m,\sin\theta\}$. Here we would like to clarify that for the entire analysis we have kept the masses of the new scalars fixed at:

\begin{equation}
m_H = 300~{\text {GeV}}; m_{H^{\pm}} =m_A= 650~{\text {GeV}}\,. 
\label{eq:sclr-mass}
\end{equation}
We perform a scan over a range of the parameter space:
\bea
m_{\psi_1}: \{1-3000\}~{\text{GeV}}; \Delta m: \{1-3000\}~ {\text{GeV}}; \sin\theta: \{0.01-0.8\}; m_{12}: \{1-500\}~{\text{GeV}}\,.
\eea
As discussed earlier in Sec.~\ref{sec:conpt}, this choice of the scalar masses is motivated from the requirement of a SFOPT. Also, we would like to remind once more that we are strictly following the alignment limit, and hence the lightest CP-even scalar resembles the 125 GeV observed Higgs. 

\begin{figure}[htb!]
$$
\includegraphics[scale=0.42]{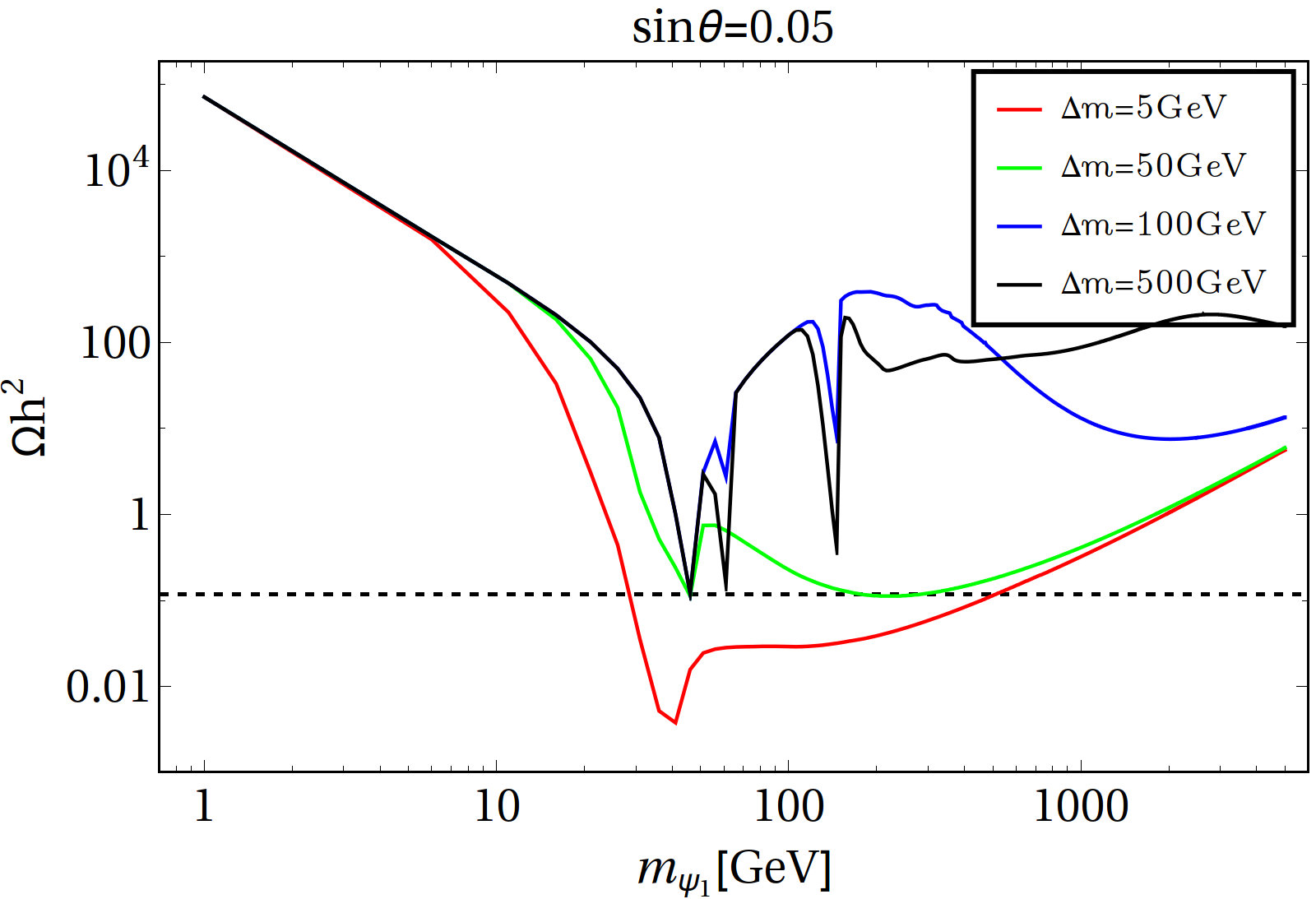}
\includegraphics[scale=0.42]{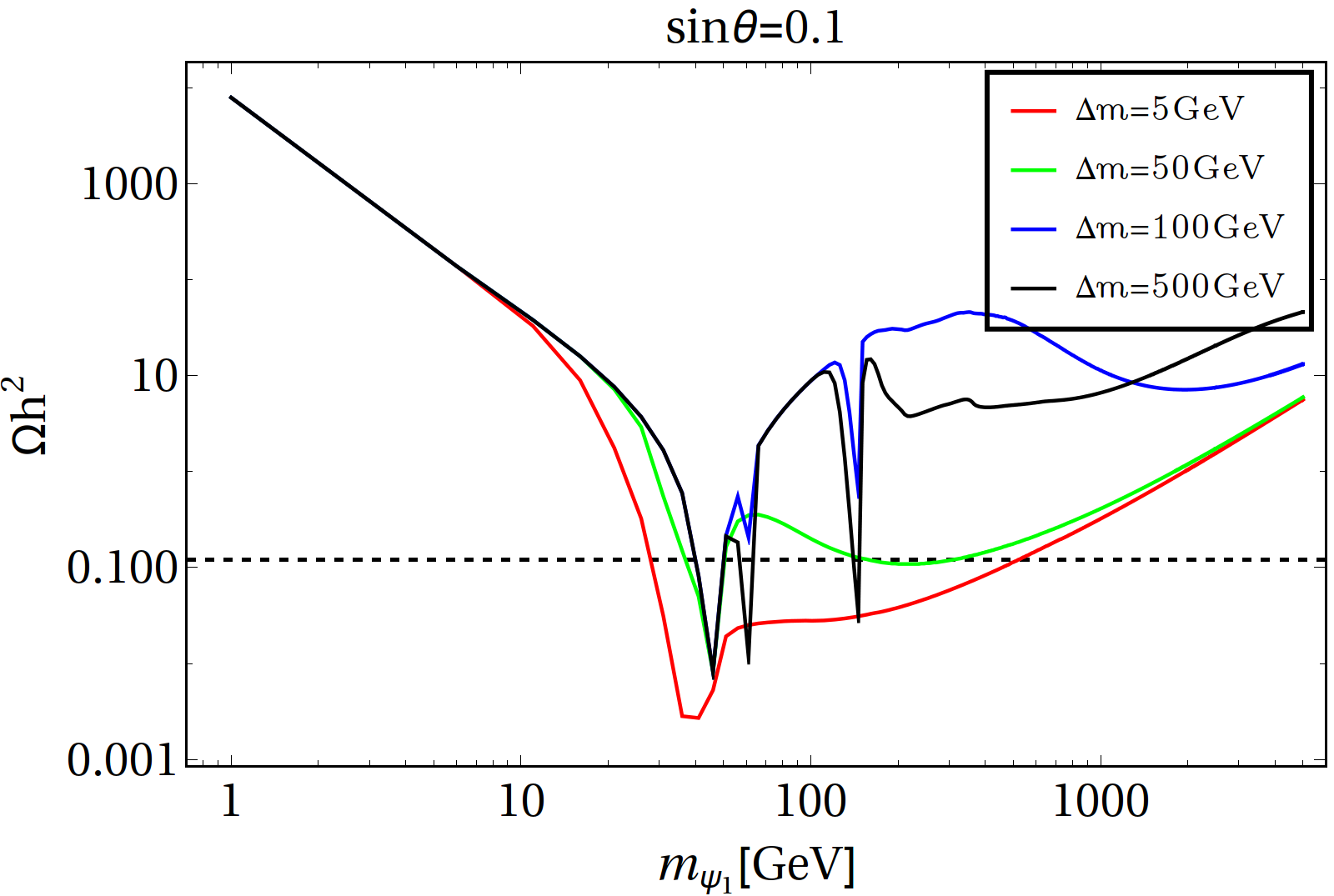}
$$
$$
\includegraphics[scale=0.42]{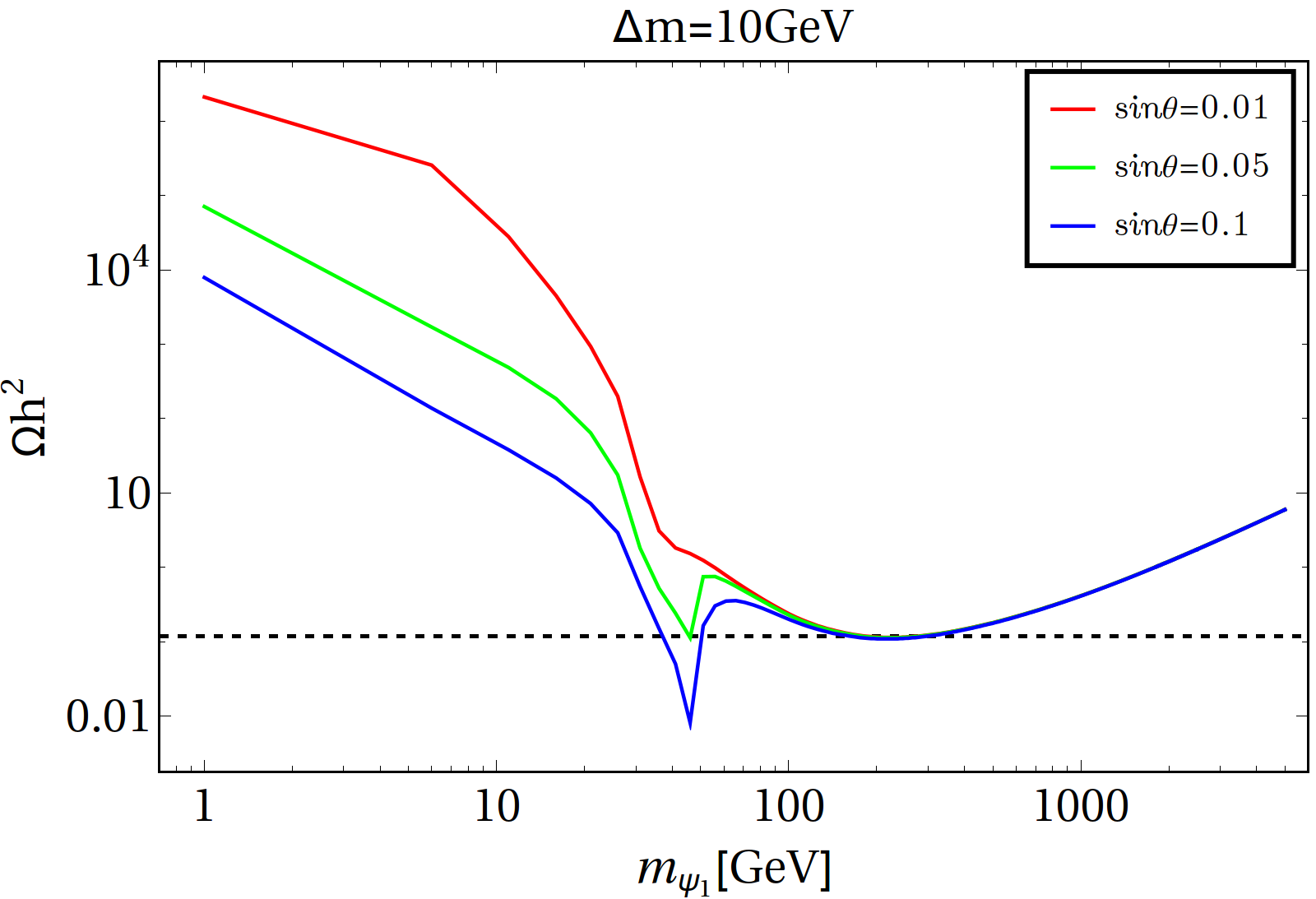}
\includegraphics[scale=0.42]{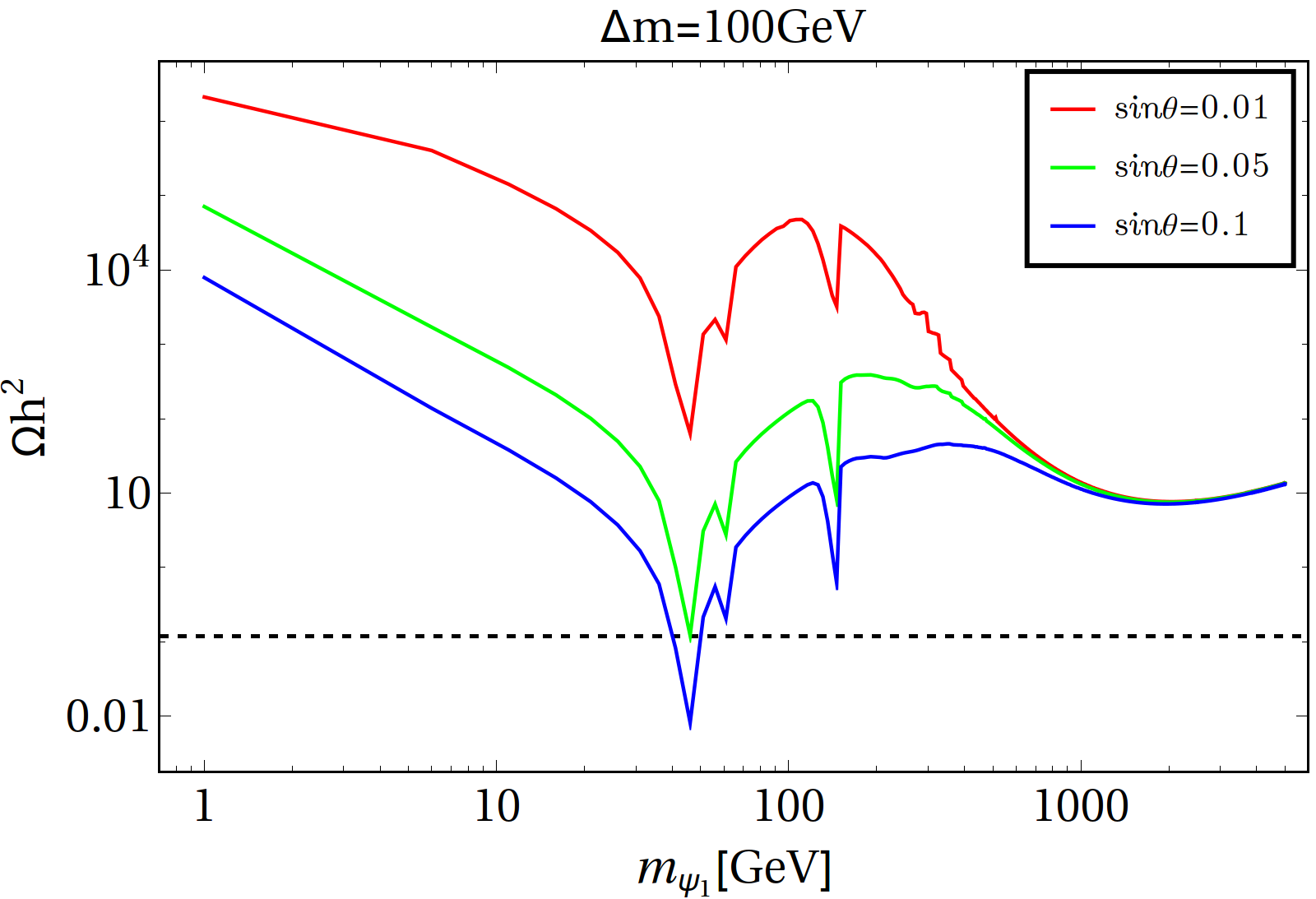}
$$
\caption{Top Left: Variation of relic abundance with DM mass for $\Delta m:\{5,50,100,500\}~\rm GeV$ in red, green, blue and black respectively for VLF mixing $\sin\theta=0.05$. Top Right: Same with $\sin\theta=0.1$. Bottom Left: Variation of DM relic abundance with $m_{\psi_1}$ for different choices of $\sin\theta:\{0.01,0.05,0.1\}$ shown in red, green and blue respectively for a fixed $\Delta m=10~\rm GeV$. Bottom Right: Same for $\Delta m=100~\rm GeV$. For all plots $\tan\beta=1.3$ has been chosen with $m_{12}=170$.}\vspace{0.2cm}
\label{fig:relic1}
\end{figure}

\begin{figure}[htb!]
$$
\includegraphics[scale=0.42]{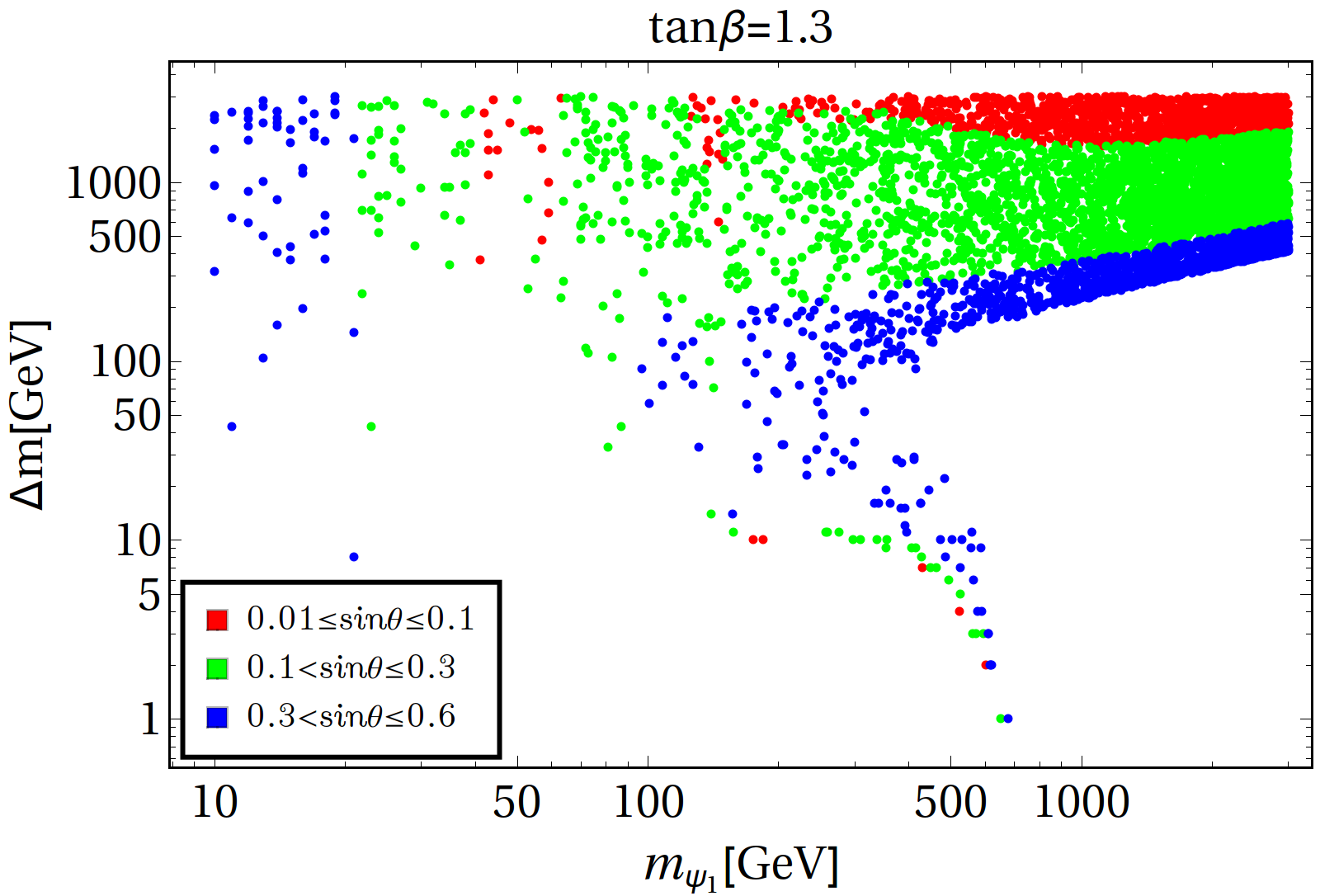}
\includegraphics[scale=0.42]{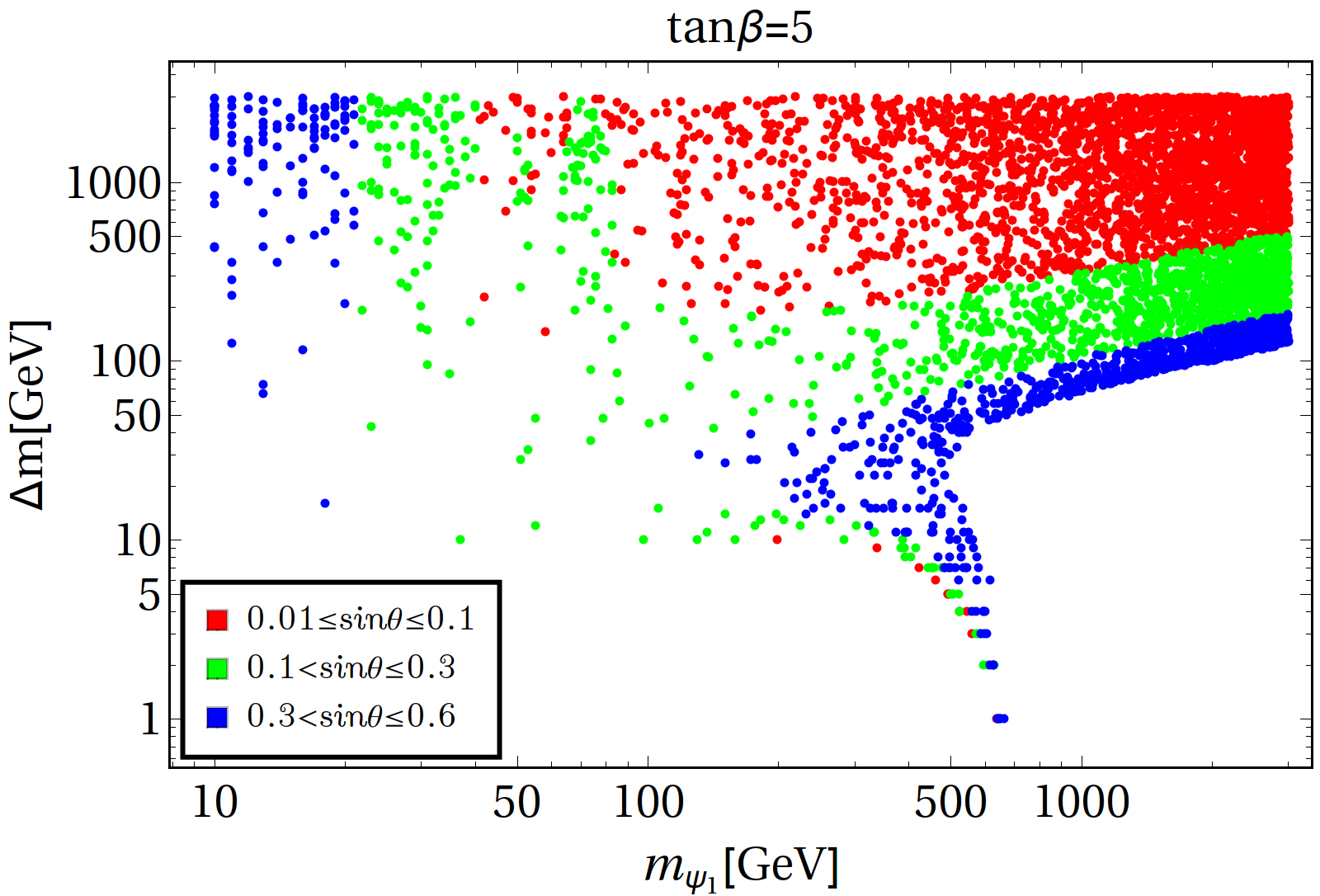}
$$
$$
\includegraphics[scale=0.42]{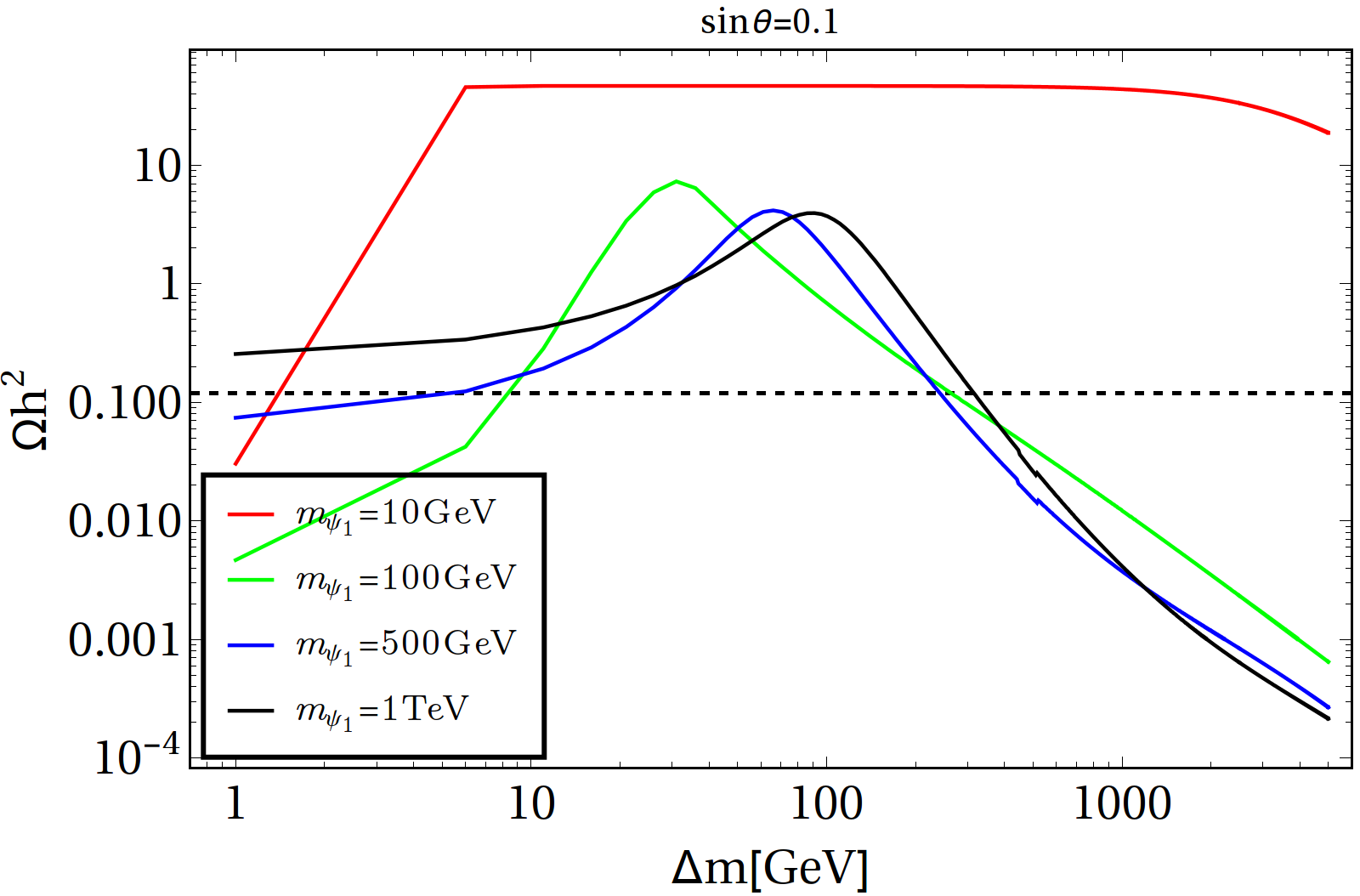}
\includegraphics[scale=0.42]{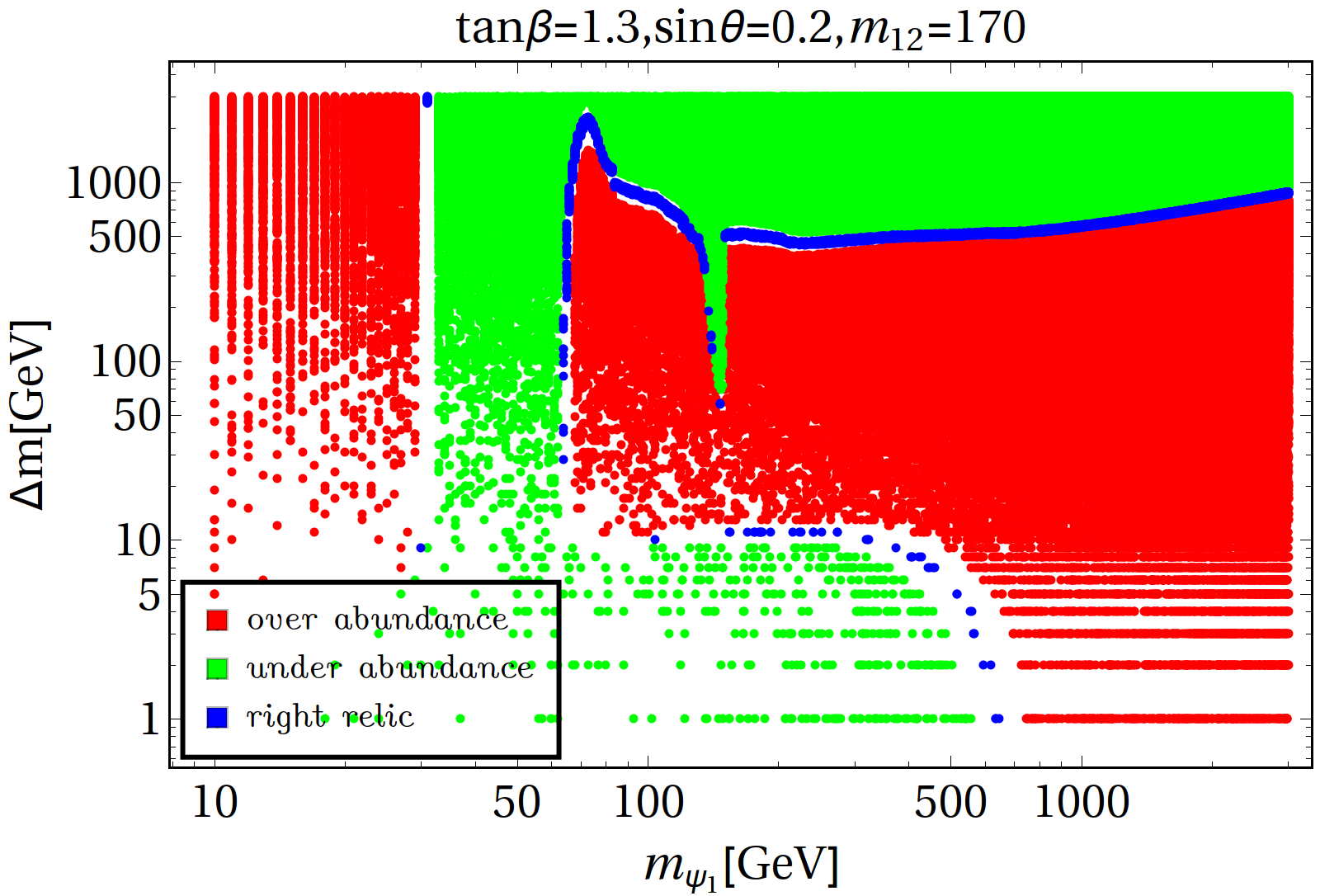}
$$
\caption{Top Left: Parameter space satisfying PLANCK observed relic abundance in $m_{\psi_1}$-$\Delta m$ plane for $0.01\leq\sin\theta\leq 0.1$ in red, $0.1<\sin\theta\leq 0.3$ in green and $0.3<\sin\theta\leq 0.6$ in blue respectively for $\tan\beta=1.3$. Top Right: Same for $\tan\beta=5$ (colour codes are unchanged). In both cases the gray shaded region below represents the neutrino floor (see text for details). Bottom Left: Variation of DM relic abundance with $\Delta m$ for a fixed $\sin\theta=0.1$ and four different choices of the DM mass: $m_{\psi_1}:\{10,100,500,1000\}~\rm GeV$. Bottom Right: Parameter space in $m_{\psi_1}$-$\Delta m$ plane for a fixed $\sin\theta=0.1$, where the over abundant, under abundant and observed abundant regions are shown respectively in red, green and blue. }\vspace{0.2cm}
\label{fig:relscan1}
\end{figure}

In Fig.~\ref{fig:relic1} we show how the relic density of the DM $\psi_1$ varies with DM mass while we choose some of the free parameters at fixed values. In the left hand side (LHS) of the top panel we show such a variation for a fixed VLF mixing $\sin\theta=0.05$ for four different values of $\Delta m:\{5,50,100,500\}~\rm GeV$ in red, green, blue and black curves respectively. Here we see a number of interesting features. First of all, for smaller $\Delta m$ (red) the DM is largely under abundant. This is due to the fact that smaller $\Delta m$ enhances co-annihilation  by making the effective annihilation cross-section large as evident from Eq.~\ref{eq:vf-ann}. This, in turn, reduces the relic abundance. For larger $\Delta m$, co-annihilation effect diminishes and we see right relic is obtained at two sharp resonances: $m_{\psi_1}=\frac{m_Z}{2}$ and $m_{\psi_1}=\frac{m_{h_1}}{2}$. There is another resonance at $m_{\psi_1}\simeq 150~\rm GeV$, which occurs due to the second Higgs at 300 GeV. Thus, for a fixed $\sin\theta$, smaller $\Delta m\simeq 5~\rm GeV$ results in co-annihilation dominantly to light quark final states. For larger $\Delta m\simeq 100~\rm GeV$, on the other hand, annihilation dominates and $WW,Zh_1, h_1h_1$ final states contribute dominantly to the relic abundance. We see the same features in the right hand side (RHS) of the top panel in Fig.~\ref{fig:relic1}. In both the plots all the curves rise with the increase in DM mass ensuring the unitarity of the model. In the bottom panel of Fig.~\ref{fig:relic1} we again show the variation of relic density of the DM with DM mass $m_{\psi_1}$, but now for a fixed $\Delta m$ and for three different $\sin\theta:\{0.01,0.05,0.1\}$ in red, green and blue respectively. Here we see, again, for small $\Delta m$ (left hand side of the bottom panel) due to co-annihilation domination, the resonances are not sharp. However, with increase in $\sin\theta$ the DM becomes under abundant. This is understandable, as larger $\sin\theta$ gives rise to larger (co-)annihilation due to $\sin^2\theta$ dependence at the vertex for gauge-mediated processes and $\sin2\theta$ dependence for scalar mediated processes (due to proportionality to the Yukawa $Y$). As a result, the relic abundance naturally decreases. On the RHS in the bottom panel of Fig.~\ref{fig:relic1} we show the same plot as that of the LHS but for larger $\Delta m=100~\rm GeV$. Now we see, as before, the resonances become important where observed relic density is achieved. For all these plots we have kept $\tan\beta=1.3$ and $m_{12}=170$, for  some other choice of $\tan\beta$ (e.g, $\tan\beta=5$) the inferences remain unaltered.

\begin{figure}[htb!]
$$
\includegraphics[scale=0.4]{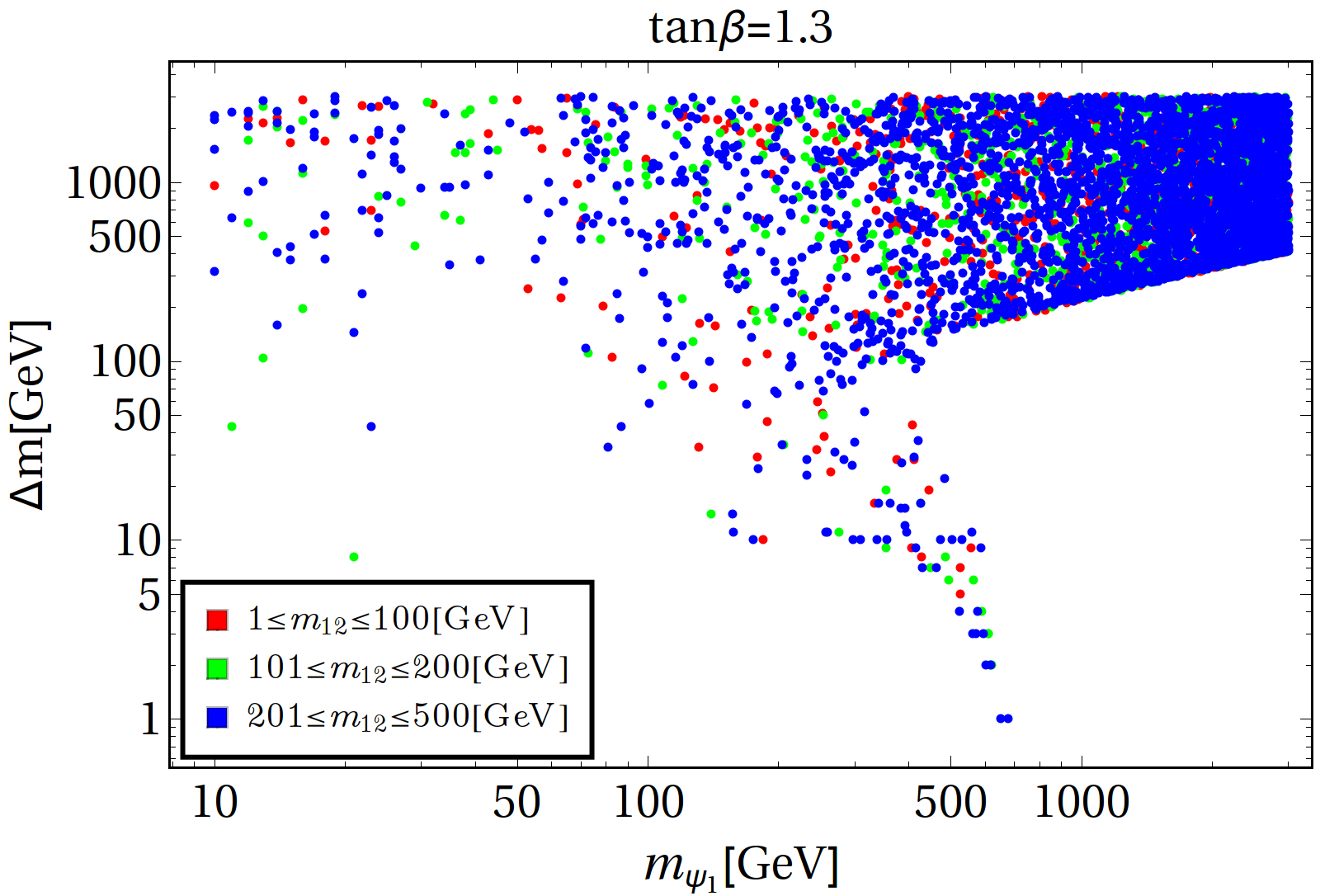}
\includegraphics[scale=0.4]{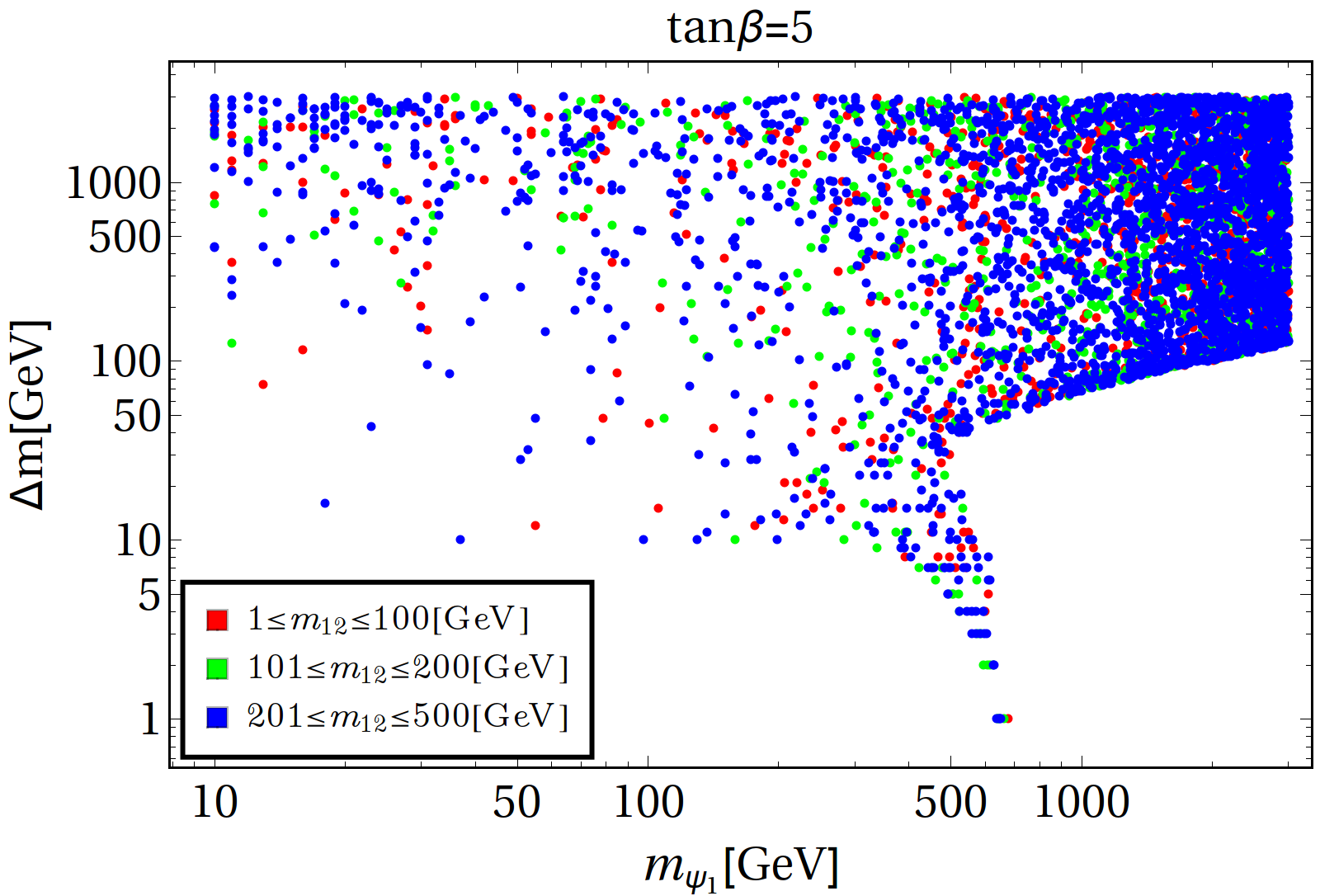}
$$
\caption{Left: Relic density allowed parameter space in the bi-dimensional plane of $m_{\psi_1}-\Delta m$, where different colours correspond to different choices of $m_{12}$: $1\leq m_{12}\leq 100~\rm GeV$ in red, $101<m_{12}\leq 200~\rm GeV$ in green and $201\leq m_{12}\leq 300~\rm GeV$ in blue for $\tan\beta=1.3$. Right: Same for $\tan\beta=5$.}\vspace{0.2cm}
\label{fig:m12rel}
\end{figure}

In top left panel of Fig.~\ref{fig:relscan1} we show the parameter space allowed by PLANCK observed relic density in $m_{\psi_1}$-$\Delta m$ plane for different choices of the VLF mixing $\sin\theta$, where $0.01\leq\sin\theta\leq 0.1$ is shown by the red points, $0.1<\sin\theta\leq 0.3$ are shown by the green points and $0.3<\sin\theta\leq 0.6$ where $\tan\beta=1.3$. Here we see, large $\Delta m$ is achieved for smaller $\sin\theta$. This can intuitively understood in the following way: the scalar mediated annihilation channels are essentially proportional to the Yukawa $Y$, which is proportional to both $\sin\theta$ and $\Delta m$. Hence a smaller $\sin\theta$ requires a larger $\Delta m$ to produce the correct abundance. On the top right panel of Fig.~\ref{fig:relscan1} the same parameter space is shown for $\tan\beta=5$. We note the same pattern here except for the fact that $0.01\leq\sin\theta\leq 0.1$ region is more populated. This is a direct consequence of Eq.~\ref{eq:vlfyuk}, which shows $\sin\theta$ needs to be reduced as $\tan\beta$ increases ($i.e., v_2$ increases) to adjust the Yukawa $Y$ such that the relic abundance is satisfied.

In order to understand the behaviour more intricately we have obtained a parameter space for a fixed $\sin\theta$ in $m_{\psi_1}$-$\Delta m$ plane as shown in the bottom right panel of Fig.~\ref{fig:relscan1}. Here we have shown three different regions corresponding to under abundance (green), over abundance (red) and right relic (blue). As we move from left to right in this plot, we first encounter over abundant regions for small DM mass $\lsim 20~\rm GeV$. This is due to the lack of annihilation channels present for the DM to produce the right relic, as the only annihilation channels are to the light quarks. As we reach $m_{\psi_1}\sim 30~\rm GeV$, co-annihilation starts playing and as a consequence, the DM becomes under abundant. Still right relic abundance is not obtained as $Y$ is small due to small $\Delta m$ and hence all the scalar mediated annihilations do not contribute significantly. Right relic is first obtained at $m_{\psi_1}\sim\frac{m_{h_1}}{2}$, due to the SM Higgs resonance. For DM mass $\sim 100~\rm GeV$ as we move from lower $\Delta m$ to higher $\Delta m$ (from bottom to top), the DM is at first under abundant due to co-annihilation domination. Then right relic abundance is achieved as the co-annihilation is correctly tuned. Immediately after that there is an over abundant region for larger $\Delta m$ as co-annihilation loses its goodness and hence the effective annihilation cross-section (Eq.~\ref{eq:vf-ann}) becomes small. Note that, for a fixed DM mass $\gsim 100~\rm GeV$ right relic abundance is reached twice: (a) Once for small $\Delta m$, where right co-annihilation gives rise to observed relic and (b) for large $\Delta m$, where $Y$ is large enough to produce correct relic via scalar-mediated channels (to gauge-boson dominated final states) as shown in the bottom left panel of Fig.~\ref{fig:relscan1}. Beyond $m_{\psi_1}\sim 500~\rm GeV$ the parameter space is largely over abundant as the suppression due to $1/m_{\psi_1}^2$ becomes significant, thus over producing the DM. Note that, right relic abundance is also obtained at the second Higgs resonance at $m_{\psi_1}\sim 150~\rm GeV$. Beyond $\Delta m\sim 200~\rm GeV$, for a fixed DM mass, the parameter space is largely under abundant as co-annihilation is completely switched off and annihilation to all possible final states are open. This produces a very large effective annihilation cross-section, making the DM completely under abundant. A cumulative effect of all these features is reflected in the upper panel of Fig.~\ref{fig:relscan1} for different choices of the VLF mixing $\sin\theta$. Before moving on to the DM direct search section, we would like to see what are the values of $m_{12}$ that satisfy right relic abundance. This is shown in Fig.~\ref{fig:m12rel} for both $\tan\beta=1.3$ (left) and $\tan\beta=5$ (right). As we see, the dependence of the relic abundance parameter space on the choice $m_{12}$ is not very strict as almost all values of $m_{12}$ is allowed by any choice of DM mass and $\Delta m$. 

\subsection{Direct detection of the dark matter}
\label{sec:dd}

\begin{figure}[htb!]
$$
\includegraphics[scale=0.45]{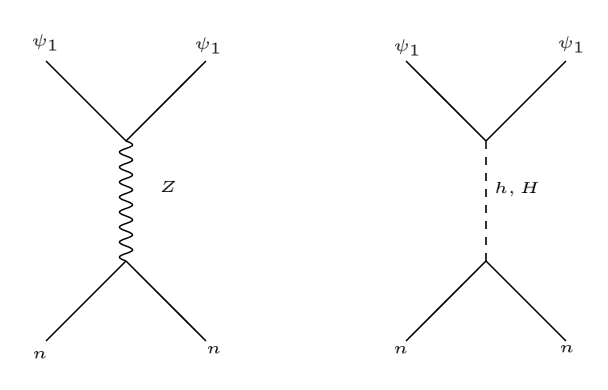} 
$$
\caption{Feynman graph showing DM-nucleon scattering via SM $Z$ (left) and SM-like Higgs ($h$) and heavier Higgs ($H$) (right). Here $n\in n,p$ stands for the nucleons.}
\label{fig:dd}
\end{figure}

In the present framework, the DM exhibits spin-independent interactions with the nuclei induced at the tree-level by the mediation of CP-even states $h_1$ (SM Higgs-like) and $h_2$ (heavier Higgs), and also by the SM $Z$-boson exchange (as in Fig.~\ref{fig:dd}). The relevant cross-section per nucleon reads~\cite{Arcadi:2018pfo}:

\bea
\sigma_{SI}^{h_{1,2}} = \frac{1}{\pi A^2}\mu_r^2 \|\mathcal{M}\|^2,
\label{eq:sigSI1}
\eea

where $A$ is the mass number of the target nucleus, $\mu_r=\frac{m_{\psi_1} m_N}{m_{\psi_1}+m_N}$ is the DM-nucleus reduced mass and $\|\mathcal{M}\|$ is the spin-averaged DM-nucleus scattering amplitude given by:

\bea
\|\mathcal{M}\| = \sum_{i=1,2}\left[Z f_p^i+\left(A-Z\right)f_n^i\right].
\label{eq:siamp1}
\eea

The effective couplings in Eq.~\ref{eq:siamp1} can be expressed as:

\bea
f_{p,n}^i = \sum_{q=u,d,s} f_{T_q}^{p,n} \alpha_q^{i} \frac{m_{p,n}}{m_q}+\frac{2}{27} f_{T_G}^{p,n}\sum_{Q=c,t,b} \alpha_Q^{i}\frac{m_{p,n}}{m_Q},
\label{eq:coupling}
\eea

with

\bea
\alpha_q^{1} = -\frac{\sqrt{2}Y \sin \theta \cos\theta\cos^2\alpha}{\sin\beta~m_{h_1}^2}\frac{m_q\cot\beta}{v_1}\\
\alpha_q^{2} = -\frac{\sqrt{2}Y \sin \theta \cos\theta\sin^2\alpha}{\sin\beta~m_H^2}\frac{m_q\cot\beta}{v_1}, 
\label{eq: alpha}
\eea

where $f_T^{p,n}$ are nucleon form factors. For $Z$-mediated spin-independent direct detection, on the other hand, one can write the scattering cross-section per nucleon as~\cite{Arcadi:2018pfo}:

\bea
\sigma_{SI}^{Z} = \frac{1}{\pi A^2}\mu_r^2 \|\mathcal{M}\|^2,
\label{eq:sigSi2}
\eea

with 

\bea
\|\mathcal{M}\| = \sqrt{2} G_F \left[Z\left(\frac{f_p}{f_n}+\left(A-Z\right)\right)\right] f_n\sin^2\theta,
\label{eq:siamp2}
\eea

\begin{figure}[htb!]
$$
\includegraphics[scale=0.42]{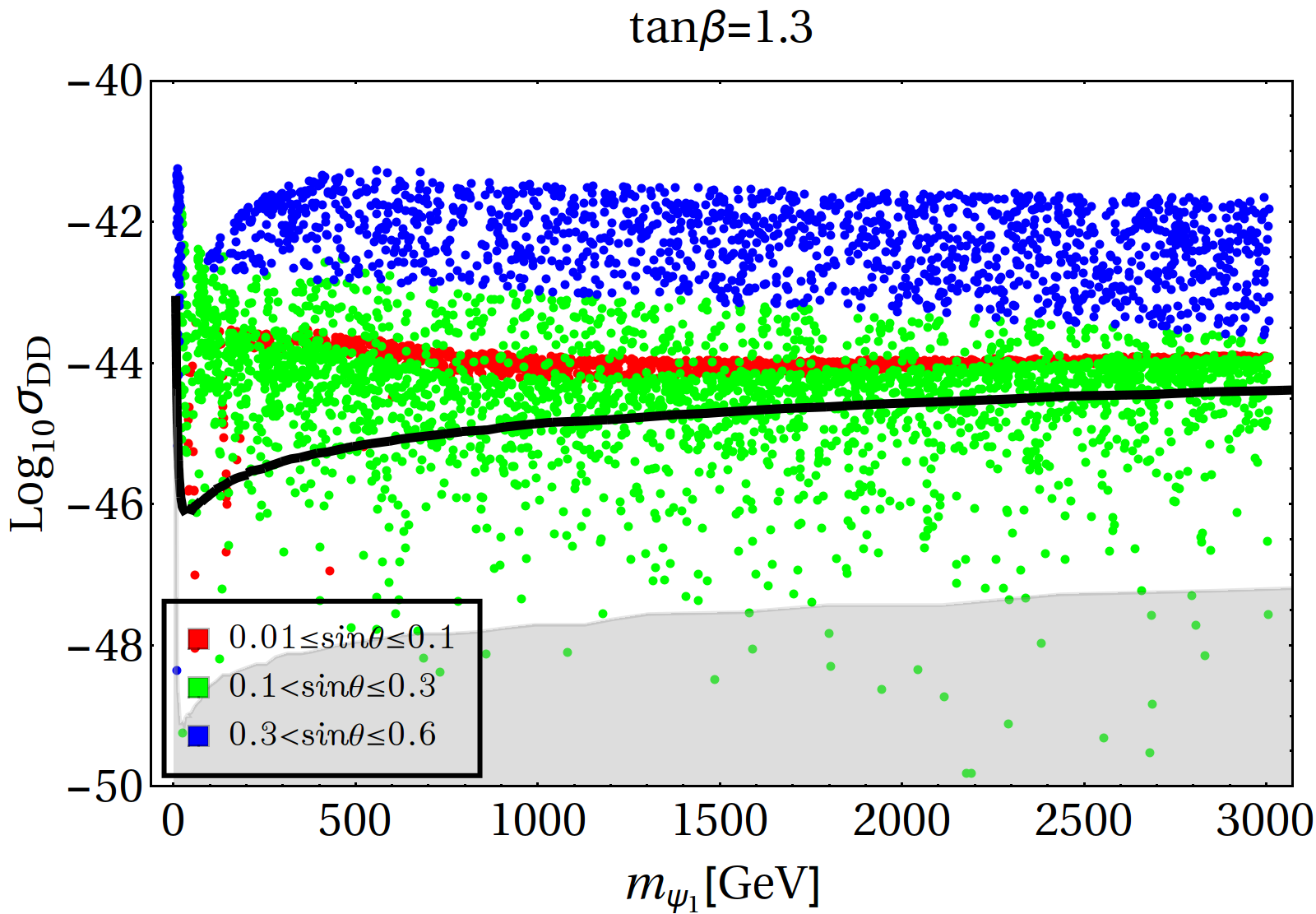}
\includegraphics[scale=0.42]{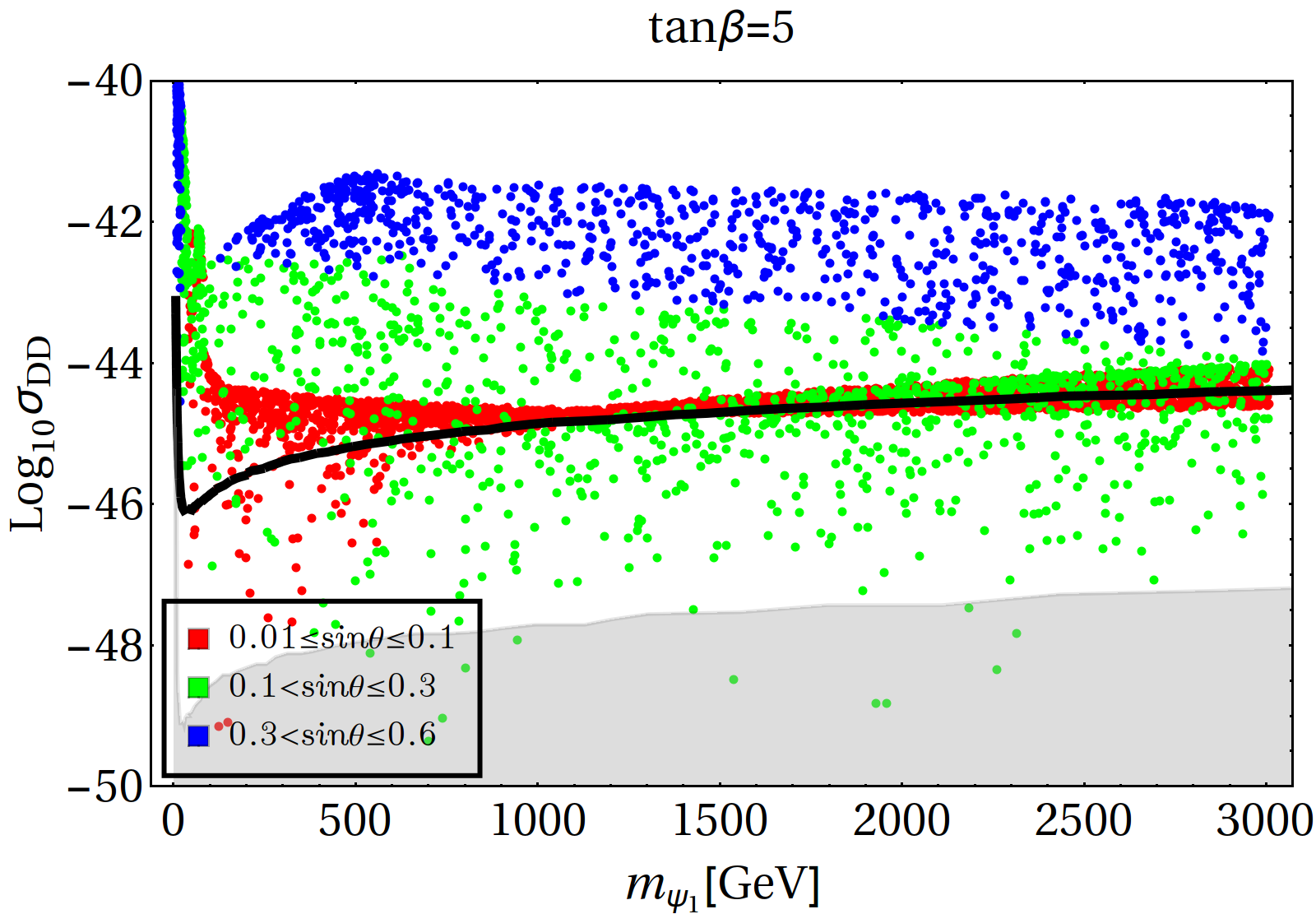}
$$
$$
\includegraphics[scale=0.32]{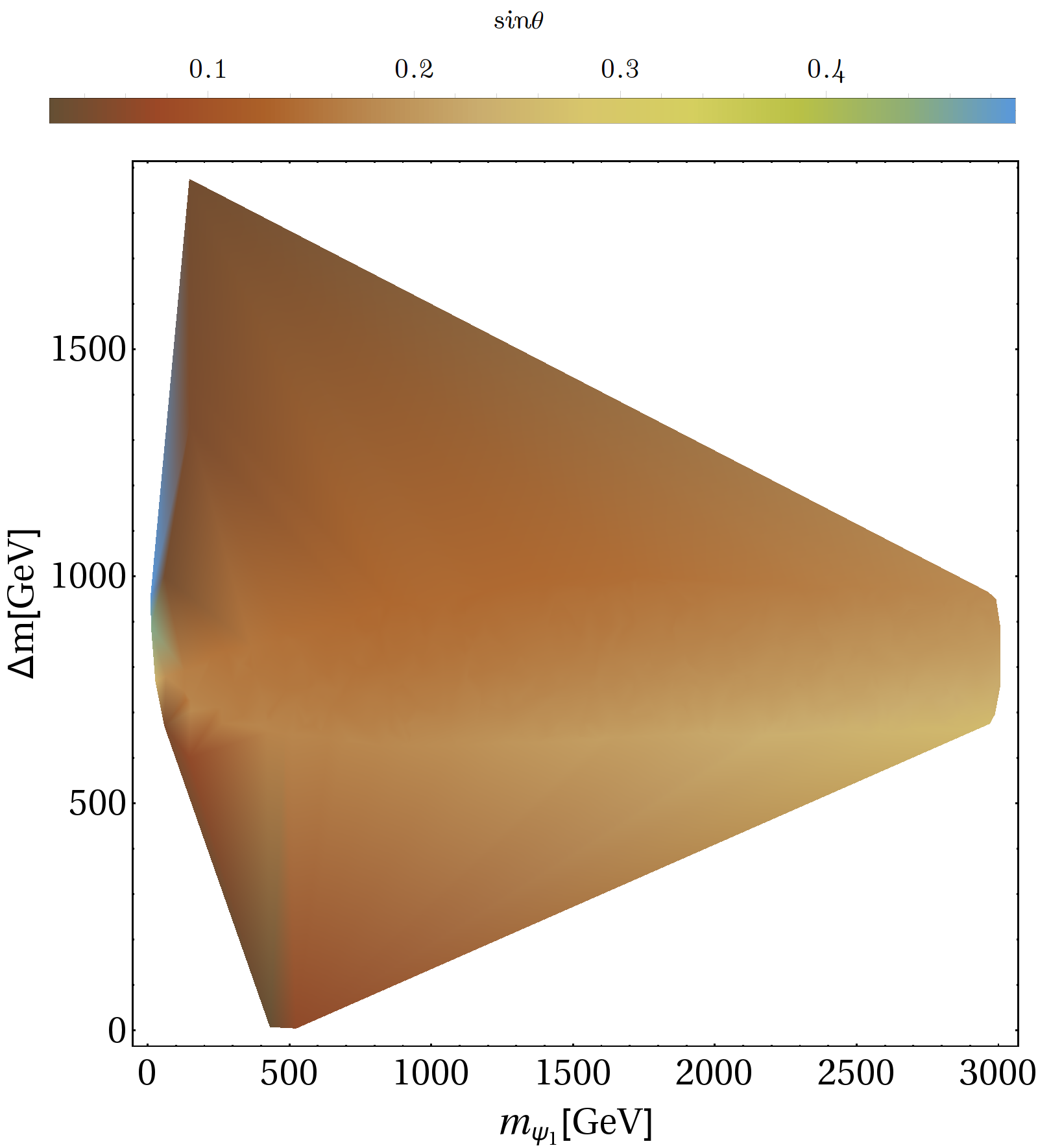}
\includegraphics[scale=0.32]{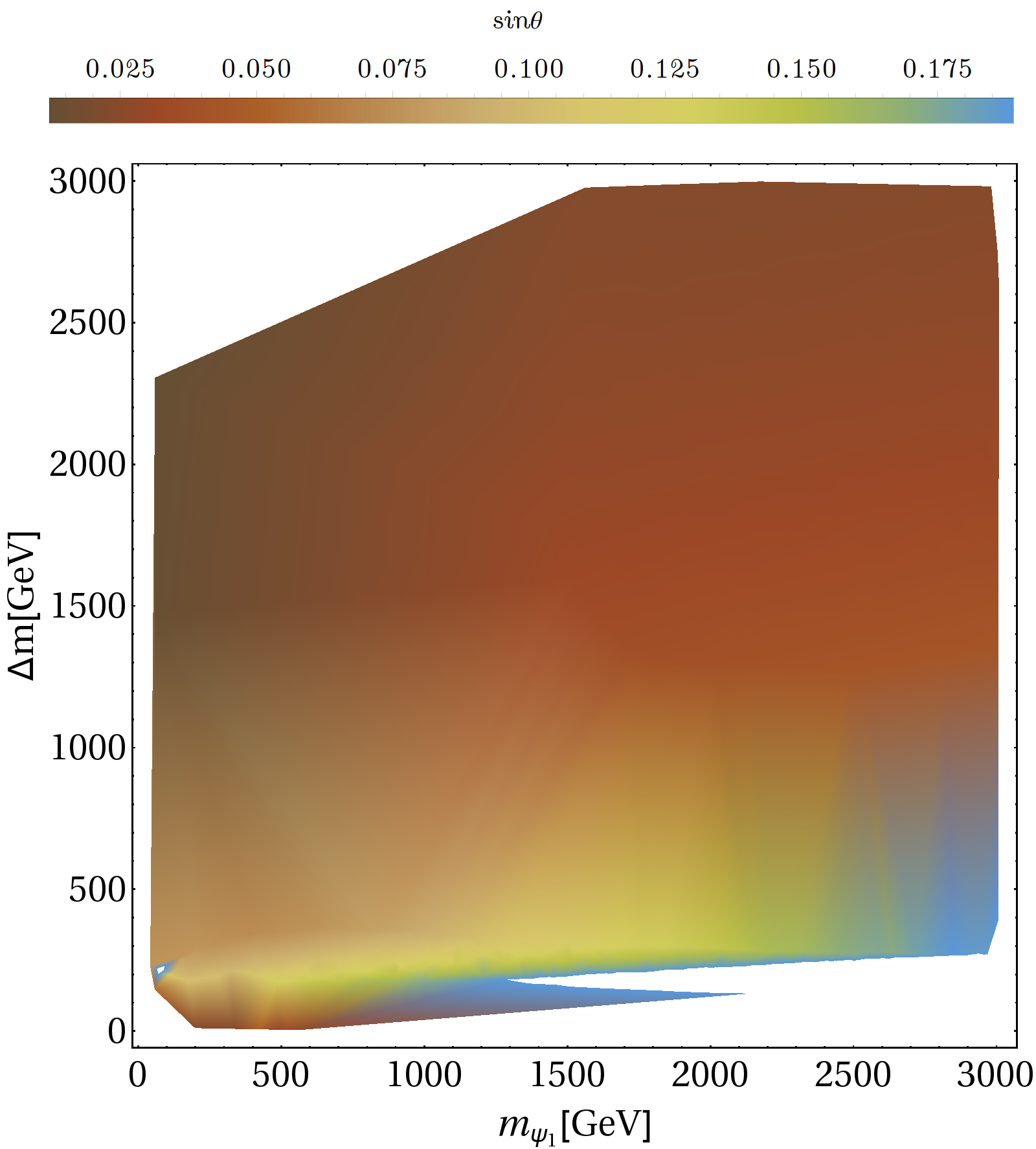}
$$
\caption{Top Left: Relic abundance allowed parameter space in direct search plane for $\tan\beta=1.3$ for different choices of VLF mixing: $0.01\leq\sin\theta\leq 0.1$ in red, $0.1<\sin\theta\leq 0.3$ in green and $0.3<\sin\theta\leq 0.6$ in blue respectively for $\tan\beta=1.3$. Top Right: Same as top left with $\tan\beta=5$ (colour codes remain unchanged). In both the plots the thick black curve is the exclusion limit from XENON1T experiment. Bottom Left: Resulting parameter space satisfying both relic abundance and spin-independent direct detection for $\tan\beta=1.3$ in $m_{\psi_1}$-$\Delta m$ plane, where the colour bar shows different values of $\sin\theta$. Bottom Right: Same as bottom left with $\tan\beta=5$. }\vspace{0.2cm}
\label{fig:ddrelic}
\end{figure}

where $f_{p,n}$ are again suitable nucleon form factors {\footnote{For the numerical values of all form factors we have relied on the default assignations of the {\tt micrOMEGAs} package. }}. For simplicity we can assume conservation of isospin {\it i.e,} $f_p/f_n=1$. Here one should note that the $Z$-mediation also gives rise to spin-dependent direct search cross-section. But the order of magnitude of such spin-dependent cross-section being extremely small compared to that of spin-independent ones we choose to ignore that.{\footnote{To give an order of estimate, the set of data $\{\sin\theta,\Delta m, m_{\psi_1}\}$=\{0.07,~366,~647\}, which gives rise to a correct relic abundance for $\tan\beta=5$, produces a spin-independent direct detection cross-section of $\sim\mathcal{O}(10^{-10})~\rm pb$, compared to a spin-dependent direct search cross-section of $\sim\mathcal{O}(10^{-50})~\rm pb$.}}. 



In the top left panel of Fig.~\ref{fig:ddrelic} we have illustrated the relic density allowed parameter space that survives present spin-independent direct detection bound from XENON1T for $\tan\beta=1.3$. We see a moderate range of $\sin\theta$'s are allowed by direct search: $0.01\lsim\sin\theta\lsim 0.3$. This is expected as the direct detection cross-section is proportional to $\sin^2\theta$ for scalar-mediation and $\sin^4\theta$ for gauge-mediation. As a consequence, smaller mixing should give rise to smaller $\sigma_{SI}$, making the DM parameter space more viable from direct search bound. The presence of the second Higgs helps in keeping the VLF mixing within a moderate limit, unlike the case in~\cite{Bhattacharya:2015qpa} where the bound on the mixing is even more stringent due to the presence of only one (SM) Higgs. This is a consequence of $\frac{\sin^2\alpha}{m_H^2}$ suppression due to the heavier Higgs and small scalar mixing in case of scalar-mediated elastic scattering. This is also possible due to some cancellation between the Higgs-mediated diagrams leading to a destructive interference that allows one to choose $\sin\theta$ as large as $\sim 0.3$ without getting disallowed by the direct search exclusion. Also note here, most of the allowed parameter space lies just above the neutrino floor~\cite{Billard:2013qya} and hence can still be probed by the future direct search experiments with improved sensitivity. In summary, constraints from the requirement of right relic abundance, together with the direct search exclusion limit allows the VLF mixing to vary within a range of $0.01\lsim\sin\theta\lsim 0.3$ for a DM mass starting from around 100 GeV upto 3 TeV. The presence of the second Higgs helps the model to evade present direct search bound and allows the parameter space to fit just above the neutrino floor, leaving the window open to either get discovered or get discarded from the very next limit on spin-independent direct search. Top right panel of Fig.~\ref{fig:ddrelic} shows the same with $\tan\beta=5$.

In the bottom left panel of Fig.~\ref{fig:ddrelic} we show the residual parameter space satisfying both relic abundance and direct detection bounds in $m_{\psi_1}$-$\Delta m$ plane with respect to the variation of the VLF mixing $\sin\theta$ for $\tan\beta=1.3$. This clearly shows that $\Delta m\lsim 1.5~\rm TeV$ (for DM mass $\sim$ 500 GeV) in order to abide by both relic abundance and spin-independent direct detection bounds for $\sin\theta\lsim 0.3$. For larger $\tan\beta$, shown in the right panel of Fig.~\ref{fig:ddrelic}, the bound on $\Delta m$ is bit more relaxed, which allows it to $\sim 2.5~\rm TeV$ but for a larger DM mass. However, the VLF mixing is rather restricted and can be as large as $\sin\theta\sim 0.1$, which helps in suitably choosing the Yukawa $Y$ via Eq.~\ref{eq:vlfyuk}. The bound on $\Delta m$ is crucial as larger $\Delta m$ results in larger missing energy, which, in turn helps the model to be separated from the SM background at the colliders as we shall explain in Sec.~\ref{sec:colpheno}. For smaller $\Delta m$ the model can still be found at the colliders via stable charged track signature.

\begin{table}[htb!]
\begin{center}
\begin{tabular}{|c|c|c|c|c|c|c|c|c|c|c|c|c|}
\hline
Benchmark   &$\sin\theta$ & $\Delta m$ & $m_{\psi_1}$ & $10^3\hat S$ & $10^3\hat T$ &$\tan\beta$ &$m_{12}$& $\sigma_{DD}$ & $\Omega h^2$   \\ [0.5ex] 
Point         &   & (GeV)   &  (GeV)     &   &   &  &(GeV)&$(cm^2)$  &   \\ [0.5ex] 
\hline\hline

BP1   & 0.06 & 966 & 143 & $4.06\times 10^{-2}$ & $2.94\times 10^{-3}$& 1.3 & 176  &  $6.14\times 10^{-47}$ & 0.122  \\
\hline
BP2   & 0.19 & 628 & 1175 & $6.96\times 10^{-1}$&$1.25\times 10^{-2}$ &  1.3 &  185 & $1.50\times 10^{-45}$ & 0.119   \\
\hline
BP3   & 0.05 & 383 & 60 &  $1.21\times 10^{-2}$& $2.19\times 10^{-4}$& 1.3  & 176  &  $5.56\times 10^{-47}$ & 0.121   \\
\hline
BP4   & 0.08 & 369 & 804 &  $1.08\times 10^{-1}$ &$1.14\times 10^{-4}$& 5 & 125 & $1.05\times 10^{-45}$ & 0.121   \\
\hline
BP5   & 0.22 & 218 & 72 &  $1.70\times 10^{-1}$ &$1.73\times 10^{-2}$& 5 &  130 & $9.77\times 10^{-48}$ & 0.121   \\
\hline
BP6   & 0.03 & 10 & 256 &  $6.81\times 10^{-3}$ &$2.03\times 10^{-11}$& 5 &  125 & $6.23\times 10^{-47}$ & 0.121   \\
\hline\hline
\end{tabular}
\end{center}
\caption {Choices of the benchmark points for collider analysis. Masses, mixings, relic density and direct search cross-sections for the DM candidate are tabulated. In each case corresponding value of $S$ and $T$ parameters are also quoted.}\vspace{0.2cm}

\label{tab:bp}
\end{table}

Before moving on to the collider analysis we have tabulated some of the benchmark points (BP) in Tab.~\ref{tab:bp}. These BPs satisfy bounds from relic abundance, direct detection and also those arising from EWPO. The BPs are listed in the decreasing order of $\Delta m$, where BP6 has the minimum $\Delta m$. We have also kept the DM mass $< 2~\rm TeV$ to have a sizeable $\psi^\pm$ production cross-section. As we shall show in Sec.~\ref{sec:colpheno}, BP(1-5) can be distinguished at the LHC as they produce huge missing energy due to large $\Delta m$ that helps to separate them from the SM background. On the other hand, due to very small $\Delta m$ (specifically as $\Delta m<m_W$), BP6 can only produce a displaced vertex at the collider or can be searched at the ILC via missing energy excess as shown in~\cite{Bhattacharya:2017sml,Bhattacharya:2018fus,Barman:2019tuo}.

\section{Collider Phenomenology}
\label{sec:colpheno}

The detailed study of collider signature for vector-like fermions can be found in~\cite{Bhattacharya:2015qpa,Bhattacharya:2017sml,Bhattacharya:2018fus,Barman:2019tuo}. As we have already seen, due to the presence of the second Higgs doublet large $\sin\theta$ can be achieved satisfying both relic density and direct detection. As a result one needs not be confined in small $\Delta m$ and a large $\Delta m$ is also viable. Such large $\Delta m$'s are favorable in order to distinguish this model at the collider from the SM background~\cite{Barman:2019aku,Barman:2019tuo}. It is to be noted that the charged component of $SU(2)_L$ doublet VLF can be produced at the LHC via SM $Z$ and photon mediation. The charged VLF can further decay via on-shell and/or off-shell $W$ (depending on whether $\Delta M\gsim80~\rm GeV$ or $\Delta m\lsim 80~\rm GeV$) to the following final states:

\begin{itemize}
 \item Hadronically quiet opposite sign dilepton (OSD) with missing energy $\left(\ell^{+}\ell^{-}+\slashed{E_T}\right)$.
 \item Single lepton, with two jets plus missing energy $\left(\ell^{\pm}+jj+\slashed{E_T}\right)$.
 \item Four jets plus missing energy $\left(jjjj+\slashed{E_T}\right)$.
\end{itemize}

\begin{figure}[htb!]
$$
\includegraphics[scale=0.32]{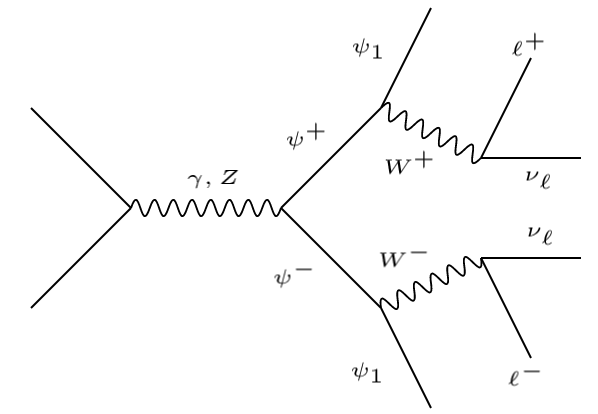} 
\includegraphics[scale=0.35]{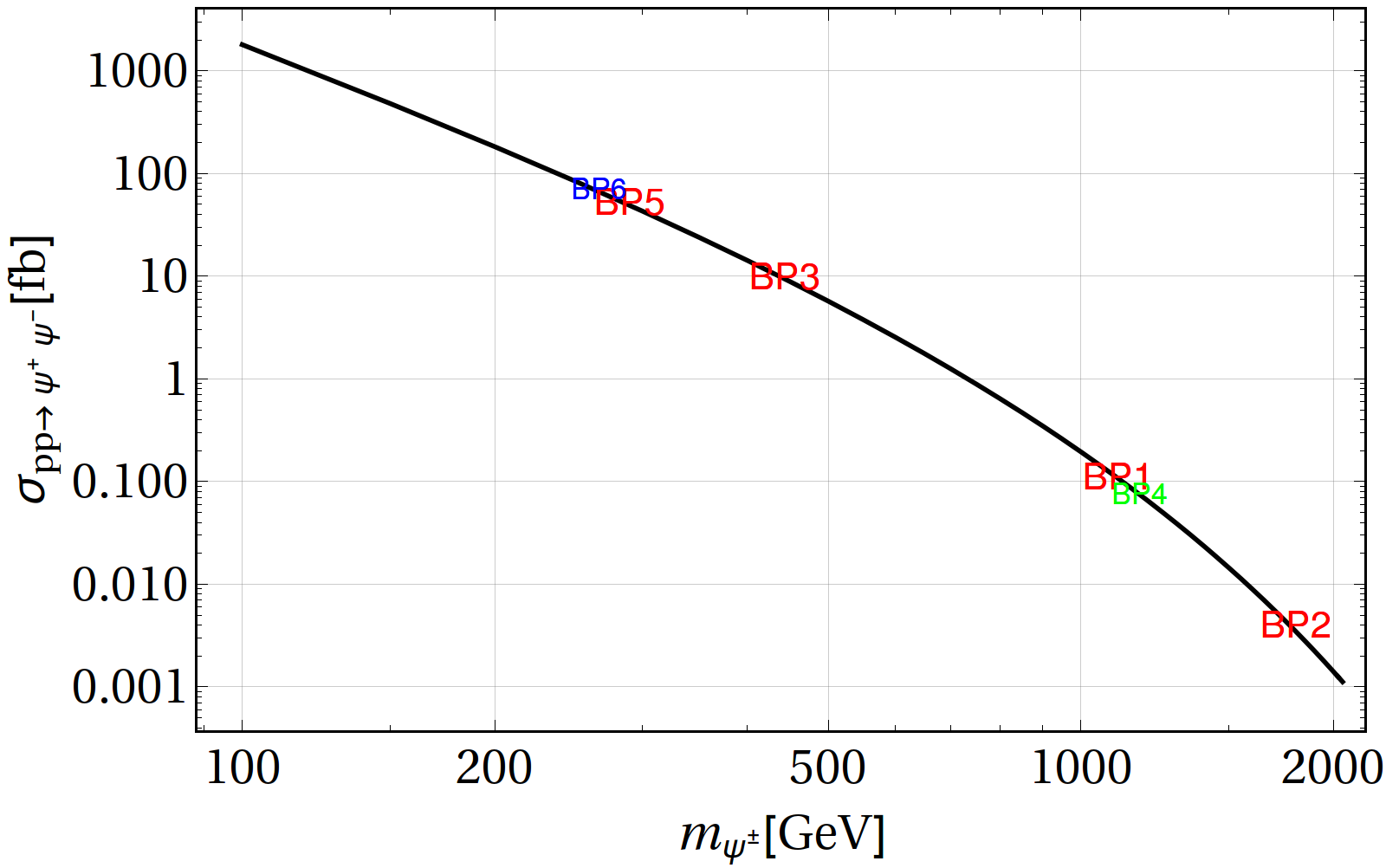} 
$$
\caption{Left: Pair production of charged VLFs and their subsequent decay to OSD+$\slashed{E_T}$ final state. Right: Variation of production cross-section of $\psi^\pm$ with $m_{\psi^\pm}$ for $\Delta m=10~\rm GeV$ and center of mass energy $\sqrt{s}=14~\rm TeV$. The production cross-sections corresponding to different BPs are also shown in red, green and blue on top of the curve.}
\label{fig:prodctn}
\end{figure}

We shall focus only on the leptonic final states (hadronically quiet dilepton) as they are much cleaner compared to others. The Feynman graph for such a process is depicted in the LHS of Fig.~\ref{fig:prodctn}. In the RHS of Fig.~\ref{fig:prodctn} we have shown the variation of pair production cross-section of the charged component of the VLF with VLF mass at $\sqrt{s}=14~\rm TeV$. As one can see, the production cross-section decreases with increase in the charged VLF mass showing the usual nature. We have also shown the position of different BPs on the same plot. As one can notice, BP6 has the highest production cross-section while BP2 has the least. This is evident from the fact that for BP6 $\Delta m=10~\rm GeV$, giving rise to charged VLF mass of $m_{\psi^\pm}=276~\rm GeV$. On the other hand BP2 has a larger $\Delta m$ but highest DM mass, which gives rise to $m_{\psi^\pm}=1803~\rm GeV$ much larger than that for BP4.

\subsection{Object reconstruction and simulation details}
\label{sec:simul}

As already mentioned, we implemented this model in {\tt {\tt LanHEP-3.3.2}} and the parton level events are generated in {\tt CalcHEP-3.7.3}~\cite{Belyaev:2012qa}. Those events are then fed to {\tt PYTHIA-6.4}~\cite{Sjostrand:2006za} for showering and hadronization. The dominant SM backgrounds that can imitate our final state are generated in {\tt MADGRAPH-2.6.6}~\cite{Alwall:2011uj}, and the corresponding production cross-sections are multiplied with appropriate $K$-factor~\cite{Alwall:2014hca} in order to match with the next to leading order (NLO) cross-sections. For all cases we have used {\tt CTEQ6l} as the parton distribution function (PDF)~\cite{Placakyte:2011az}. Now, in order to mimic the collider environment, all the leptons, jets and unclustered objects have been reconstructed assuming the following criteria:

\begin{itemize}
 \item[i)] {\it Lepton ($l=e,\mu$):} Leptons are identified with a minimum transverse momentum $p_T>20$ GeV and pseudorapidity $|\eta|<2.5$. Two leptons can be distinguished as separate objects if their mutual distance in the $\eta-\phi$ plane is $\Delta R=\sqrt{\left(\Delta\eta\right)^2+\left(\Delta\phi\right)^2}\ge 0.2$, while the separation between a lepton and a jet needs to be $\Delta R\ge 0.4$.
 
 \item[ii)] {\it Jets ($j$):} All the partons within $\Delta R=0.4$ from the jet initiator cell are included to form the jets using the cone jet algorithm {\tt PYCELL} built in {\tt PYTHIA}. We demand $p_T>20$ GeV for a clustered object to be considered as jet. Jets are isolated from unclustered objects if $\Delta R>0.4$.  
 
 
 \item[iii)] {\it Unclustered Objects:}  All the final state objects which are neither clustered to form jets, nor identified as leptons, belong to this category. Particles with $0.5<p_T<20$ GeV and $|\eta|<5$ are considered as unclustered. Although unclustered objects do not intervene with our signal definition, but they are important in constructing the missing energy of the events.
 
 \item [iv)] {\it Missing Energy ($\slashed{E}_T$):} The transverse momentum of all the missing particles (those are not registered in the detector) can be estimated from the momentum imbalance in the transverse direction associated to the visible particles. Missing energy (MET) is thus defined as:
 
 \bea
 \slashed{E}_T = -\sqrt{(\sum_{\ell,j} p_x)^2+(\sum_{\ell,j} p_y)^2},
 \eea
 
 where the sum runs over all visible objects that include the leptons, jets and the unclustered components. 
 
 \item [v)] {\it Invariant dilepton mass $\left(m_{\ell\ell}\right)$}: We can construct the invariant dilepton mass variable for two opposite sign leptons by defining:
 \bea
 m_{\ell\ell}^2 = \left(p_{\ell^{+}}+p_{\ell^{-}}\right)^2.
 \eea
 Invariant mass of OSD events, if created from a single parent, peak at the parent mass, for example, $Z$ boson. As the signal events (Fig.~\ref{fig:prodctn}) do not arise from a single parent particle, invariant mass cut plays key role in eliminating the $Z$ mediated SM background. 
 
 \item [vi)] $H_T$: $H_T$ is defined as the scalar sum of all isolated jets and lepton $p_T$'s:
 \bea
 H_T = \sum_{\ell,j} p_T. 	
 \eea
 For our signal the sum only includes the two leptons that are present in the final state. 
 \end{itemize}

 We shall use different cuts on these observables depending on their distribution patterns to separate the signal from the SM backgrounds. Thus we can predict the significance as a function of the integrated luminosity. These are discussed in the following sections.
 
\subsection{Event rates and signal significance}
\label{sec:event}

\begin{table}[htb!]
\begin{center}
\begin{tabular}{|c|c|c|c|c|c|c|c|c|c|}
\hline
Benchmark & $\sigma_{\psi^+\psi^-}$ & $\slashed{E_T}$ &  & $\sigma^{\rm OSD}$  \\ [0.5ex] 
Points   & (fb)  & (GeV)  &  & (fb)    \\ [0.5ex] 
\hline\hline

BP1 & & $>100$ &  & $2.51\times 10^{-2}$ \\
& $1.12\times 10^{-1}$ &$>200$ && $1.97\times 10^{-2}$\\
& &$>300$ && $1.39\times10^{-2}$\\
\cline{5-5}
\cline{1-3}
BP2&  & $>100$ & & $7.43\times 10^{-4}$  \\
& $3.94\times 10^{-3}$ &$>200$ &  & $5.46\times 10^{-4}$  \\
& &$>300$ && $3.47\times 10^{-4}$  \\
\cline{5-5}
\cline{1-3}
BP3& & $>100$ & & $5.60\times 10^{-1}$ \\
& 9.95 &$>200$ & $H_T>$ 300 GeV & $2.01\times 10^{-1}$ \\
&&$>300$ && $6.94\times 10^{-2}$ \\
\cline{5-5}
\cline{1-3}
BP4& & $>100$ & & $8.19\times 10^{-3}$ \\
& $7.77\times 10^{-2}$ &$>200$ &  & $3.75\times 10^{-3}$ \\
&&$>300$ && $1.23\times 10^{-3}$ \\
\cline{5-5}
\cline{1-3}
BP5& & $>100$ & & $5.95\times 10^{-1}$ \\
& 52.26 &$>200$ &  & $1.94\times 10^{-1}$ \\
&&$>300$ && $5.22\times 10^{-2}$ \\
\hline
\end{tabular}
\end{center}
\caption {Variation of final state signal cross-section with MET cut for a fixed cut on $H_T>300~\rm GeV$. All simulations are done at $\sqrt{s}=14~\rm TeV$.  } 
\label{tab:sig1}
\end{table}

\begin{figure}[htb!]
$$
\includegraphics[scale=0.35]{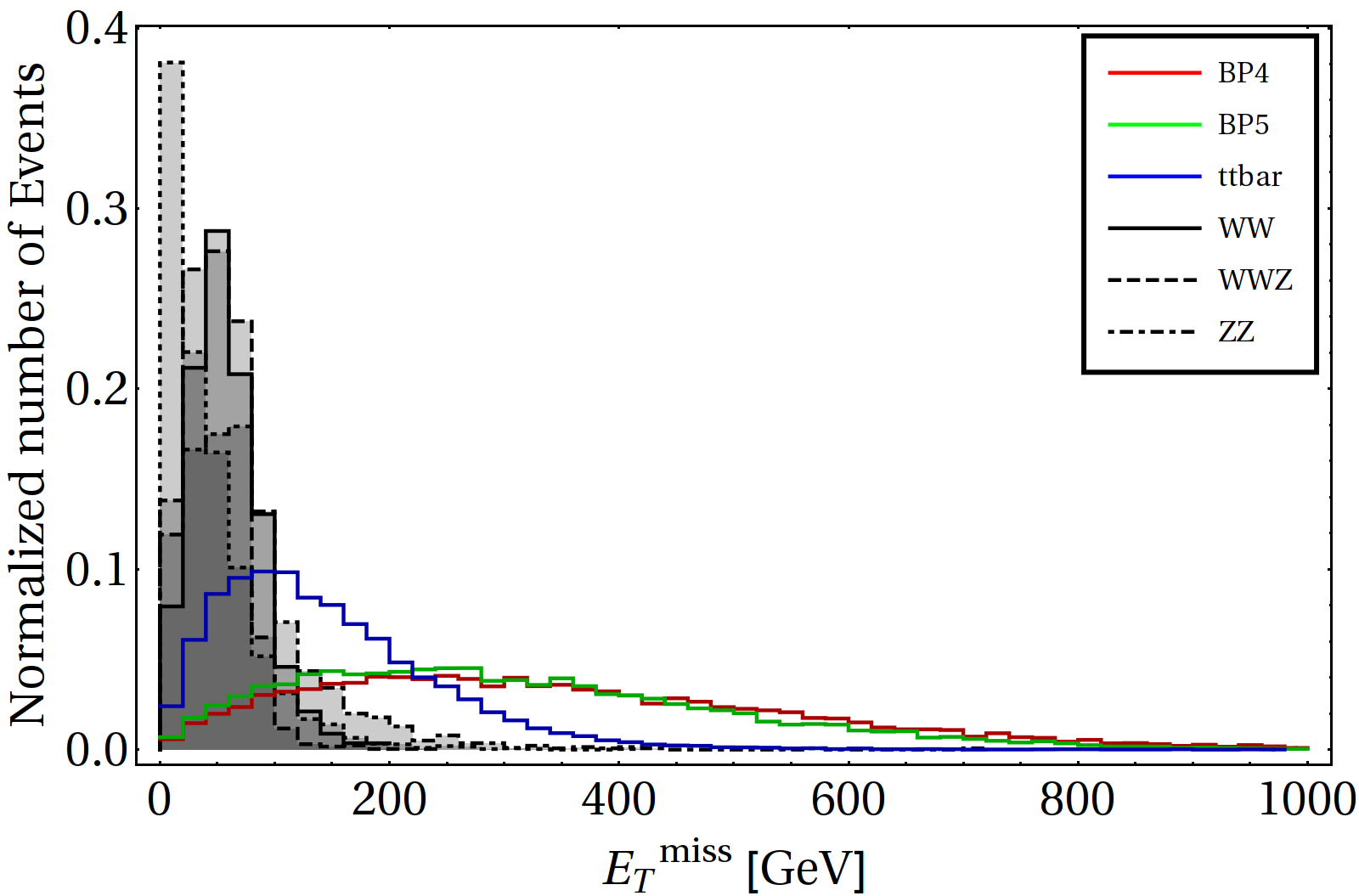} 
\includegraphics[scale=0.35]{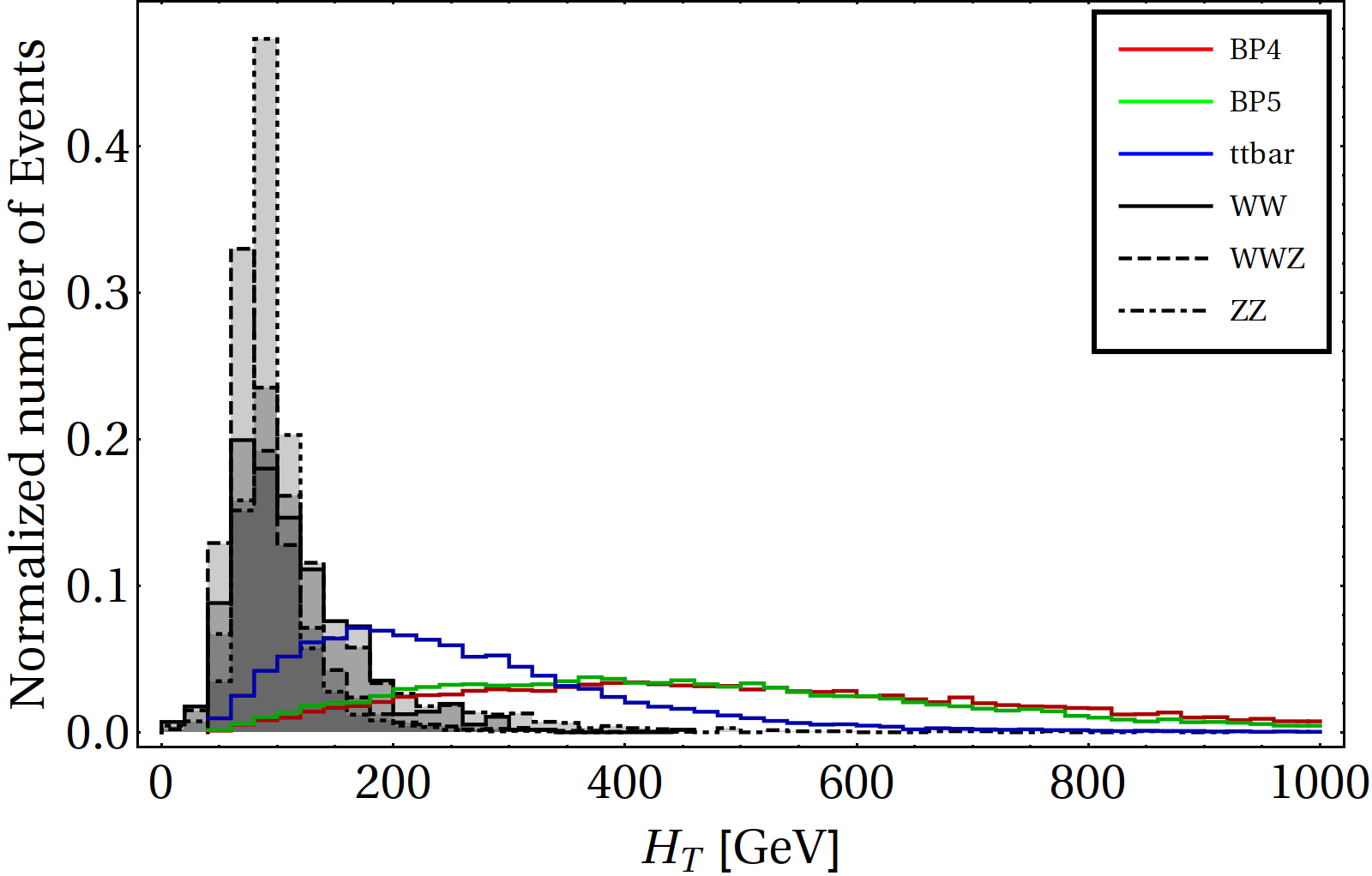} 
$$
$$
\includegraphics[scale=0.35]{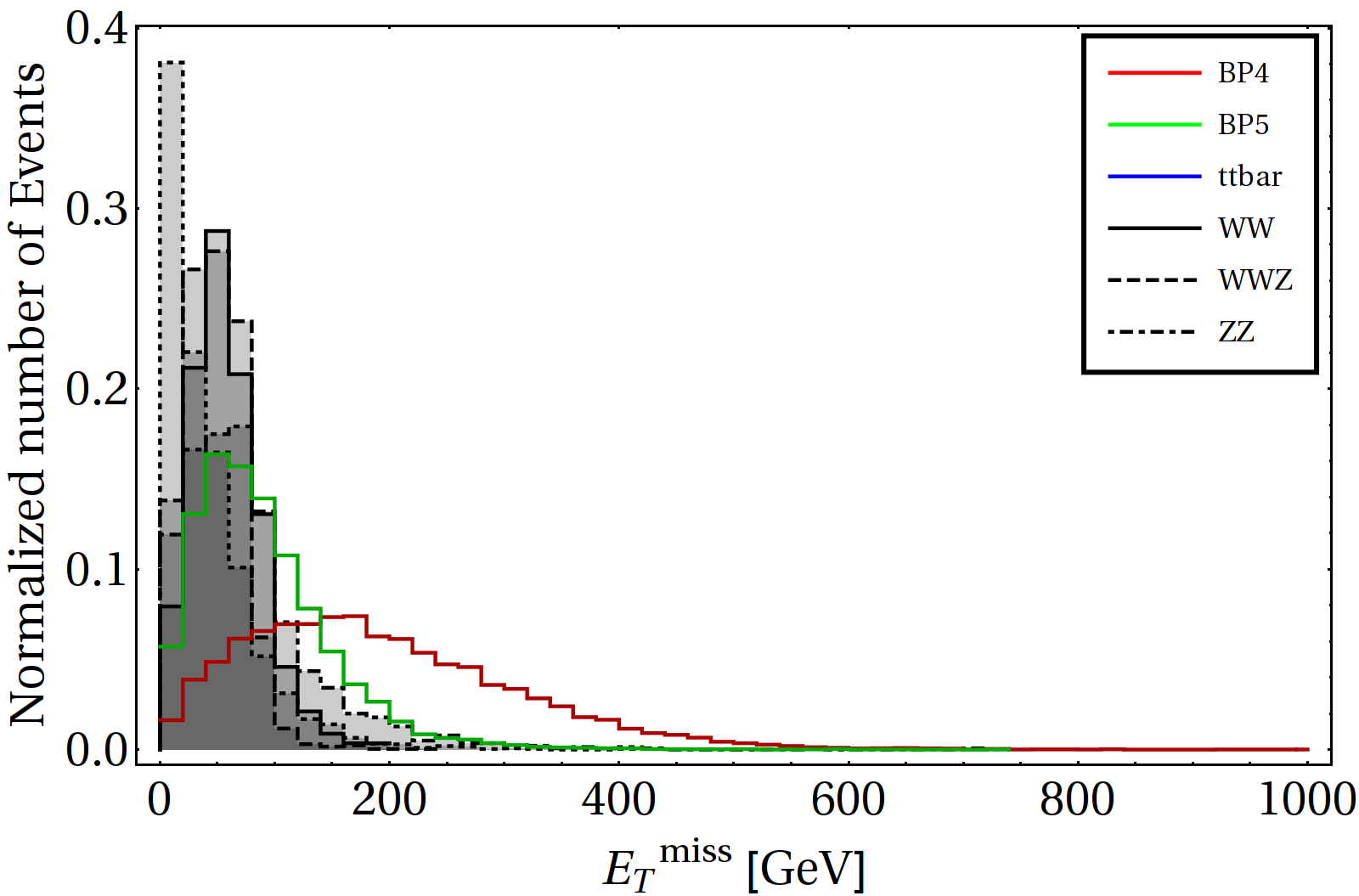} 
\includegraphics[scale=0.35]{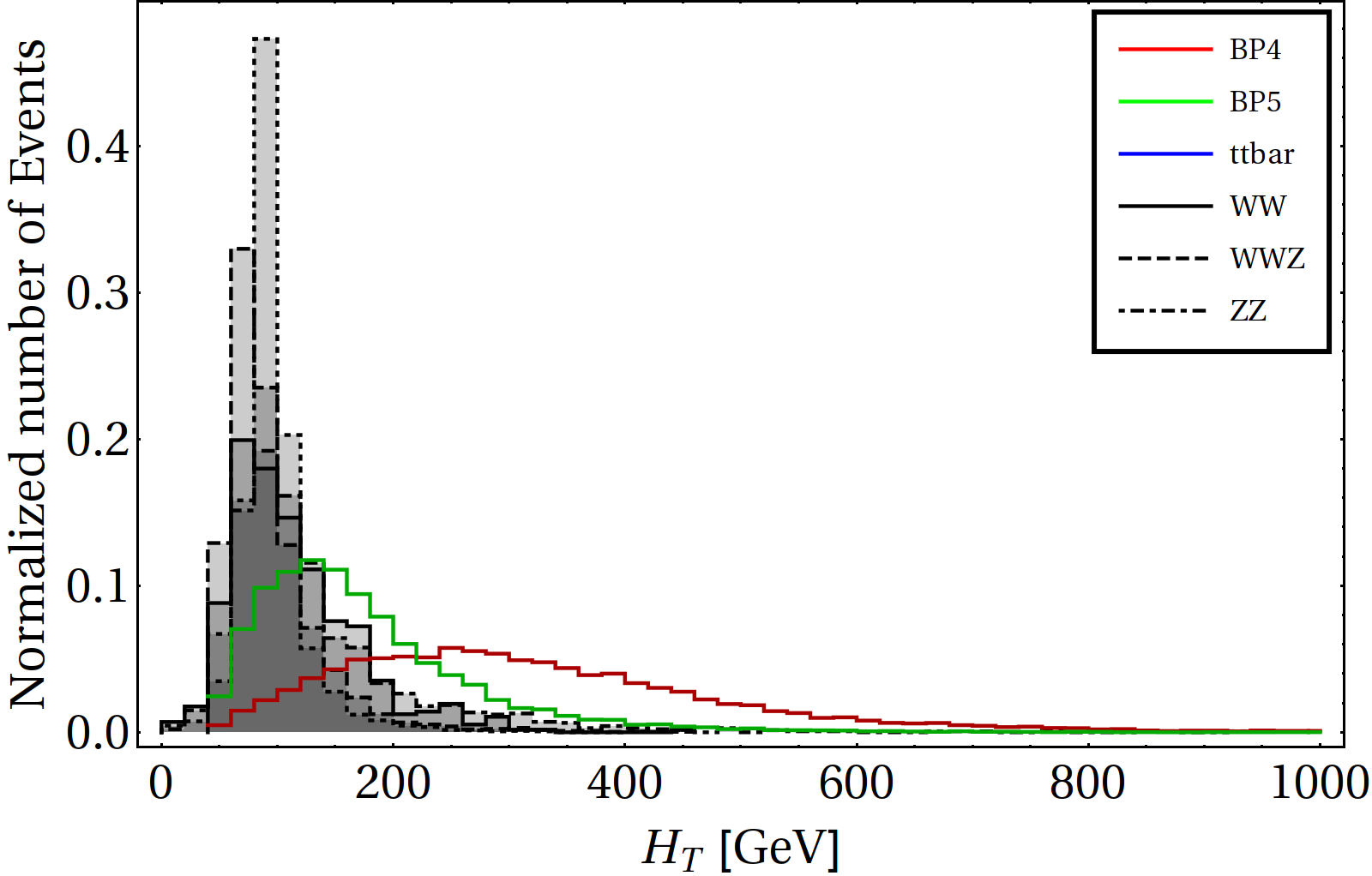} 
$$
\caption{Top Left: Distribution of normalized number of signal and background events with MET for BP(1,2,3). Top Right: Distribution of normalized number of events with $H_T$ for BP(1,2,3). Bottom Left: Same as top left but for BP(4,5). Bottom Right: Same as top right but for BP(4,5). In all cases the black and gray histograms correspond to the dominant SM backgrounds. All simulations are done at $\sqrt{s}=14~\rm TeV$ with {\tt CTEQ6l} as parton distribution function. Note that we do not show the corresponding distributions for BP6 since due to very small $\Delta m$ it is indistinguishable from the SM backgrounds.}\vspace{0.2cm}
\label{fig:metht}
\end{figure}

\begin{table}[htb!]
\begin{center}
\begin{tabular}{|c|c|c|c|c|c|c|c|c|c|}
\hline
Processes & $\sigma_{\rm production}$ & $\slashed{E_T}$ & & $\sigma^{\rm OSD}$  \\ [0.5ex] 
 & (pb) & (GeV)  &&  (fb)    \\ [0.5ex] 
\hline\hline

$t\bar t$ & & $>100$ && 0 \\
 & 814.64  & $>200$ && 0 \\
&& $>300$ && 0 \\
\cline{5-5}
\cline{1-3}
$W^+W^-$ && $>100$ &  & 2.99 \\
& 99.98 & $>200$ && 1.49 \\
&& $>300$ && 0 \\
\cline{5-5}
\cline{1-3}
$W^+W^-Z$ && $>100$ & $H_T>$ 300 GeV & 0.039 \\
& 0.15 & $>200$ && 0.024 \\
&& $>300$ && 0.012 \\
\cline{5-5}
\cline{1-3}
$ZZ$ && $>100$ && 0\\
& 14.01 &$>200$&& 0 \\
&&$>300$&& 0 \\
\hline
\hline
\end{tabular}
\end{center}
\caption {Variation of final state SM background cross-section with MET cut for a fixed cut on $H_T>250~\rm GeV$. All simulations are done at $\sqrt{s}=14~\rm TeV$.} 
\label{tab:bck}
\end{table}

In Fig.~\ref{fig:metht} we have shown the distribution of normalised number of events with respect to MET (LHS) and $H_T$ (RHS) for all the chosen BPs. In the same plot we have also shown the distribution from dominant SM backgrounds that can mimic our signal. For the SM the only source of MET are the SM neutrinos, which are almost massless with respect to centre of mass energy of the collider. As a result, the MET and $H_T$ distribution for SM peaks up at a lower value, while for the model on top of the SM neutrinos MET arises from the DM $\psi_1$ which is massive, and hence corresponding distribution for the signals are much flattened. The notable feature in these plots is the fact that for larger $\Delta m$ the signal distributions are well separated from that of the background. This is due to the fact that the peak of the MET distribution is determined by how much of $p_T$ is being carried away by the missing particle ({\it i.e,} the DM), which in turn depends on the mass difference of charged and neutral component of the VLF {\it i.e, $\Delta m$}. Hence for larger $\Delta m$ the DM carries away most of the $p_T$ making the distribution much flatter, while for smaller $\Delta m$ the distribution peaks up at lower value as the produced DM particles are not boosted enough. As a consequence, BP1 and BP2 have the most flattened distribution, while BP(3,4,5) are increasingly overwhelmed by the SM background. Since BP6 has the smallest $\Delta m$, we refrain from showing this in the plots as BP6 will be inseparable from the SM backgrounds. From these distributions it is quite evident that with a cut on MET $\gsim 200~\rm GeV$ and on $H_T\gsim 300~\rm GeV$ one can get rid of the SM backgrounds retaining most of the signals. This is also reflected in Tab.~\ref{tab:bp} and Tab.~\ref{tab:bck} where we have tabulated the cross-section corresponding to final state OSD+$\slashed{E_T}$ events for the BPs and for the SM backgrounds respectively. In order to understand the effectiveness of the choice of our cuts we have shown the cut flow for an increasing choice of $\slashed{E_T}$, keeping the $H_T$ fixed. Note that, for signal (Tab.~\ref{tab:bp}) the final state cross-section in each case gradually diminishes with increase in the MET cut {\it i.e,} harder the cut lesser is the cross-section. The same is also true for the SM backgrounds as shown in Tab.~\ref{tab:bck}. Note that, for the backgrounds, the most dominant one {\it i.e,} $t\bar t$ is completely killed by the zero jet vetto. Amongst the rest $WW$ and $WWZ$ still exist for $\slashed{E_T}>200~\rm GeV$, but can still be put to zero by a harder MET cut. $ZZ$, on the other hand, is again completely eliminated thanks to the invariant mass cut over a window of $|m_Z\pm 50|~\rm GeV$ around the central value of $Z$ mass. Thus, a wise choice of the cuts on the observables can help to completely get rid of the SM backgrounds while retaining most of the signal. 

\begin{figure}[htb!]
$$
\includegraphics[scale=0.45]{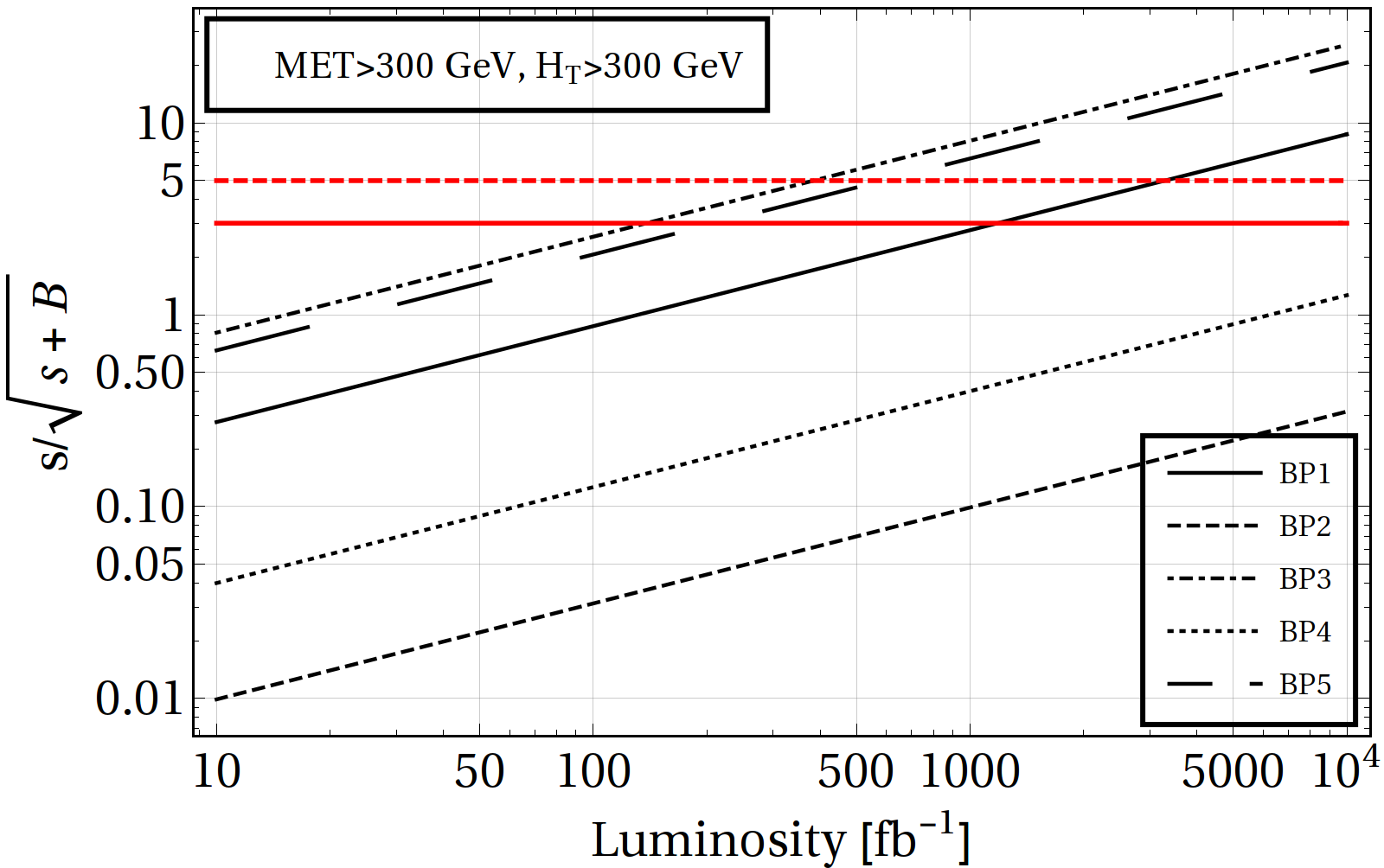} 
$$
\caption{Significance plotted against integrated luminosity for the three benchmarks in Tab.~\ref{tab:bp}. The solid, dotted and dotdashed black lines correspond different benchmark points. The solid red and dashed red lines denote the 3$\sigma$ and 5$\sigma$ confidence respectively.}\vspace{0.2cm}
\label{fig:signi}
\end{figure}

This can be translated into the signal significance for this model which is shown in Fig.~\ref{fig:signi}. Here we have plotted the significance of the chosen BPs with respect to the integrated luminosity. In order to minimize the SM background we have employed $\slashed{E_T}>300~\rm GeV$ and $H_T>300~\rm GeV$ to compute the significance. We see, of all the BPs, BP(2,4) has the least possibility to be probed even with a luminosity as high as $3000~\rm fb^{-1}$. A large $\Delta m$ in one way helps to distinguish BP(2,4) from the SM background, but due to very small cross-section at the production level the final state cross-section becomes even smaller. This makes BP(2,4) least significant. For BP(1,3,5), on the other hand, $\Delta m$ is optimum such that they can be separated from the background because of their larger missing energy, and at the same time the cross-section is large enough so that a 5$\sigma$ significance can be achieved. Thus, we see, BP3 and BP5 reach a discovery limit at a lower luminosity $\sim 500~\rm fb^{-1}$, while BP1 needs $\sim 1000~\rm fb^{-1}$ for a 5$\sigma$ reach. Finally, we would like to mention that for all benchmarks with $\Delta m\lsim m_W$ ({\it e.g,} BP6) one can find a stable charged track due to the off-shell decay of the heavy charged VLF $\psi^\pm$ via $W$-boson or they can also be probed at the ILC with a lower cut on MET. These scenarios have already been thoroughly investigated in~\cite{Barman:2019tuo,Bhattacharya:2017sml}, hence we do not discuss here further.

\section{Strong First-Order Electroweak Phase transition and Gravitational Wave Signals}
\label{sec:gw}

In this section we would like to show the possibility of generation of stochastic gravitational wave (GW) from a strong first-order phase transition (SFOPT). The frequency of such GWs are well within the reach of the proposed GW detectors. The occurrence of a SFOPT and subsequent GW generation in context with 2HDM has already been thoroughly studied~\cite{Wang:2019pet,Bernon:2017jgv}. In the context of our present model we explore how such a detectable GW signal can be an alternate search strategy for singlet-doublet DM. The dynamics of the SFOPT is completely determined by the parameters of the scalar potential as we shall see in the following sections. It is interesting to note that the choice of the scalar potential parameters agrees well with both the DM phenomenology and the GW generation, thus giving us a handle to probe the dark sector beyond DM and collider search experiments.

\subsection{Finite Temperature Effective Potential}
\label{sec:veff}

In order to explore the electroweak phase transition (EWPT) in 2HDM we need to include temperature corrections with the tree-level potential. In general, the finite temperature effective potential at a temperature $T$ can be expressed as~ \cite{Wainwright:2011kj}

\begin{equation}\label{eq:41}
V_{\text{eff}}=V_{\text{}tree}+V_{1-loop}^{T=0}+V_{1-loop}^{T\neq0},
\end{equation}
where $V_{\text{}tree}$, $V_{1-loop}^{T=0}$ and $V_{1-loop}^{T\neq0}$ are the tree-level potential at zero temperature, the Coleman-Weinberg one-loop effective potential at zero temperature and one-loop effective potential at finite temperature respectively. The tree-level potential $V_{\text{}tree}$ can be obtained from Eq.~\ref{treepot} by replacing the fields $\Phi_1$ and $\Phi_2$ with their classical fields $v_1$ and $v_2$ which is given by
\bea
V_{\text{}tree} &=&
\dfrac{1}{2} m^2_{11}\, v_1^2
+\dfrac{1}{2} m^2_{22}\, v_2^2 -
\dfrac{1}{2} m^2_{12}\, v_1 v_2
+\dfrac{1}{8} \lambda_1 v_1^4
+\dfrac{1}{8} \lambda_2 v_2^4
+\dfrac{1}{4} \left(\lambda_3 +\lambda_4 +\lambda_5\right)\, v_1^2 v_2^2 .
\label{tpot}
\eea

The Coleman-Weinberg one-loop effective potential at zero temperature $V_{1-loop}^{T=0}$ can be written as \cite{Wainwright:2011kj, Coleman:1973jx}
\begin{equation}\label{eq:CW}
V_{1-loop}^{T=0}=\pm \dfrac{1}{64\pi^2} \sum_{i} n_i m_i^4 \left[ \log\dfrac{m_i^2}{Q^2}-C_i\right],
\end{equation}
where the `+' sign corresponds to bosons and `-' sign corresponds to fermions. The sum $i$ is over the  Goldstone bosons $G^\pm$, $G$, $A$, $H^{\pm}$, Higgs bosons $h$, $H$, gauge bosons $W^\pm$, $Z$ and the top fermion $t$. The field-dependent squared masses $m_i^2$ for the top quark and gauge bosons at $T=0$ are given by

\begin{equation}\label{eq:fdm1}
m_t^2=\dfrac{1}{2}y_t^2 v_2^2/s_\beta^2 \,\,,
\end{equation}

\begin{equation}\label{eq:fdm2}
m_W^2=\dfrac{1}{4}g^2\left(v_1^2+v_2^2\right)\,\,,
\end{equation}

\begin{equation}\label{eq:fdm3}
m_Z^2=\dfrac{1}{4}\left(g^2+{g^{\prime}}^2\right)\left(v_1^2+v_2^2\right)\,\,,
\end{equation}
where $y_t$, $g$ and $g^{\prime}$ are the top Yukawa coupling, $SU(2)_L$ and $U(1)_Y$ gauge couplings of the SM respectively.

The field-dependent squared masses $m_i^2$ at $T=0$ for the scalar bosons can be obtained by diagonalizing the following matrices
\bea
\quad
m_{h,H}^2=
\begin{pmatrix} 
m_{11}^2+\dfrac{3}{2}\lambda_1 v_1^2+ \dfrac{1}{2}\left(\lambda_3 +\lambda_4 +\lambda_5\right) v_2^2 & -\dfrac{1}{2}m_{12}^2 +\left(\lambda_3 +\lambda_4 +\lambda_5\right) v_1 v_2  \\
-\dfrac{1}{2}m_{12}^2 +\left(\lambda_3 +\lambda_4 +\lambda_5\right) v_1 v_2 & m_{22}^2+\dfrac{3}{2}\lambda_2 v_2^2+ \dfrac{1}{2}\left(\lambda_3 +\lambda_4 +\lambda_5\right) v_1^2
\end{pmatrix}
\quad,
\label{eq:fdm4}
\eea

\bea
\quad
m_{G,A}^2=
\begin{pmatrix} 
m_{11}^2+\dfrac{1}{2}\lambda_1 v_1^2+ \dfrac{1}{2} \lambda_3 v_2^2 & -\dfrac{1}{2}m_{12}^2 + \dfrac{1}{2}\left(\lambda_4 +\lambda_5\right) v_1 v_2  \\
-\dfrac{1}{2}m_{12}^2 +\dfrac{1}{2}\left(\lambda_4 +\lambda_5\right) v_1 v_2 & m_{22}^2+\dfrac{1}{2}\lambda_2 v_2^2+ \dfrac{1}{2} \lambda_3 v_1^2
\end{pmatrix}
\quad,
\label{eq:fdm5}
\eea

\bea
\quad
m_{G^\pm,H^\pm}^2=
\begin{pmatrix} 
2 m_{11}^2+\lambda_1 v_1^2+\lambda_3 v_2^2 & -m_{12}^2 +\left(\lambda_4 +\lambda_5\right) v_1 v_2  \\
-m_{12}^2 +\left(\lambda_4 +\lambda_5\right) v_1 v_2 & 2 m_{22}^2+\lambda_2 v_2^2+\lambda_3 v_1^2
\end{pmatrix}
\quad.
\label{eq:fdm6}
\eea 

Here we have applied Landau gauge, where the Goldstones are massless at zero temperature ($T=0$)  but at finite temperature ($T\neq0$) they acquire a mass~\cite{Basler:2016obg}. In Eq. \ref{eq:CW}, $C_i$'s the are renormalisation-scheme-dependent numerical constant, $Q$ is a renormalizable scale and $n_i$'s are the number of degrees of freedom (DOF). For the gauge bosons ($W, \hspace{1mm} Z$) $C_{W,Z}=5/6$ and for the other particle species $C_{h,H,G,A,H^+,H^-,G^+,G^-,t}=3/2$ with the corresponding DOFs: $n_{W^{\pm}}=6$, $n_Z=3$, $n_{h,H,G,A,H^+,H^-,G^+,G^-}=1$ and $n_t=12$. The one-loop finite temperature effective potential $V_{1-loop}^{T\neq0}$ (Eq. (\ref{eq:CW})) reads \cite{Wainwright:2011kj}
\begin{equation}\label{eq:43}
V_{1-loop}^{T\neq0}=\dfrac{T^2}{2\pi^2} \sum_{i} n_i J_{\pm}\left[\dfrac{m_i^2}{T^2}\right],
\end{equation}
where the function $J_{\pm}$ are
\begin{equation}\label{eq:44}
J_{\pm}\left(\dfrac{m_i^2}{T^2}\right)=\pm \int_0^{\infty} dy \hspace{1mm} y^2 \log\left(1\mp e^{-\sqrt{y^2+\dfrac{m_i^2}{T^2}}}\right).
\end{equation}
In the finite temperature effective potential $V_{1-loop}^{T\neq0}$ we include the temperature corrected terms to the boson masses by following Daisy resummation method \cite{Arnold:1992rz}. In the Daisy resummation method the thermal masses  are \cite{Huang:2017rzf}, \cite{Blinov:2015vma, Vieu:2018nfq, Gil:2012ya}
$\mu_{1}^2(T)=m_{11}^2+c_1 T^2$ and $\mu_{2}^2(T)=m_{22}^2+c_2 T^2$,
where 
\begin{equation}
c_1=\dfrac{3\lambda_1+2\lambda_3+\lambda_4}{12}+\dfrac{3g^2+{g^{\prime}}^2}{16}+\dfrac{y_t^2}{4}\,\,,
\end{equation}

\begin{equation}
c_2=\dfrac{3\lambda_2+2\lambda_3+\lambda_4}{12}+\dfrac{3g^2+{g^{\prime}}^2}{16}\,\,.
\end{equation}

\subsection{Gravitational Wave from SFOPT}\label{sec:gwformalism}

The central idea of first order phase transition (FOPT) is the bubble nucleation of a true vacuum state (from several metastable states) at a temperature commonly known as the nucleation temperature. The bubbles produced in this process can be of different sizes: small and large. The smaller bubbles tend to collapse, whereas the larger bubbles tend to expand after attaining the criticality. These bubbles of critical size then collide with each other and their spherical symmetry is thus broken. This initiates the phase transition and subsequent production of the GW. The bubble nucleation rate per unit volume at a temperature $T$ can be expressed as~\cite{Linde:1981zj}

\begin{equation}\label{eq:45}
\Gamma=\Gamma_0\left(T \right) e^{-S_3\left(T \right)/T}
\end{equation}

where $\Gamma_0\left(T \right) \propto T^4$ and $S_3\left(T \right)$ denotes the Euclidean action of the critical bubble~\cite{Linde:1981zj}:

\begin{equation}\label{eq:46}
S_{3}=4\pi \int dr \hspace{1mm}r^{2} \left[ \dfrac{1}{2} \left(\partial_{r} \vec{\phi} \right)^2 +V_{eff}\right],
\end{equation}
where 
$V_{eff}$ is the effective finite temperature potential (Eq. (\ref{eq:41})). Bubble nucleation occurs at the nucleation temperature $T_n$ which satisfies the condition:  $S_3\left(T_n \right)/T_n\approx 140$ \cite{Wainwright:2011kj}.

As mentioned in the Sec.~\ref{sec:intro}, GWs are produced from the FOPT majorly via three mechanisms namely, bubble collisions \cite{Kosowsky:1991ua,Paul:2019pgt,Kosowsky:1992vn,Huber:2008hg,Kosowsky:1992rz,Kamionkowski:1993fg,Caprini:2007xq}, sound wave \cite{Hindmarsh:2013xza,Giblin:2013kea,Giblin:2014qia,Hindmarsh:2015qta} and turbulence in the plasma \cite{Caprini:2006jb,Kahniashvili:2008pf,Kahniashvili:2008pe,Kahniashvili:2009mf,Caprini:2009yp}. The total GW intensity $\Omega_{\text{GW}}{\rm h}^2$ as a function of frequency can be expressed as the sum of the contributions from the individual components\cite{Kosowsky:1991ua,Paul:2019pgt,Kosowsky:1992vn,Huber:2008hg,Kosowsky:1992rz,Kamionkowski:1993fg,Caprini:2007xq,Hindmarsh:2013xza,Giblin:2013kea,Giblin:2014qia,Hindmarsh:2015qta,Caprini:2006jb,Kahniashvili:2008pf,Kahniashvili:2008pe,Kahniashvili:2009mf,Caprini:2009yp}: 
\begin{equation}\label{eq:47}
\Omega_{\text{GW}}{\rm h}^2=\Omega_{\text{col}}{\rm h}^2+ \Omega_{\text{SW}}{\rm h}^2+ \Omega_{\text{turb}}{\rm h}^2.
\end{equation}

The component from the bubbles collision $\Omega_{\text{col}}{\rm h}^2$ is given by (for an analytic and more accurate derivation see~\cite{Jinno:2016vai,Jinno:2017fby}):

\begin{equation}\label{eq:48}
\Omega_{\text{col}}{\rm h}^2=1.67\times 10^{-5} \left(\dfrac{\beta^{\prime}}{H} \right) ^{-2} \dfrac{0.11 v_{w}^3}{0.42+v_{w}^2} \left(\dfrac{\kappa \alpha^{\prime}}{1+\alpha^{\prime}}\right)^2 \left(\dfrac{g_*}{100}\right)^{-\frac{1}{3}}\dfrac{3.8 \left(\dfrac{f}{f_{col}}\right)^{2.8}}{1+2.8 \left(\dfrac{f}{f_{\text{col}}}\right)^{3.8}}\,\,\,,
\end{equation}
where the parameter $\beta^{\prime}$
\begin{equation}\label{eq:49}
\beta^{\prime}=\left[H T \dfrac{d}{dT}\left( \dfrac{S_3}{T}\right) \right]\bigg|_{T_n},
\end{equation}

where $T_n$ is the nucleation temperature and $H_n$ is the Hubble parameter at $T_n$. The most general expression of the bubble wall velocity $v_w$ can be written as ~\footnote{A more detailed discussion on the choice of $v_w$ can be found in \cite{Kozaczuk:2015owa}} \cite{Kamionkowski:1993fg,Steinhardt:1981ct}

\begin{equation}\label{eq:50}
v_w=\dfrac{1/\sqrt{3}+\sqrt{{\alpha^{\prime}}^2+2\alpha^{\prime}/3}}{1+\alpha^{\prime}} .
\end{equation}
The parameter $\kappa$ in Eq. (\ref{eq:48}) is the fraction of latent heat deposited in a thin shell which can be expressed as:

\begin{equation}\label{eq:51}
\kappa=1-\dfrac{\alpha^{\prime}_{\infty}}{\alpha^{\prime}},
\end{equation}

with \cite{Shajiee:2018jdq,Caprini:2015zlo}

\begin{equation}\label{eq:52}
\alpha^{\prime}_{\infty}=\dfrac{30}{24\pi^2 g_{*}} \left(\dfrac{v_n}{T_n} \right)^2 \left[6 \left( \dfrac{m_W}{v}\right)^2 +3\left( \dfrac{m_Z}{v}\right)^2 +6\left( \dfrac{m_{t}}{v}\right)^2\right]\,\,,
\end{equation}

where $v_n$ represents the vacuum expectation value of Higgs at $T_n$ and $m_W$, $m_Z$ and $m_t$ are the masses of W, Z and top quarks respectively. The parameter $\alpha^{\prime}$, which is defined as the ratio of vacuum energy density $\rho_{\text{vac}}$ released by the electroweak phase transition to the background energy density of the plasma $\rho_*^{\text{rad}}$ at $T_n$, has the form:
\begin{equation}\label{eq:53}
\alpha^{\prime}=\left[\dfrac{\rho_{\text{vac}}}{\rho^*_{\text{rad}}}\right]\bigg|_{T_n}.
\end{equation}
with
\begin{equation}\label{eq:54}
\rho_{\text{vac}}=\left[\left(V_{\text{eff}}^{\text{high}}-T\dfrac{dV_{\text{eff}}^{\text{high}}}{dT} \right)-\left(V_{\text{eff}}^{\text{low}}-T\dfrac{dV_{\text{eff}}^{\text{low}}}{dT} \right)\right],
\end{equation}
and
\begin{equation}\label{eq:55}
\rho^*_{\text{rad}}=\dfrac{g_* \pi^2 T_n^4}{30}.
\end{equation}
The quantity $f_\text{col}$ in Eq. (\ref{eq:48}) is the peak frequency produced by the bubble collisions and reads:
\begin{equation}\label{eq:56}
f_{\text{col}}=16.5\times10^{-6}\hspace{1mm} \text{Hz} \left( \dfrac{0.62}{v_{w}^2-0.1 v_w+1.8}\right)\left(\dfrac{\beta^{\prime}}{H} \right) \left(\dfrac{T_n}{100 \hspace{1mm} \text{GeV}} \right) \left(\dfrac{g_*}{100}\right)^{\frac{1}{6}}.
\end{equation}

The sound wave (SW) component of the gravitational wave (Eq. (\ref{eq:47})) is given by
\begin{equation}\label{eq:57}
\Omega_{\text{SW}}{\rm h}^2=2.65\times 10^{-6} \left(\dfrac{\beta^{\prime}}{H} \right) ^{-1} v_{w} \left(\dfrac{\kappa_{v} \alpha^{\prime}}{1+\alpha^{\prime}}\right)^2 \left(\dfrac{g_*}{100}\right)^{-\frac{1}{3}}\left(\dfrac{f}{f_{\text{SW}}}\right)^{3} \left[\dfrac{7}{4+3 \left(\dfrac{f}{f_{\text{SW}}}\right)^{2}}\right]^{\frac{7}{2}},
\end{equation}
where $\kappa_v$ is the faction of latent heat transformed into the bulk motion of the fluid which can be expressed as
\begin{equation}\label{eq:58}
\kappa_v=\dfrac{\alpha^{\prime}_{\infty}}{\alpha^{\prime}}\left[ \dfrac{\alpha^{\prime}_{\infty}}{0.73+0.083\sqrt{\alpha^{\prime}_{\infty}}+\alpha^{\prime}_{\infty}}\right].
\end{equation}
In Eq. (\ref{eq:57}) $f_{\text{SW}}$ denotes the peak frequency produced by the sound wave mechanisms which has the following form
\begin{equation}\label{eq:59}
f_{\text{SW}}=1.9\times10^{-5}\hspace{1mm} \text{Hz} \left( \dfrac{1}{v_{w}}\right)\left(\dfrac{\beta^{\prime}}{H} \right) \left(\dfrac{T_n}{100 \hspace{1mm} \text{GeV}} \right) \left(\dfrac{g_*}{100}\right)^{\frac{1}{6}}.
\end{equation}

To check the contribution of sound wave component to the total GW intensity we need to estimate the suppression factor $H R_*/\bar{U}_f$, where $\bar{U}_f$ denotes the root-mean-square (RMS) fluid velocity and $R_*$ denotes the mean bubble separation~\cite{Caprini:2015zlo,Ellis:2018mja,Ellis:2019oqb}. If the calculated suppression factor $H R_*/\bar{U}_f$ of a given model comes out to be $>1$ , then the sound wave lasts more than a Hubble time, otherwise it is an overestimate to the GW signal.

\begin{table}[htb!]
\centering
\begin{tabular}{|l|c|c|c|c|c|c|c|c|c|c|c|r|}
\hline
BP&$m_{11}^2$&$m_{22}^2$&$m_{12}$&$\lambda_1$&$\lambda_2$&$\lambda_3$&$\lambda_4$&$\lambda_5$&$\tan\beta$\\
&in GeV$^2$&in GeV$^2$&in GeV &&&&&&\\
 \hline
I&27511.8&18531.5&176&6.05&2.00&6.63&-8.27&-8.27&1.3\\
\hline
II&36668.7&18503.2&185&0.79&0.45&11.54&-5.80&-5.80&1.3\\
\hline
III&70301.2&-4698.75&125&3.87&0.26&11.37&-5.63&-5.63&5\\
\hline
IV&76676.2&-4443.75&130&1.13&0.26&11.26&-5.52&-5.52&5\\
\hline
\end{tabular}
\caption{Choice of the benchmark points (BPs) allowed by DM phenomenology to investigate the phase transition properties and production of GW.}\label{t2}\vspace{0.2cm}
\end{table}

Finally, the component from the turbulence in the plasma $\Omega_{\text{turb}}{\rm h}^2$ is given by:

\begin{equation}\label{eq:60}
\Omega_{\text{turb}}{\rm h}^2=3.35\times 10^{-4} \left(\dfrac{\beta^{\prime}}{H} \right) ^{-1} v_{w} \left(\dfrac{\epsilon \kappa_v \alpha^{\prime}}{1+\alpha^{\prime}}\right)^{\frac{3}{2}} \left(\dfrac{g_*}{100}\right)^{-\frac{1}{3}} \dfrac{\left(\dfrac{f}{f_{\text{turb}}}\right)^{3}\left( 1+\dfrac{f}{f_{\text{turb}}}\right)^{-\frac{11}{3}}}{\left(1+\dfrac{8\pi f}{h_{*}}\right)},
\end{equation}
where $\epsilon=0.1$ and $f_{\text{turb}}$ denotes the peak frequency produced by the turbulence mechanism which takes the form

\begin{equation}\label{eq:61}
f_{\text{turb}}=2.7\times10^{-5}\hspace{1mm} \text{Hz} \left( \dfrac{1}{v_{w}}\right)\left(\dfrac{\beta^{\prime}}{H} \right) \left(\dfrac{T_n}{100 \hspace{1mm} \text{GeV}} \right) \left(\dfrac{g_*}{100}\right)^{\frac{1}{6}}.
\end{equation}
In Eq. (\ref{eq:60}) the parameter $h_*$ can be written as
\begin{equation}\label{eq:62}
h_{*}=16.5\times10^{-6}\hspace{1mm} \text{Hz} \left(\dfrac{T_n}{100 \hspace{1mm} \text{GeV}} \right) \left(\dfrac{g_*}{100}\right)^{\frac{1}{6}}.
\end{equation}

Eq.~(\ref{eq:47})-(\ref{eq:62}) are used for calculating the gravitational wave intensity. In order to study the phase transition properties and production of GW in the present DM model, we choose four BPs (Table \ref{t2}) from the viable model parameter space. An EWPT takes place at the nucleation temperature where a high phase and a low phase is separated by a potential barrier. The strength of the phase transition depends on the ratio of the VEVs $v_c=\sqrt{<v_1>^2+<v_2>^2}$, measured at the critical temperature, to the critical temperature $T_c$ at which two degenerate minima exist. If the condition $\xi=v_c/T_c>1$ is satisfied then the strong first-order phase transition (SFOPT) is said to be occured \cite{Bernon:2017jgv}.

\begin{table}[htb!]
\centering
\begin{tabular}{|l|c|c|c|c|c|c|c|c|r|}
\hline
BP&$T_c$&$\xi$&$v_n$&$T_{n}$&$\alpha^{\prime}$&$\dfrac{\beta^{\prime}}{H}$\\
 &(GeV) && (GeV) & (GeV) &&\\
 \hline
I&71.36&1.71&125.73&66.86&0.26&3527\\
\hline
II&49.34&1.74&88.82&46.22&0.26&3571.17\\
\hline
III&62.39&1.17&75.87&61.33&0.12&14078.9\\
\hline
IV&58.88&1.21&73.74&57.79&0.13&13190.8\\
\hline
\end{tabular}
\caption{Thermal parameters associated with the strong first-order electroweak phase transition (SFOEWPT) for the chosen four benchmark points (BPs).}\label{t3}\vspace{0.2cm}
\end{table}

The BPs in Tab.~\ref{t2} are chosen in such a way that they not only satisfy constraints from the DM sector, but obeys the condition for SFOPT as well. Thus, one should note here, the choices of $m_{12}$ and $\tan\beta$ in Tab.~\ref{t2} are in accordance with those in Tab.~\ref{tab:bp}. In Table \ref{t3} we present our calculated $\xi$ value corresponding to the benchmarks in Tab.~\ref{t2}. In Table \ref{t3} we also tabulate other thermal parameters: $\{v_n, \hspace{1mm}T_c,\hspace{1mm}T_n,\hspace{1mm}\alpha^{\prime},\hspace{1mm}\beta^{\prime}/H\}$ for each of the BPs which are further used for calculation of GW intensity.  From Table \ref{t3}, one can see that the nucleation temperature $T_n$ is smaller than the critical temperature $T_c$ for each of the BPs. The renormalizable scale $Q=246.22$ GeV is fixed for the calculation.

\begin{figure}[htb!]
$$
\includegraphics[scale=0.45]{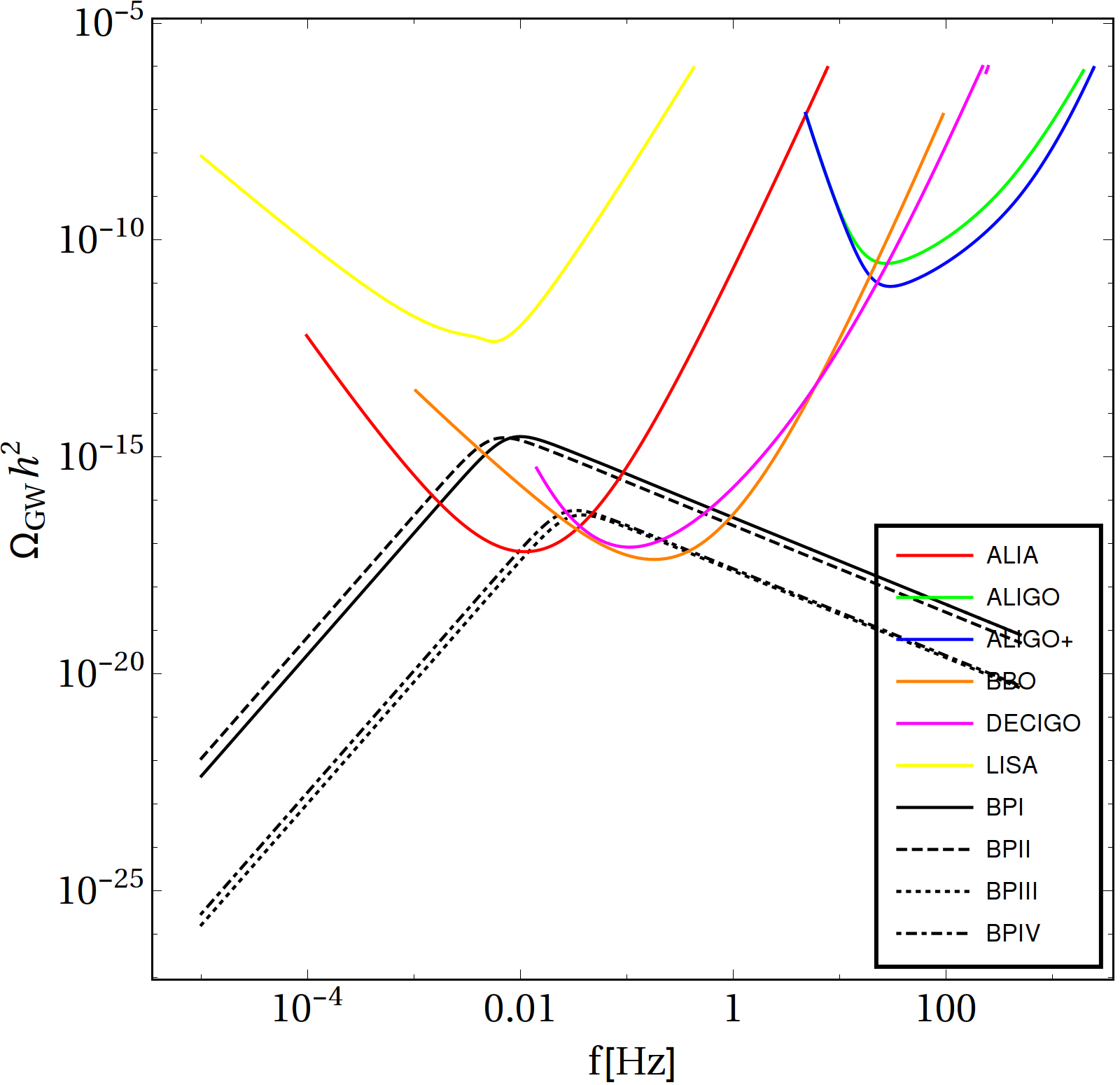}
$$
 \caption{Variation of GW intensity as a function of frequency for the chosen four BPs with the sensitivity curves of ALIA, BBO, DECIGO, aLIGO, aLIGO+ and LISA detectors.}
\label{fig:6}
\end{figure}

We calculate the GW intensity using Eq.~(\ref{eq:47})-(\ref{eq:62}). For computing the  intensity we need to first estimate the thermal parameters related to FOPT. In order to find them we have used {\tt Cosmotransition} package \cite{Wainwright:2011kj}. The tree level potential (Eq.~\ref{tpot}) is given as an input to this package, and resulting thermal parameters are tabulated in Table \ref{t3}. The GW intensity mainly depends on several factors {\it e.g,} nucleation temperature $T_n$, bubble wall velocity $v_{w}$, strength of the FOPT $\alpha^{\prime}$ and the parameter $\beta^{\prime}$. As mentioned in Sec.~\ref{sec:gwformalism}, the sound wave contribution to the total GW intensity depends on the suppression factor $H R_*/\bar{U}_f$ depending on whether it lasts more than a Hubble time or not. We estimate the suppression factor $H R_*/\bar{U}_f$ following~\cite{Caprini:2015zlo,Ellis:2018mja,Ellis:2019oqb} and found it to be $<$ 1 for all the BPs. Due to this fact, following~\cite{Caprini:2015zlo,Ellis:2018mja,Ellis:2019oqb}, we include the suppression factor $H R_*/\bar{U}_f$ to the sound wave component of the GW intensity. In Figure \ref{fig:6} we have plotted and compared the GW intensities for the chosen benchmarks (Table \ref{t2}) as a function of frequency against the power-law-integrated sensitivity {\footnote{For an alternative approach see~\cite{Alanne:2019bsm}}} curves for future GW detectors such as ALIA, BBO, DECIGO, aLIGO, aLIGO+ and LISA following Ref.~\cite{Thrane:2013oya,Dev:2019njv}. The frequencies at which the GW intensities acquire maximum value are $10^{-2}$ Hz, $7 \times 10^{-3}$ Hz, $3.80 \times 10^{-2}$ Hz and $3.40 \times 10^{-2}$ Hz for BPI, BPII, BPIII, and BPIV respectively. As it is evident from Figure \ref{fig:6}, the GW intensities for all the BPs (BPI-BPIV) lie within the sensitivity curves of ALIA, BBO and DECIGO. The upshot is thus to note that fact that the benchmark values of $m_{12}$ and $\tan\beta$ for DM phenomenology agrees well with that of a successful SFPOT leading to production of detectable GW signal.


\section{Conclusions}
\label{sec:concl}

In this paper we have proposed a singlet-doublet fermionic dark matter (DM) model, where the lightest   fermion (odd under an imposed discrete symmetry $\mathcal{Z}_2^{'}$ ), emerging as a singlet-doublet admixture of vectorlike fermions (VLF), can be a viable DM candidate ($\psi_1$). We extend the model with a second Higgs doublet where the second Higgs is odd under another discrete symmetry $\mathcal{Z}_2$. The imposition of two different discrete symmetries is necessary to prevent the occurrence of the flavour changing neutral current (FCNC) at tree-level, while  allowing a strong first order phase transition within a consistent DM framework. The DM in this case is a weakly interacting massive particle that undergoes freeze-out to yield the PLANCK observed relic abundance. This is achieved via annihilation of the DM with itself, also via its co-annihilation with its massive component ($\psi_2$) and with the charged component ($\psi^\pm$). As the doublet carries a $SU(2)_L$ charge, hence on top of scalar mediation, the annihilation channels are also SM gauge mediated. The presence of the $Z$-mediated direct search puts a strong bound on the model parameter space allowing the VLF mixing $\sin\theta\lsim 0.3$ both for $\tan\beta=1.3$ and $\tan\beta=5$ for DM mass up to $\gsim\rm 1~\rm TeV$. The presence the second Higgs makes the direct detection bounds less stringent. This is typically due to (a) $\sin^2\alpha/m_H^2$ suppression from the heavier Higgs with small scalar mixing and (b) some destructive interference between the two scalar mediated diagrams (the so called ``blind spot'') that offers some breathing space in the direct detection parameter space. Thus one can still achieve a moderate $\sin\theta$ in contrast with singlet-doublet models with only SM Higgs. The model thus lives over a large parameter space satisfying both relic abundance and spin-independent direct search bounds. 
 
We have then explored possible signatures that this model can give rise to at the LHC. As the leptonic channels are cleaner, we have studied the hadronically quiet dilepton final states (HQ2L) where we see, a substantial signal significance is achievable (for an integrated luminosity $\mathcal{L}\sim 300~\rm fb^{-1}$) by a judicious choice of cuts on the missing energy (MET) and $H_T$. This is again possible because of the presence of the second Higgs doublet that allows large $\sin\theta$ satisfying both relic abundance and direct search, which, in turn, allows a large $\Delta m\left(\equiv m_{\psi^\pm}-m_{\psi_1}\right)\sim 1~\rm TeV$. Larger $\Delta m$ is the key to distinguish the signal from the dominant SM backgrounds exploiting hard cuts on MET and $H_T$. For $\Delta m\lsim m_W$ the model may be probed via displaced vertex signature due to the off-shell decay of the charged VLF to SM leptons and neutrino.

We finally have looked into the possibility of getting gravitational wave (GW) via a strong first-order phase transition (SFOPT) due to the extended scalar sector. We have found for both $\tan\beta=\{1.3,5\}$ the model is capable of producing detectable GW signal modulo we tune the parameter $m_{12}$ accordingly. The probability of getting a SFOPT is thus very sensitive to the choice of $m_{12}$. We see, for some benchmark values of $m_{12}$ and $\tan\beta$, one can have detectable GW signal via SFPOT while satisfying all DM constraints. This model thus leaves us a with the window of probing such singlet-doublet DM models via GW detectors even if other experiments give rise to null results.  


%


\acknowledgements

BB and AP would like to acknowledge Workshop in High Energy Physics Phenomenology (WHEPP), IIT Guwahati, where a part of the work was completed. BB would also like to acknowledge fruitful discussions with Eung Jin Chun. AP would like to thank Biswajit Banerjee of SINP, Kolkata for helping in modifying the {\tt Cosmotransition} package. ADB thanks Rishav Roshan for useful discussions. We would like to thank Ryusuke Jinno for comments on GW analysis and providing useful references. The authors acknowledge A. Elbakyan for resources.

\appendix
\section{Appendix}
\subsection{Invisible Higgs and Z-decay}
\label{sec:invdecay}


The SM Higgs can decay to $\psi_1$ pairs. Now,  the combination of SM channels yields an observed (expected) upper limit on the Higgs branching fraction of 0.24 at 95 \%\ CL~\cite{Khachatryan:2016whc} with a total decay width $\Gamma=4.07\times 10^{-3}~\rm GeV$. On the other hand, SM $Z$ boson can also decay to DM pairs and hence constrained from observation: $\Gamma_{inv}^{Z}=499\pm 1.5~\rm MeV$~\cite{PhysRevD.98.030001}. So, if $Z$ is allowed to decay into $\psi_1\psi_1$ pair, the decay width should not be more than 1.5 MeV. 

\begin{table}[htb!]
\begin{center}
\begin{tabular}{|c|c|c|c|c|c|c|c|c|c|}
\hline
Benchmark & $Br_{inv}^{higgs}$ &  $\Gamma_{inv}^{Z}$ (MeV)  \\ [0.5ex] 
Point     &                    &             (MeV)  \\ [0.5ex] 
\hline\hline 

BP1 & {\it NA} & {\it NA} \\
\hline
BP2 & {\it NA} & {\it NA} \\
\hline
BP3 & $\sim 10^{-6}$ & {\it NA}  \\
\hline
BP4 & {\it NA} & {\it NA} \\
\hline
BP5 & {\it NA} & {\it NA} \\
\hline
BP6 & {\it NA} & {\it NA} \\
\hline\hline
\end{tabular}
\end{center}
\caption {Invisible Higgs branching ratio and invisible $Z$ decay width for different benchmark points tabulated in Tab.~\ref{tab:bp}. Only for BP3 the constraint from invisible Higgs branching ratio is applicable, which is $\mathcal{O}(10^{-6})$ for $\tan\beta=1.3$ or $\tan\beta=5$. {\it NA} stands for `Not Applicable` for cases where $m_{\psi_1}>m_Z(m_h)/2$.} 
\label{tab:invdecay}
\end{table}

Since either $\Delta m$ (for BP(1,2,3,4,5,6)) or $m_{\psi_1}$ is much larger than 100 GeV for all the benchmarks, hence neither $Z$ nor $h$ can decay to the heavier mass eigenstate $\psi_2$. Therefore, the expressions for $H_1\to\psi_1\psi_1$ and $Z\to\psi_1\psi_1$ decay widths are given by:

\bea
\Gamma_{inv}^{higgs}\left(H_1\to\psi_1\psi_1\right) = \left(\frac{Y^2\sin^2\theta\cos^2\theta\sin^2\alpha}{8\pi}\right)m_{h_1}\left(1-\frac{4 m_{\psi_1}^2}{m_{h_1}^2}\right)^{3/2}
\eea

\bea
\Gamma_{inv}^{Z}\left(Z\to\psi_1\psi_1\right) = \frac{m_Z}{48\pi} \frac{e^2\sin^4\theta}{\sin^2\theta_W\cos^2\theta_W}\left(1+\frac{m_{\psi_1}^2}{m_Z^2}\right)\sqrt{1-\frac{4 m_{\psi_1}^2}{m_Z^2}}.
\eea

Note that for none of the benchmark points in Tab.~\ref{tab:bp}, except BP3, either of the constraints from Higgs invisible decay branching or $Z$-boson invisible decay branching is applicable. Since for BP3 the DM mass is 60 GeV, it is possible for the SM Higgs to decay to a pair of $\psi_1$. However, due to small VLF mixing such a decay is well within the measured invisible decay rate of SM Higgs.


\subsection{Relevant Feynmann Diagrams for DM (co-)annihilation}
\label{sec:feyn-diag}
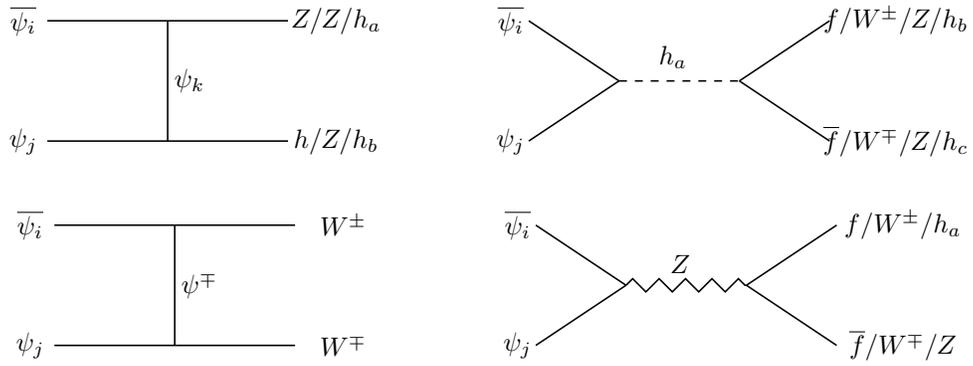
\begin{figure}[htb!]
\begin{center}
    \begin{tikzpicture}[line width=0.6 pt, scale=0.8]
        \draw[solid] (-3,1.0)--(-1.0,1.0);
        \draw[solid] (-3,-1.0)--(-1.0,-1.0);
        \draw[solid] (-1.0,1.0)--(-1.0,-1.0);
        \draw[solid] (-1.0,1.0)--(1.0,1.0);
        \draw[solid] (-1.0,-1.0)--(1.0,-1.0);
        \node at (-3.4,1.0) {$\overline{\psi_i}$};
        \node at (-3.4,-1.0) {$\psi_j$};
        \node [right] at (-1.05,0.0) {$\psi_k$};
        \node at (1.8,1.0) {$Z/Z/h_{a}$};
        \node at (1.8,-1.0) {$h/Z/h_{b}$};
        \draw[solid] (5.0,1.0)--(6.5,0.0);
        \draw[solid] (5.0,-1.0)--(6.5,0.0);
        \draw[dashed] (6.5,0.0)--(8.5,0.0);
        \draw[solid] (8.5,0.0)--(10.0,1.0);
        \draw[solid] (8.5,0.0)--(10.0,-1.0);
        \node at (4.7,1.0) {$\overline{\psi_i}$};
        \node at (4.7,-1.0) {$\psi_j$};
        \node [above] at (7.4,0.05) {$h_{a}$};
        \node at (11.1,1.0) {$f/W^\pm/Z/h_{b}$};
        \node at (11.1,-1.0) {$\overline{f}/W^\mp/Z/h_{c}$};
     \end{tikzpicture}
 \end{center}
\begin{center}
    \begin{tikzpicture}[line width=0.6 pt, scale=0.8]
        \draw[solid] (-3,1.0)--(-1.0,1.0);
        \draw[solid] (-3,-1.0)--(-1.0,-1.0);
        \draw[solid](-1.0,1.0)--(-1.0,-1.0);
        \draw[solid] (-1.0,1.0)--(1.0,1.0);
        \draw[solid] (-1.0,-1.0)--(1.0,-1.0);
        \node at (-3.4,1.0) {$\overline{\psi_i}$};
        \node at (-3.4,-1.0) {$\psi_j$};
        \node [right] at (-1.05,0.0) {$\psi^\mp$};
        \node at (1.8,1.0) {$W^\pm$};
        \node at (1.8,-1.0) {$W^\mp$};
        \draw[solid] (5.0,1.0)--(6.5,0.0);
        \draw[solid] (5.0,-1.0)--(6.5,0.0);
        \draw[snake] (6.5,0.0)--(8.5,0.0);
        \draw[solid] (8.5,0.0)--(10.0,1.0);
        \draw[solid] (8.5,0.0)--(10.0,-1.0);
        \node at (4.7,1.0) {$\overline{\psi_i}$};
        \node at (4.7,-1.0) {$\psi_j$};
        \node [above] at (7.4,0.05) {$Z$};
        \node at (11.1,1.0) {$f/W^\pm/h_a$};
        \node at (11.1,-1.0) {$\overline{f}/W^\mp/Z$};
     \end{tikzpicture}
 \end{center}
\caption{ Annihilation ($i=j$) and Co-annihilation ($i\neq j$) type number changing processes for Vector 
like fermionic DM in the model. Here $i,j,k=1,2$;~ $a,b,c=1,2$~and $f$ stands for SM fermions. }
\label{fd:an-coan}
 \end{figure}
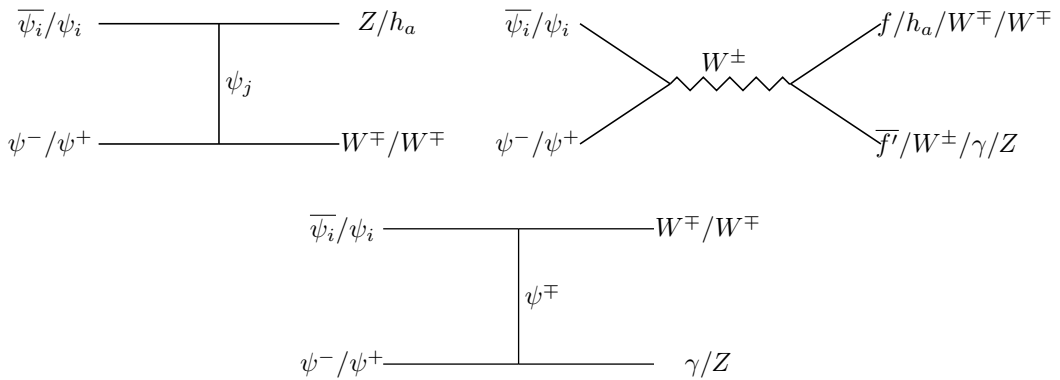
\begin{figure}[htb!]
\begin{center}
    \begin{tikzpicture}[line width=0.6 pt, scale=0.8]
        \draw[solid] (-3,1.0)--(-1.0,1.0);
        \draw[solid] (-3,-1.0)--(-1.0,-1.0);
        \draw[solid] (-1.0,1.0)--(-1.0,-1.0);
        \draw[solid] (-1.0,1.0)--(1.0,1.0);
        \draw[solid] (-1.0,-1.0)--(1.0,-1.0);
        \node at (-3.8,1.0) {$\overline{\psi_i}/\psi_i$};
        \node at (-3.8,-1.0) {$\psi^-/\psi^+$};
        \node [right] at (-1.05,0.0) {$\psi_j$};
        \node at (1.8,1.0) {$Z/h_a$};
        \node at (1.9,-1.0) {$W^\mp/W^\mp$};
        \draw[solid] (5.0,1.0)--(6.5,0.0);
        \draw[solid] (5.0,-1.0)--(6.5,0.0);
        \draw[snake] (6.5,0.0)--(8.5,0.0);
        \draw[solid] (8.5,0.0)--(10.0,1.0);
        \draw[solid] (8.5,0.0)--(10.0,-1.0);
        \node at (4.3,1.0) {$\overline{\psi_i}/\psi_i$};
        \node at (4.3,-1.0) {$\psi^-/\psi^+$};
        \node [above] at (7.4,0.05) {$W^\pm$};
        \node at (11.4,1.0) {$f/h_a/W^\mp/W^\mp$};
        \node at (11.1,-1.0) {$\overline{f^\prime}/W^\pm/\gamma/Z$};
     \end{tikzpicture}
 \end{center}
\begin{center}
    \begin{tikzpicture}[line width=0.5 pt, scale=0.9]
        \draw[solid] (-3,1.0)--(-1.0,1.0);
        \draw[solid] (-3,-1.0)--(-1.0,-1.0);
        \draw[solid](-1.0,1.0)--(-1.0,-1.0);
        \draw[solid] (-1.0,1.0)--(1.0,1.0);
        \draw[solid] (-1.0,-1.0)--(1.0,-1.0);
        \node at (-3.6,1.0) {$\overline{\psi_i}/\psi_i$};
        \node at (-3.6,-1.0) {$\psi^-/\psi^+$};
        \node [right] at (-1.05,0.0) {$\psi^\mp$};
        \node at (1.8,1.0) {$W^\mp/W^\mp$};
        \node at (1.8,-1.0) {$\gamma/Z$};
     \end{tikzpicture}
 \end{center}
\caption{ Feynmann diagrams for co-annihilation type number changing processes of $\psi_i ~(i=1,2)$ with the 
charged component $\psi^\pm$ to SM particles. Here $f$ and $f^\prime$ stand for SM fermions ($f \neq f^\prime$). }
\label{co-ann-2}
 \end{figure}
\begin{figure}[htb!]
\begin{center}
    \begin{tikzpicture}[line width=0.6 pt, scale=0.8]
        \draw[solid] (-3,1.0)--(-1.0,1.0);
        \draw[solid] (-3,-1.0)--(-1.0,-1.0);
        \draw[solid] (-1.0,1.0)--(-1.0,-1.0);
        \draw[solid] (-1.0,1.0)--(1.0,1.0);
        \draw[solid] (-1.0,-1.0)--(1.0,-1.0);
        \node at (-3.4,1.0) {$\psi^+$};
        \node at (-3.4,-1.0) {$\psi^-$};
        \node [right] at (-1.05,0.0) {$\psi_i$};
        \node at (1.8,1.0) {$W^+$};
        \node at (1.8,-1.0) {$W^-$};
        \draw[solid] (5.0,1.0)--(6.5,0.0);
        \draw[solid] (5.0,-1.0)--(6.5,0.0);
        \draw[snake] (6.5,0.0)--(8.5,0.0);
        \draw[solid] (8.5,0.0)--(10.0,1.0);
        \draw[solid] (8.5,0.0)--(10.0,-1.0);
        \node at (4.6,1.0) {$\psi^+$};
        \node at (4.6,-1.0) {$\psi^-$};
        \node [above] at (7.4,0.05) {$\gamma/Z$};
        \node at (11.1,1.0) {$f/W^+$};
        \node at (11.1,-1.0) {$\overline{f}/W^-$};
     \end{tikzpicture}
 \end{center}
\begin{center}
    \begin{tikzpicture}[line width=0.6 pt, scale=0.8]
        \draw[solid] (-3,1.0)--(-1.0,1.0);
        \draw[solid] (-3,-1.0)--(-1.0,-1.0);
        \draw[solid](-1.0,1.0)--(-1.0,-1.0);
        \draw[solid] (-1.0,1.0)--(1.0,1.0);
        \draw[solid] (-1.0,-1.0)--(1.0,-1.0);
        \node at (-3.4,1.0) {$\psi^+$};
        \node at (-3.4,-1.0) {$\psi^-$};
        \node [right] at (-1.05,0.0) {$\psi^-$};
        \node at (1.8,1.0) {$\gamma/Z$};
        \node at (1.8,-1.0) {$\gamma/Z$};
        \draw[solid] (5.0,1.0)--(6.5,0.0);
        \draw[solid] (5.0,-1.0)--(6.5,0.0);
        \draw[snake] (6.5,0.0)--(8.5,0.0);
        \draw[solid] (8.5,0.0)--(10.0,1.0);
        \draw[solid] (8.5,0.0)--(10.0,-1.0);
        \node at (4.6,1.0) {$\psi^+$};
        \node at (4.6,-1.0) {$\psi^-$};
        \node [above] at (7.4,0.05) {$Z$};
        \node at (10.6,1.0) {$h_a$};
        \node at (10.6,-1.0) {$Z$};
     \end{tikzpicture}
 \end{center}
\caption{Feynmann diagrams for charged fermionic DM, $\psi^\pm$ annihilation to SM particles in final states. Here $a=1,2$ . }
\label{co-ann-3}
 \end{figure}
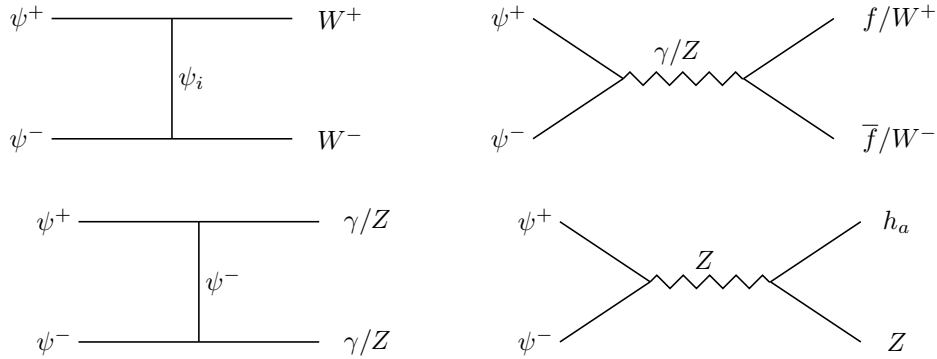
\bibliography{ref}

\begin{thebibliography}{151}%
\makeatletter
\providecommand \@ifxundefined [1]{%
 \@ifx{#1\undefined}
}%
\providecommand \@ifnum [1]{%
 \ifnum #1\expandafter \@firstoftwo
 \else \expandafter \@secondoftwo
 \fi
}%
\providecommand \@ifx [1]{%
 \ifx #1\expandafter \@firstoftwo
 \else \expandafter \@secondoftwo
 \fi
}%
\providecommand \natexlab [1]{#1}%
\providecommand \enquote  [1]{``#1''}%
\providecommand \bibnamefont  [1]{#1}%
\providecommand \bibfnamefont [1]{#1}%
\providecommand \citenamefont [1]{#1}%
\providecommand \href@noop [0]{\@secondoftwo}%
\providecommand \href [0]{\begingroup \@sanitize@url \@href}%
\providecommand \@href[1]{\@@startlink{#1}\@@href}%
\providecommand \@@href[1]{\endgroup#1\@@endlink}%
\providecommand \@sanitize@url [0]{\catcode `\\12\catcode `\$12\catcode
  `\&12\catcode `\#12\catcode `\^12\catcode `\_12\catcode `\%12\relax}%
\providecommand \@@startlink[1]{}%
\providecommand \@@endlink[0]{}%
\providecommand \url  [0]{\begingroup\@sanitize@url \@url }%
\providecommand \@url [1]{\endgroup\@href {#1}{\urlprefix }}%
\providecommand \urlprefix  [0]{URL }%
\providecommand \Eprint [0]{\href }%
\providecommand \doibase [0]{http://dx.doi.org/}%
\providecommand \selectlanguage [0]{\@gobble}%
\providecommand \bibinfo  [0]{\@secondoftwo}%
\providecommand \bibfield  [0]{\@secondoftwo}%
\providecommand \translation [1]{[#1]}%
\providecommand \BibitemOpen [0]{}%
\providecommand \bibitemStop [0]{}%
\providecommand \bibitemNoStop [0]{.\EOS\space}%
\providecommand \EOS [0]{\spacefactor3000\relax}%
\providecommand \BibitemShut  [1]{\csname bibitem#1\endcsname}%
\let\auto@bib@innerbib\@empty
\bibitem [{\citenamefont {Zwicky}(1933)}]{Zwicky:1933gu}%
  \BibitemOpen
  \bibfield  {author} {\bibinfo {author} {\bibfnamefont {F.}~\bibnamefont
  {Zwicky}},\ }\href {\doibase 10.1007/s10714-008-0707-4} {\bibfield  {journal}
  {\bibinfo  {journal} {Helv. Phys. Acta}\ }\textbf {\bibinfo {volume} {6}},\
  \bibinfo {pages} {110} (\bibinfo {year} {1933})},\ \bibinfo {note} {[Gen.
  Rel. Grav.41,207(2009)]}\BibitemShut {NoStop}%
\bibitem [{\citenamefont {Rubin}\ and\ \citenamefont
  {Ford}(1970)}]{Rubin:1970zza}%
  \BibitemOpen
  \bibfield  {author} {\bibinfo {author} {\bibfnamefont {V.~C.}\ \bibnamefont
  {Rubin}}\ and\ \bibinfo {author} {\bibfnamefont {W.~K.}\ \bibnamefont {Ford},
  \bibfnamefont {Jr.}},\ }\href {\doibase 10.1086/150317} {\bibfield  {journal}
  {\bibinfo  {journal} {Astrophys. J.}\ }\textbf {\bibinfo {volume} {159}},\
  \bibinfo {pages} {379} (\bibinfo {year} {1970})}\BibitemShut {NoStop}%
\bibitem [{\citenamefont {Clowe}\ \emph {et~al.}(2006)\citenamefont {Clowe},
  \citenamefont {Bradac}, \citenamefont {Gonzalez}, \citenamefont {Markevitch},
  \citenamefont {Randall}, \citenamefont {Jones},\ and\ \citenamefont
  {Zaritsky}}]{Clowe:2006eq}%
  \BibitemOpen
  \bibfield  {author} {\bibinfo {author} {\bibfnamefont {D.}~\bibnamefont
  {Clowe}}, \bibinfo {author} {\bibfnamefont {M.}~\bibnamefont {Bradac}},
  \bibinfo {author} {\bibfnamefont {A.~H.}\ \bibnamefont {Gonzalez}}, \bibinfo
  {author} {\bibfnamefont {M.}~\bibnamefont {Markevitch}}, \bibinfo {author}
  {\bibfnamefont {S.~W.}\ \bibnamefont {Randall}}, \bibinfo {author}
  {\bibfnamefont {C.}~\bibnamefont {Jones}}, \ and\ \bibinfo {author}
  {\bibfnamefont {D.}~\bibnamefont {Zaritsky}},\ }\href {\doibase
  10.1086/508162} {\bibfield  {journal} {\bibinfo  {journal} {Astrophys. J.}\
  }\textbf {\bibinfo {volume} {648}},\ \bibinfo {pages} {L109} (\bibinfo {year}
  {2006})},\ \Eprint {http://arxiv.org/abs/astro-ph/0608407}
  {arXiv:astro-ph/0608407 [astro-ph]} \BibitemShut {NoStop}%
\bibitem [{\citenamefont {Massey}\ \emph {et~al.}(2010)\citenamefont {Massey},
  \citenamefont {Kitching},\ and\ \citenamefont {Richard}}]{Massey:2010hh}%
  \BibitemOpen
  \bibfield  {author} {\bibinfo {author} {\bibfnamefont {R.}~\bibnamefont
  {Massey}}, \bibinfo {author} {\bibfnamefont {T.}~\bibnamefont {Kitching}}, \
  and\ \bibinfo {author} {\bibfnamefont {J.}~\bibnamefont {Richard}},\ }\href
  {\doibase 10.1088/0034-4885/73/8/086901} {\bibfield  {journal} {\bibinfo
  {journal} {Rept. Prog. Phys.}\ }\textbf {\bibinfo {volume} {73}},\ \bibinfo
  {pages} {086901} (\bibinfo {year} {2010})},\ \Eprint
  {http://arxiv.org/abs/1001.1739} {arXiv:1001.1739 [astro-ph.CO]} \BibitemShut
  {NoStop}%
\bibitem [{\citenamefont {Aghanim}\ \emph {et~al.}(2018)\citenamefont {Aghanim}
  \emph {et~al.}}]{Aghanim:2018eyx}%
  \BibitemOpen
  \bibfield  {author} {\bibinfo {author} {\bibfnamefont {N.}~\bibnamefont
  {Aghanim}} \emph {et~al.} (\bibinfo {collaboration} {Planck}),\ }\href@noop
  {} {\  (\bibinfo {year} {2018})},\ \Eprint {http://arxiv.org/abs/1807.06209}
  {arXiv:1807.06209 [astro-ph.CO]} \BibitemShut {NoStop}%
\bibitem [{\citenamefont {Kolb}\ and\ \citenamefont
  {Turner}(1990)}]{Kolb:1990vq}%
  \BibitemOpen
  \bibfield  {author} {\bibinfo {author} {\bibfnamefont {E.~W.}\ \bibnamefont
  {Kolb}}\ and\ \bibinfo {author} {\bibfnamefont {M.~S.}\ \bibnamefont
  {Turner}},\ }\href@noop {} {\bibfield  {journal} {\bibinfo  {journal} {Front.
  Phys.}\ }\textbf {\bibinfo {volume} {69}},\ \bibinfo {pages} {1} (\bibinfo
  {year} {1990})}\BibitemShut {NoStop}%
\bibitem [{\citenamefont {Arcadi}\ \emph {et~al.}(2018)\citenamefont {Arcadi},
  \citenamefont {Dutra}, \citenamefont {Ghosh}, \citenamefont {Lindner},
  \citenamefont {Mambrini}, \citenamefont {Pierre}, \citenamefont {Profumo},\
  and\ \citenamefont {Queiroz}}]{Arcadi:2017kky}%
  \BibitemOpen
  \bibfield  {author} {\bibinfo {author} {\bibfnamefont {G.}~\bibnamefont
  {Arcadi}}, \bibinfo {author} {\bibfnamefont {M.}~\bibnamefont {Dutra}},
  \bibinfo {author} {\bibfnamefont {P.}~\bibnamefont {Ghosh}}, \bibinfo
  {author} {\bibfnamefont {M.}~\bibnamefont {Lindner}}, \bibinfo {author}
  {\bibfnamefont {Y.}~\bibnamefont {Mambrini}}, \bibinfo {author}
  {\bibfnamefont {M.}~\bibnamefont {Pierre}}, \bibinfo {author} {\bibfnamefont
  {S.}~\bibnamefont {Profumo}}, \ and\ \bibinfo {author} {\bibfnamefont
  {F.~S.}\ \bibnamefont {Queiroz}},\ }\href {\doibase
  10.1140/epjc/s10052-018-5662-y} {\bibfield  {journal} {\bibinfo  {journal}
  {Eur. Phys. J.}\ }\textbf {\bibinfo {volume} {C78}},\ \bibinfo {pages} {203}
  (\bibinfo {year} {2018})},\ \Eprint {http://arxiv.org/abs/1703.07364}
  {arXiv:1703.07364 [hep-ph]} \BibitemShut {NoStop}%
\bibitem [{\citenamefont {Akerib}\ \emph {et~al.}(2017)\citenamefont {Akerib}
  \emph {et~al.}}]{Akerib:2017kat}%
  \BibitemOpen
  \bibfield  {author} {\bibinfo {author} {\bibfnamefont {D.~S.}\ \bibnamefont
  {Akerib}} \emph {et~al.} (\bibinfo {collaboration} {LUX}),\ }\href {\doibase
  10.1103/PhysRevLett.118.251302} {\bibfield  {journal} {\bibinfo  {journal}
  {Phys. Rev. Lett.}\ }\textbf {\bibinfo {volume} {118}},\ \bibinfo {pages}
  {251302} (\bibinfo {year} {2017})},\ \Eprint
  {http://arxiv.org/abs/1705.03380} {arXiv:1705.03380 [astro-ph.CO]}
  \BibitemShut {NoStop}%
\bibitem [{\citenamefont {Tan}\ \emph {et~al.}(2016)\citenamefont {Tan} \emph
  {et~al.}}]{Tan:2016zwf}%
  \BibitemOpen
  \bibfield  {author} {\bibinfo {author} {\bibfnamefont {A.}~\bibnamefont
  {Tan}} \emph {et~al.} (\bibinfo {collaboration} {PandaX-II}),\ }\href
  {\doibase 10.1103/PhysRevLett.117.121303} {\bibfield  {journal} {\bibinfo
  {journal} {Phys. Rev. Lett.}\ }\textbf {\bibinfo {volume} {117}},\ \bibinfo
  {pages} {121303} (\bibinfo {year} {2016})},\ \Eprint
  {http://arxiv.org/abs/1607.07400} {arXiv:1607.07400 [hep-ex]} \BibitemShut
  {NoStop}%
\bibitem [{\citenamefont {Cui}\ \emph {et~al.}(2017)\citenamefont {Cui} \emph
  {et~al.}}]{Cui:2017nnn}%
  \BibitemOpen
  \bibfield  {author} {\bibinfo {author} {\bibfnamefont {X.}~\bibnamefont
  {Cui}} \emph {et~al.} (\bibinfo {collaboration} {PandaX-II}),\ }\href@noop {}
  {\  (\bibinfo {year} {2017})},\ \Eprint {http://arxiv.org/abs/1708.06917}
  {arXiv:1708.06917 [astro-ph.CO]} \BibitemShut {NoStop}%
\bibitem [{\citenamefont {Aprile}\ \emph {et~al.}(2016)\citenamefont {Aprile}
  \emph {et~al.}}]{Aprile:2015uzo}%
  \BibitemOpen
  \bibfield  {author} {\bibinfo {author} {\bibfnamefont {E.}~\bibnamefont
  {Aprile}} \emph {et~al.} (\bibinfo {collaboration} {XENON}),\ }\href
  {\doibase 10.1088/1475-7516/2016/04/027} {\bibfield  {journal} {\bibinfo
  {journal} {JCAP}\ }\textbf {\bibinfo {volume} {1604}},\ \bibinfo {pages}
  {027} (\bibinfo {year} {2016})},\ \Eprint {http://arxiv.org/abs/1512.07501}
  {arXiv:1512.07501 [physics.ins-det]} \BibitemShut {NoStop}%
\bibitem [{\citenamefont {Aprile}\ \emph {et~al.}(2018)\citenamefont {Aprile}
  \emph {et~al.}}]{Aprile:2018dbl}%
  \BibitemOpen
  \bibfield  {author} {\bibinfo {author} {\bibfnamefont {E.}~\bibnamefont
  {Aprile}} \emph {et~al.} (\bibinfo {collaboration} {XENON}),\ }\href@noop {}
  {\  (\bibinfo {year} {2018})},\ \Eprint {http://arxiv.org/abs/1805.12562}
  {arXiv:1805.12562 [astro-ph.CO]} \BibitemShut {NoStop}%
\bibitem [{\citenamefont {Hall}\ \emph {et~al.}(2010)\citenamefont {Hall},
  \citenamefont {Jedamzik}, \citenamefont {March-Russell},\ and\ \citenamefont
  {West}}]{Hall:2009bx}%
  \BibitemOpen
  \bibfield  {author} {\bibinfo {author} {\bibfnamefont {L.~J.}\ \bibnamefont
  {Hall}}, \bibinfo {author} {\bibfnamefont {K.}~\bibnamefont {Jedamzik}},
  \bibinfo {author} {\bibfnamefont {J.}~\bibnamefont {March-Russell}}, \ and\
  \bibinfo {author} {\bibfnamefont {S.~M.}\ \bibnamefont {West}},\ }\href
  {\doibase 10.1007/JHEP03(2010)080} {\bibfield  {journal} {\bibinfo  {journal}
  {JHEP}\ }\textbf {\bibinfo {volume} {03}},\ \bibinfo {pages} {080} (\bibinfo
  {year} {2010})},\ \Eprint {http://arxiv.org/abs/0911.1120} {arXiv:0911.1120
  [hep-ph]} \BibitemShut {NoStop}%
\bibitem [{\citenamefont {Hochberg}\ \emph {et~al.}(2014)\citenamefont
  {Hochberg}, \citenamefont {Kuflik}, \citenamefont {Volansky},\ and\
  \citenamefont {Wacker}}]{Hochberg:2014dra}%
  \BibitemOpen
  \bibfield  {author} {\bibinfo {author} {\bibfnamefont {Y.}~\bibnamefont
  {Hochberg}}, \bibinfo {author} {\bibfnamefont {E.}~\bibnamefont {Kuflik}},
  \bibinfo {author} {\bibfnamefont {T.}~\bibnamefont {Volansky}}, \ and\
  \bibinfo {author} {\bibfnamefont {J.~G.}\ \bibnamefont {Wacker}},\ }\href
  {\doibase 10.1103/PhysRevLett.113.171301} {\bibfield  {journal} {\bibinfo
  {journal} {Phys. Rev. Lett.}\ }\textbf {\bibinfo {volume} {113}},\ \bibinfo
  {pages} {171301} (\bibinfo {year} {2014})},\ \Eprint
  {http://arxiv.org/abs/1402.5143} {arXiv:1402.5143 [hep-ph]} \BibitemShut
  {NoStop}%
\bibitem [{\citenamefont {Kuflik}\ \emph {et~al.}(2017)\citenamefont {Kuflik},
  \citenamefont {Perelstein}, \citenamefont {Lorier},\ and\ \citenamefont
  {Tsai}}]{Kuflik:2017iqs}%
  \BibitemOpen
  \bibfield  {author} {\bibinfo {author} {\bibfnamefont {E.}~\bibnamefont
  {Kuflik}}, \bibinfo {author} {\bibfnamefont {M.}~\bibnamefont {Perelstein}},
  \bibinfo {author} {\bibfnamefont {N.~R.-L.}\ \bibnamefont {Lorier}}, \ and\
  \bibinfo {author} {\bibfnamefont {Y.-D.}\ \bibnamefont {Tsai}},\ }\href
  {\doibase 10.1007/JHEP08(2017)078} {\bibfield  {journal} {\bibinfo  {journal}
  {JHEP}\ }\textbf {\bibinfo {volume} {08}},\ \bibinfo {pages} {078} (\bibinfo
  {year} {2017})},\ \Eprint {http://arxiv.org/abs/1706.05381} {arXiv:1706.05381
  [hep-ph]} \BibitemShut {NoStop}%
\bibitem [{\citenamefont {Davoudiasl}\ and\ \citenamefont
  {Mohlabeng}(2019)}]{Davoudiasl:2019xeb}%
  \BibitemOpen
  \bibfield  {author} {\bibinfo {author} {\bibfnamefont {H.}~\bibnamefont
  {Davoudiasl}}\ and\ \bibinfo {author} {\bibfnamefont {G.}~\bibnamefont
  {Mohlabeng}},\ }\href@noop {} {\  (\bibinfo {year} {2019})},\ \Eprint
  {http://arxiv.org/abs/1912.05572} {arXiv:1912.05572 [hep-ph]} \BibitemShut
  {NoStop}%
\bibitem [{\citenamefont {Lin}(2019)}]{Lin:2019uvt}%
  \BibitemOpen
  \bibfield  {author} {\bibinfo {author} {\bibfnamefont {T.}~\bibnamefont
  {Lin}},\ }\bibfield  {booktitle} {\emph {\bibinfo {booktitle} {{Proceedings,
  Theoretical Advanced Study Institute in Elementary Particle Physics: Theory
  in an Era of Data (TASI 2018): Boulder, Colorado, USA, June 4-29, 2018}}},\
  }\href {\doibase 10.22323/1.333.0009} {\bibfield  {journal} {\bibinfo
  {journal} {PoS}\ }\textbf {\bibinfo {volume} {333}},\ \bibinfo {pages} {009}
  (\bibinfo {year} {2019})},\ \Eprint {http://arxiv.org/abs/1904.07915}
  {arXiv:1904.07915 [hep-ph]} \BibitemShut {NoStop}%
\bibitem [{\citenamefont {Branco}\ \emph {et~al.}(2012)\citenamefont {Branco},
  \citenamefont {Ferreira}, \citenamefont {Lavoura}, \citenamefont {Rebelo},
  \citenamefont {Sher},\ and\ \citenamefont {Silva}}]{Branco:2011iw}%
  \BibitemOpen
  \bibfield  {author} {\bibinfo {author} {\bibfnamefont {G.~C.}\ \bibnamefont
  {Branco}}, \bibinfo {author} {\bibfnamefont {P.~M.}\ \bibnamefont
  {Ferreira}}, \bibinfo {author} {\bibfnamefont {L.}~\bibnamefont {Lavoura}},
  \bibinfo {author} {\bibfnamefont {M.~N.}\ \bibnamefont {Rebelo}}, \bibinfo
  {author} {\bibfnamefont {M.}~\bibnamefont {Sher}}, \ and\ \bibinfo {author}
  {\bibfnamefont {J.~P.}\ \bibnamefont {Silva}},\ }\href {\doibase
  10.1016/j.physrep.2012.02.002} {\bibfield  {journal} {\bibinfo  {journal}
  {Phys. Rept.}\ }\textbf {\bibinfo {volume} {516}},\ \bibinfo {pages} {1}
  (\bibinfo {year} {2012})},\ \Eprint {http://arxiv.org/abs/1106.0034}
  {arXiv:1106.0034 [hep-ph]} \BibitemShut {NoStop}%
\bibitem [{\citenamefont {Mader}\ \emph {et~al.}(2012)\citenamefont {Mader},
  \citenamefont {Park}, \citenamefont {Pruna}, \citenamefont {Stockinger},\
  and\ \citenamefont {Straessner}}]{Mader:2012pm}%
  \BibitemOpen
  \bibfield  {author} {\bibinfo {author} {\bibfnamefont {W.}~\bibnamefont
  {Mader}}, \bibinfo {author} {\bibfnamefont {J.-h.}\ \bibnamefont {Park}},
  \bibinfo {author} {\bibfnamefont {G.~M.}\ \bibnamefont {Pruna}}, \bibinfo
  {author} {\bibfnamefont {D.}~\bibnamefont {Stockinger}}, \ and\ \bibinfo
  {author} {\bibfnamefont {A.}~\bibnamefont {Straessner}},\ }\href {\doibase
  10.1007/JHEP01(2014)006, 10.1007/JHEP09(2012)125} {\bibfield  {journal}
  {\bibinfo  {journal} {JHEP}\ }\textbf {\bibinfo {volume} {09}},\ \bibinfo
  {pages} {125} (\bibinfo {year} {2012})},\ \bibinfo {note} {[Erratum:
  JHEP01,006(2014)]},\ \Eprint {http://arxiv.org/abs/1205.2692}
  {arXiv:1205.2692 [hep-ph]} \BibitemShut {NoStop}%
\bibitem [{\citenamefont {Chen}(2013)}]{Chen:2013qda}%
  \BibitemOpen
  \bibfield  {author} {\bibinfo {author} {\bibfnamefont {C.-Y.}\ \bibnamefont
  {Chen}},\ }in\ \href
  {http://www.slac.stanford.edu/econf/C1307292/docs/submittedArxivFiles/1308.3487.pdf}
  {\emph {\bibinfo {booktitle} {{Proceedings, 2013 Community Summer Study on
  the Future of U.S. Particle Physics: Snowmass on the Mississippi (CSS2013):
  Minneapolis, MN, USA, July 29-August 6, 2013}}}}\ (\bibinfo {year} {2013})\
  \Eprint {http://arxiv.org/abs/1308.3487} {arXiv:1308.3487 [hep-ph]}
  \BibitemShut {NoStop}%
\bibitem [{\citenamefont {Bhattacharyya}\ and\ \citenamefont
  {Das}(2016)}]{Bhattacharyya:2015nca}%
  \BibitemOpen
  \bibfield  {author} {\bibinfo {author} {\bibfnamefont {G.}~\bibnamefont
  {Bhattacharyya}}\ and\ \bibinfo {author} {\bibfnamefont {D.}~\bibnamefont
  {Das}},\ }\href {\doibase 10.1007/s12043-016-1252-4} {\bibfield  {journal}
  {\bibinfo  {journal} {Pramana}\ }\textbf {\bibinfo {volume} {87}},\ \bibinfo
  {pages} {40} (\bibinfo {year} {2016})},\ \Eprint
  {http://arxiv.org/abs/1507.06424} {arXiv:1507.06424 [hep-ph]} \BibitemShut
  {NoStop}%
\bibitem [{\citenamefont {Basler}\ \emph
  {et~al.}(2018{\natexlab{a}})\citenamefont {Basler}, \citenamefont {Ferreira},
  \citenamefont {Mühlleitner},\ and\ \citenamefont {Santos}}]{Basler:2017nzu}%
  \BibitemOpen
  \bibfield  {author} {\bibinfo {author} {\bibfnamefont {P.}~\bibnamefont
  {Basler}}, \bibinfo {author} {\bibfnamefont {P.~M.}\ \bibnamefont
  {Ferreira}}, \bibinfo {author} {\bibfnamefont {M.}~\bibnamefont
  {Mühlleitner}}, \ and\ \bibinfo {author} {\bibfnamefont {R.}~\bibnamefont
  {Santos}},\ }\href {\doibase 10.1103/PhysRevD.97.095024} {\bibfield
  {journal} {\bibinfo  {journal} {Phys. Rev.}\ }\textbf {\bibinfo {volume}
  {D97}},\ \bibinfo {pages} {095024} (\bibinfo {year} {2018}{\natexlab{a}})},\
  \Eprint {http://arxiv.org/abs/1710.10410} {arXiv:1710.10410 [hep-ph]}
  \BibitemShut {NoStop}%
\bibitem [{\citenamefont {Gori}\ \emph {et~al.}(2017)\citenamefont {Gori},
  \citenamefont {Haber},\ and\ \citenamefont {Santos}}]{Gori:2017qwg}%
  \BibitemOpen
  \bibfield  {author} {\bibinfo {author} {\bibfnamefont {S.}~\bibnamefont
  {Gori}}, \bibinfo {author} {\bibfnamefont {H.~E.}\ \bibnamefont {Haber}}, \
  and\ \bibinfo {author} {\bibfnamefont {E.}~\bibnamefont {Santos}},\ }\href
  {\doibase 10.1007/JHEP06(2017)110} {\bibfield  {journal} {\bibinfo  {journal}
  {JHEP}\ }\textbf {\bibinfo {volume} {06}},\ \bibinfo {pages} {110} (\bibinfo
  {year} {2017})},\ \Eprint {http://arxiv.org/abs/1703.05873} {arXiv:1703.05873
  [hep-ph]} \BibitemShut {NoStop}%
\bibitem [{\citenamefont {Mahbubani}\ and\ \citenamefont
  {Senatore}(2006)}]{Mahbubani:2005pt}%
  \BibitemOpen
  \bibfield  {author} {\bibinfo {author} {\bibfnamefont {R.}~\bibnamefont
  {Mahbubani}}\ and\ \bibinfo {author} {\bibfnamefont {L.}~\bibnamefont
  {Senatore}},\ }\href {\doibase 10.1103/PhysRevD.73.043510} {\bibfield
  {journal} {\bibinfo  {journal} {Phys. Rev.}\ }\textbf {\bibinfo {volume}
  {D73}},\ \bibinfo {pages} {043510} (\bibinfo {year} {2006})},\ \Eprint
  {http://arxiv.org/abs/hep-ph/0510064} {arXiv:hep-ph/0510064 [hep-ph]}
  \BibitemShut {NoStop}%
\bibitem [{\citenamefont {D'Eramo}(2007)}]{DEramo:2007anh}%
  \BibitemOpen
  \bibfield  {author} {\bibinfo {author} {\bibfnamefont {F.}~\bibnamefont
  {D'Eramo}},\ }\href {\doibase 10.1103/PhysRevD.76.083522} {\bibfield
  {journal} {\bibinfo  {journal} {Phys. Rev.}\ }\textbf {\bibinfo {volume}
  {D76}},\ \bibinfo {pages} {083522} (\bibinfo {year} {2007})},\ \Eprint
  {http://arxiv.org/abs/0705.4493} {arXiv:0705.4493 [hep-ph]} \BibitemShut
  {NoStop}%
\bibitem [{\citenamefont {Enberg}\ \emph {et~al.}(2007)\citenamefont {Enberg},
  \citenamefont {Fox}, \citenamefont {Hall}, \citenamefont {Papaioannou},\ and\
  \citenamefont {Papucci}}]{Enberg:2007rp}%
  \BibitemOpen
  \bibfield  {author} {\bibinfo {author} {\bibfnamefont {R.}~\bibnamefont
  {Enberg}}, \bibinfo {author} {\bibfnamefont {P.~J.}\ \bibnamefont {Fox}},
  \bibinfo {author} {\bibfnamefont {L.~J.}\ \bibnamefont {Hall}}, \bibinfo
  {author} {\bibfnamefont {A.~Y.}\ \bibnamefont {Papaioannou}}, \ and\ \bibinfo
  {author} {\bibfnamefont {M.}~\bibnamefont {Papucci}},\ }\href {\doibase
  10.1088/1126-6708/2007/11/014} {\bibfield  {journal} {\bibinfo  {journal}
  {JHEP}\ }\textbf {\bibinfo {volume} {11}},\ \bibinfo {pages} {014} (\bibinfo
  {year} {2007})},\ \Eprint {http://arxiv.org/abs/0706.0918} {arXiv:0706.0918
  [hep-ph]} \BibitemShut {NoStop}%
\bibitem [{\citenamefont {Cohen}\ \emph {et~al.}(2012)\citenamefont {Cohen},
  \citenamefont {Kearney}, \citenamefont {Pierce},\ and\ \citenamefont
  {Tucker-Smith}}]{Cohen:2011ec}%
  \BibitemOpen
  \bibfield  {author} {\bibinfo {author} {\bibfnamefont {T.}~\bibnamefont
  {Cohen}}, \bibinfo {author} {\bibfnamefont {J.}~\bibnamefont {Kearney}},
  \bibinfo {author} {\bibfnamefont {A.}~\bibnamefont {Pierce}}, \ and\ \bibinfo
  {author} {\bibfnamefont {D.}~\bibnamefont {Tucker-Smith}},\ }\href {\doibase
  10.1103/PhysRevD.85.075003} {\bibfield  {journal} {\bibinfo  {journal} {Phys.
  Rev.}\ }\textbf {\bibinfo {volume} {D85}},\ \bibinfo {pages} {075003}
  (\bibinfo {year} {2012})},\ \Eprint {http://arxiv.org/abs/1109.2604}
  {arXiv:1109.2604 [hep-ph]} \BibitemShut {NoStop}%
\bibitem [{\citenamefont {Cheung}\ and\ \citenamefont
  {Sanford}(2014)}]{Cheung:2013dua}%
  \BibitemOpen
  \bibfield  {author} {\bibinfo {author} {\bibfnamefont {C.}~\bibnamefont
  {Cheung}}\ and\ \bibinfo {author} {\bibfnamefont {D.}~\bibnamefont
  {Sanford}},\ }\href {\doibase 10.1088/1475-7516/2014/02/011} {\bibfield
  {journal} {\bibinfo  {journal} {JCAP}\ }\textbf {\bibinfo {volume} {1402}},\
  \bibinfo {pages} {011} (\bibinfo {year} {2014})},\ \Eprint
  {http://arxiv.org/abs/1311.5896} {arXiv:1311.5896 [hep-ph]} \BibitemShut
  {NoStop}%
\bibitem [{\citenamefont {Restrepo}\ \emph {et~al.}(2015)\citenamefont
  {Restrepo}, \citenamefont {Rivera}, \citenamefont {Sánchez-Peláez},
  \citenamefont {Zapata},\ and\ \citenamefont {Tangarife}}]{Restrepo:2015ura}%
  \BibitemOpen
  \bibfield  {author} {\bibinfo {author} {\bibfnamefont {D.}~\bibnamefont
  {Restrepo}}, \bibinfo {author} {\bibfnamefont {A.}~\bibnamefont {Rivera}},
  \bibinfo {author} {\bibfnamefont {M.}~\bibnamefont {Sánchez-Peláez}},
  \bibinfo {author} {\bibfnamefont {O.}~\bibnamefont {Zapata}}, \ and\ \bibinfo
  {author} {\bibfnamefont {W.}~\bibnamefont {Tangarife}},\ }\href {\doibase
  10.1103/PhysRevD.92.013005} {\bibfield  {journal} {\bibinfo  {journal} {Phys.
  Rev.}\ }\textbf {\bibinfo {volume} {D92}},\ \bibinfo {pages} {013005}
  (\bibinfo {year} {2015})},\ \Eprint {http://arxiv.org/abs/1504.07892}
  {arXiv:1504.07892 [hep-ph]} \BibitemShut {NoStop}%
\bibitem [{\citenamefont {Calibbi}\ \emph {et~al.}(2015)\citenamefont
  {Calibbi}, \citenamefont {Mariotti},\ and\ \citenamefont
  {Tziveloglou}}]{Calibbi:2015nha}%
  \BibitemOpen
  \bibfield  {author} {\bibinfo {author} {\bibfnamefont {L.}~\bibnamefont
  {Calibbi}}, \bibinfo {author} {\bibfnamefont {A.}~\bibnamefont {Mariotti}}, \
  and\ \bibinfo {author} {\bibfnamefont {P.}~\bibnamefont {Tziveloglou}},\
  }\href {\doibase 10.1007/JHEP10(2015)116} {\bibfield  {journal} {\bibinfo
  {journal} {JHEP}\ }\textbf {\bibinfo {volume} {10}},\ \bibinfo {pages} {116}
  (\bibinfo {year} {2015})},\ \Eprint {http://arxiv.org/abs/1505.03867}
  {arXiv:1505.03867 [hep-ph]} \BibitemShut {NoStop}%
\bibitem [{\citenamefont {Cynolter}\ \emph {et~al.}(2016)\citenamefont
  {Cynolter}, \citenamefont {Kovács},\ and\ \citenamefont
  {Lendvai}}]{Cynolter:2015sua}%
  \BibitemOpen
  \bibfield  {author} {\bibinfo {author} {\bibfnamefont {G.}~\bibnamefont
  {Cynolter}}, \bibinfo {author} {\bibfnamefont {J.}~\bibnamefont {Kovács}}, \
  and\ \bibinfo {author} {\bibfnamefont {E.}~\bibnamefont {Lendvai}},\ }\href
  {\doibase 10.1142/S0217732316500139} {\bibfield  {journal} {\bibinfo
  {journal} {Mod. Phys. Lett.}\ }\textbf {\bibinfo {volume} {A31}},\ \bibinfo
  {pages} {1650013} (\bibinfo {year} {2016})},\ \Eprint
  {http://arxiv.org/abs/1509.05323} {arXiv:1509.05323 [hep-ph]} \BibitemShut
  {NoStop}%
\bibitem [{\citenamefont {Bhattacharya}\ \emph
  {et~al.}(2016{\natexlab{a}})\citenamefont {Bhattacharya}, \citenamefont
  {Sahoo},\ and\ \citenamefont {Sahu}}]{Bhattacharya:2015qpa}%
  \BibitemOpen
  \bibfield  {author} {\bibinfo {author} {\bibfnamefont {S.}~\bibnamefont
  {Bhattacharya}}, \bibinfo {author} {\bibfnamefont {N.}~\bibnamefont {Sahoo}},
  \ and\ \bibinfo {author} {\bibfnamefont {N.}~\bibnamefont {Sahu}},\ }\href
  {\doibase 10.1103/PhysRevD.93.115040} {\bibfield  {journal} {\bibinfo
  {journal} {Phys. Rev.}\ }\textbf {\bibinfo {volume} {D93}},\ \bibinfo {pages}
  {115040} (\bibinfo {year} {2016}{\natexlab{a}})},\ \Eprint
  {http://arxiv.org/abs/1510.02760} {arXiv:1510.02760 [hep-ph]} \BibitemShut
  {NoStop}%
\bibitem [{\citenamefont {Dutta~Banik}\ \emph {et~al.}(2016)\citenamefont
  {Dutta~Banik}, \citenamefont {Majumdar},\ and\ \citenamefont
  {Biswas}}]{Banik:2015aya}%
  \BibitemOpen
  \bibfield  {author} {\bibinfo {author} {\bibfnamefont {A.}~\bibnamefont
  {Dutta~Banik}}, \bibinfo {author} {\bibfnamefont {D.}~\bibnamefont
  {Majumdar}}, \ and\ \bibinfo {author} {\bibfnamefont {A.}~\bibnamefont
  {Biswas}},\ }\href {\doibase 10.1140/epjc/s10052-016-4159-9} {\bibfield
  {journal} {\bibinfo  {journal} {Eur. Phys. J.}\ }\textbf {\bibinfo {volume}
  {C76}},\ \bibinfo {pages} {346} (\bibinfo {year} {2016})},\ \Eprint
  {http://arxiv.org/abs/1506.05665} {arXiv:1506.05665 [hep-ph]} \BibitemShut
  {NoStop}%
\bibitem [{\citenamefont {Bhattacharya}\ \emph
  {et~al.}(2017{\natexlab{a}})\citenamefont {Bhattacharya}, \citenamefont
  {Karmakar}, \citenamefont {Sahu},\ and\ \citenamefont
  {Sil}}]{Bhattacharya:2016rqj}%
  \BibitemOpen
  \bibfield  {author} {\bibinfo {author} {\bibfnamefont {S.}~\bibnamefont
  {Bhattacharya}}, \bibinfo {author} {\bibfnamefont {B.}~\bibnamefont
  {Karmakar}}, \bibinfo {author} {\bibfnamefont {N.}~\bibnamefont {Sahu}}, \
  and\ \bibinfo {author} {\bibfnamefont {A.}~\bibnamefont {Sil}},\ }\href
  {\doibase 10.1007/JHEP05(2017)068} {\bibfield  {journal} {\bibinfo  {journal}
  {JHEP}\ }\textbf {\bibinfo {volume} {05}},\ \bibinfo {pages} {068} (\bibinfo
  {year} {2017}{\natexlab{a}})},\ \Eprint {http://arxiv.org/abs/1611.07419}
  {arXiv:1611.07419 [hep-ph]} \BibitemShut {NoStop}%
\bibitem [{\citenamefont {Bhattacharya}\ \emph
  {et~al.}(2016{\natexlab{b}})\citenamefont {Bhattacharya}, \citenamefont
  {Karmakar}, \citenamefont {Sahu},\ and\ \citenamefont
  {Sil}}]{Bhattacharya:2016lts}%
  \BibitemOpen
  \bibfield  {author} {\bibinfo {author} {\bibfnamefont {S.}~\bibnamefont
  {Bhattacharya}}, \bibinfo {author} {\bibfnamefont {B.}~\bibnamefont
  {Karmakar}}, \bibinfo {author} {\bibfnamefont {N.}~\bibnamefont {Sahu}}, \
  and\ \bibinfo {author} {\bibfnamefont {A.}~\bibnamefont {Sil}},\ }\href
  {\doibase 10.1103/PhysRevD.93.115041} {\bibfield  {journal} {\bibinfo
  {journal} {Phys. Rev.}\ }\textbf {\bibinfo {volume} {D93}},\ \bibinfo {pages}
  {115041} (\bibinfo {year} {2016}{\natexlab{b}})},\ \Eprint
  {http://arxiv.org/abs/1603.04776} {arXiv:1603.04776 [hep-ph]} \BibitemShut
  {NoStop}%
\bibitem [{\citenamefont {Bhattacharya}\ \emph
  {et~al.}(2017{\natexlab{b}})\citenamefont {Bhattacharya}, \citenamefont
  {Sahoo},\ and\ \citenamefont {Sahu}}]{Bhattacharya:2017sml}%
  \BibitemOpen
  \bibfield  {author} {\bibinfo {author} {\bibfnamefont {S.}~\bibnamefont
  {Bhattacharya}}, \bibinfo {author} {\bibfnamefont {N.}~\bibnamefont {Sahoo}},
  \ and\ \bibinfo {author} {\bibfnamefont {N.}~\bibnamefont {Sahu}},\ }\href
  {\doibase 10.1103/PhysRevD.96.035010} {\bibfield  {journal} {\bibinfo
  {journal} {Phys. Rev.}\ }\textbf {\bibinfo {volume} {D96}},\ \bibinfo {pages}
  {035010} (\bibinfo {year} {2017}{\natexlab{b}})},\ \Eprint
  {http://arxiv.org/abs/1704.03417} {arXiv:1704.03417 [hep-ph]} \BibitemShut
  {NoStop}%
\bibitem [{\citenamefont {Bhattacharya}\ \emph {et~al.}(2018)\citenamefont
  {Bhattacharya}, \citenamefont {Ghosh}, \citenamefont {Sahoo},\ and\
  \citenamefont {Sahu}}]{Bhattacharya:2018fus}%
  \BibitemOpen
  \bibfield  {author} {\bibinfo {author} {\bibfnamefont {S.}~\bibnamefont
  {Bhattacharya}}, \bibinfo {author} {\bibfnamefont {P.}~\bibnamefont {Ghosh}},
  \bibinfo {author} {\bibfnamefont {N.}~\bibnamefont {Sahoo}}, \ and\ \bibinfo
  {author} {\bibfnamefont {N.}~\bibnamefont {Sahu}},\ }\href@noop {} {\
  (\bibinfo {year} {2018})},\ \Eprint {http://arxiv.org/abs/1812.06505}
  {arXiv:1812.06505 [hep-ph]} \BibitemShut {NoStop}%
\bibitem [{\citenamefont {Bhattacharya}\ \emph {et~al.}(2019)\citenamefont
  {Bhattacharya}, \citenamefont {Ghosh},\ and\ \citenamefont
  {Sahu}}]{Bhattacharya:2018cgx}%
  \BibitemOpen
  \bibfield  {author} {\bibinfo {author} {\bibfnamefont {S.}~\bibnamefont
  {Bhattacharya}}, \bibinfo {author} {\bibfnamefont {P.}~\bibnamefont {Ghosh}},
  \ and\ \bibinfo {author} {\bibfnamefont {N.}~\bibnamefont {Sahu}},\ }\href
  {\doibase 10.1007/JHEP02(2019)059} {\bibfield  {journal} {\bibinfo  {journal}
  {JHEP}\ }\textbf {\bibinfo {volume} {02}},\ \bibinfo {pages} {059} (\bibinfo
  {year} {2019})},\ \Eprint {http://arxiv.org/abs/1809.07474} {arXiv:1809.07474
  [hep-ph]} \BibitemShut {NoStop}%
\bibitem [{\citenamefont {Dutta~Banik}\ \emph {et~al.}(2018)\citenamefont
  {Dutta~Banik}, \citenamefont {Saha},\ and\ \citenamefont
  {Sil}}]{DuttaBanik:2018emv}%
  \BibitemOpen
  \bibfield  {author} {\bibinfo {author} {\bibfnamefont {A.}~\bibnamefont
  {Dutta~Banik}}, \bibinfo {author} {\bibfnamefont {A.~K.}\ \bibnamefont
  {Saha}}, \ and\ \bibinfo {author} {\bibfnamefont {A.}~\bibnamefont {Sil}},\
  }\href {\doibase 10.1103/PhysRevD.98.075013} {\bibfield  {journal} {\bibinfo
  {journal} {Phys. Rev.}\ }\textbf {\bibinfo {volume} {D98}},\ \bibinfo {pages}
  {075013} (\bibinfo {year} {2018})},\ \Eprint
  {http://arxiv.org/abs/1806.08080} {arXiv:1806.08080 [hep-ph]} \BibitemShut
  {NoStop}%
\bibitem [{\citenamefont {Arcadi}(2018)}]{Arcadi:2018pfo}%
  \BibitemOpen
  \bibfield  {author} {\bibinfo {author} {\bibfnamefont {G.}~\bibnamefont
  {Arcadi}},\ }\href {\doibase 10.1140/epjc/s10052-018-6327-6} {\bibfield
  {journal} {\bibinfo  {journal} {Eur. Phys. J.}\ }\textbf {\bibinfo {volume}
  {C78}},\ \bibinfo {pages} {864} (\bibinfo {year} {2018})},\ \Eprint
  {http://arxiv.org/abs/1804.04930} {arXiv:1804.04930 [hep-ph]} \BibitemShut
  {NoStop}%
\bibitem [{\citenamefont {Barman}\ \emph
  {et~al.}(2019{\natexlab{a}})\citenamefont {Barman}, \citenamefont
  {Bhattacharya}, \citenamefont {Ghosh}, \citenamefont {Kadam},\ and\
  \citenamefont {Sahu}}]{Barman:2019tuo}%
  \BibitemOpen
  \bibfield  {author} {\bibinfo {author} {\bibfnamefont {B.}~\bibnamefont
  {Barman}}, \bibinfo {author} {\bibfnamefont {S.}~\bibnamefont
  {Bhattacharya}}, \bibinfo {author} {\bibfnamefont {P.}~\bibnamefont {Ghosh}},
  \bibinfo {author} {\bibfnamefont {S.}~\bibnamefont {Kadam}}, \ and\ \bibinfo
  {author} {\bibfnamefont {N.}~\bibnamefont {Sahu}},\ }\href@noop {} {\
  (\bibinfo {year} {2019}{\natexlab{a}})},\ \Eprint
  {http://arxiv.org/abs/1902.01217} {arXiv:1902.01217 [hep-ph]} \BibitemShut
  {NoStop}%
\bibitem [{\citenamefont {Barman}\ \emph
  {et~al.}(2019{\natexlab{b}})\citenamefont {Barman}, \citenamefont {Borah},
  \citenamefont {Ghosh},\ and\ \citenamefont {Saha}}]{Barman:2019aku}%
  \BibitemOpen
  \bibfield  {author} {\bibinfo {author} {\bibfnamefont {B.}~\bibnamefont
  {Barman}}, \bibinfo {author} {\bibfnamefont {D.}~\bibnamefont {Borah}},
  \bibinfo {author} {\bibfnamefont {P.}~\bibnamefont {Ghosh}}, \ and\ \bibinfo
  {author} {\bibfnamefont {A.~K.}\ \bibnamefont {Saha}},\ }\href {\doibase
  10.1007/JHEP10(2019)275} {\bibfield  {journal} {\bibinfo  {journal} {JHEP}\
  }\textbf {\bibinfo {volume} {10}},\ \bibinfo {pages} {275} (\bibinfo {year}
  {2019}{\natexlab{b}})},\ \Eprint {http://arxiv.org/abs/1907.10071}
  {arXiv:1907.10071 [hep-ph]} \BibitemShut {NoStop}%
\bibitem [{\citenamefont {Cheung}\ \emph {et~al.}(2013)\citenamefont {Cheung},
  \citenamefont {Hall}, \citenamefont {Pinner},\ and\ \citenamefont
  {Ruderman}}]{Cheung:2012qy}%
  \BibitemOpen
  \bibfield  {author} {\bibinfo {author} {\bibfnamefont {C.}~\bibnamefont
  {Cheung}}, \bibinfo {author} {\bibfnamefont {L.~J.}\ \bibnamefont {Hall}},
  \bibinfo {author} {\bibfnamefont {D.}~\bibnamefont {Pinner}}, \ and\ \bibinfo
  {author} {\bibfnamefont {J.~T.}\ \bibnamefont {Ruderman}},\ }\href {\doibase
  10.1007/JHEP05(2013)100} {\bibfield  {journal} {\bibinfo  {journal} {JHEP}\
  }\textbf {\bibinfo {volume} {05}},\ \bibinfo {pages} {100} (\bibinfo {year}
  {2013})},\ \Eprint {http://arxiv.org/abs/1211.4873} {arXiv:1211.4873
  [hep-ph]} \BibitemShut {NoStop}%
\bibitem [{\citenamefont {Huang}\ and\ \citenamefont
  {Wagner}(2014)}]{Huang:2014xua}%
  \BibitemOpen
  \bibfield  {author} {\bibinfo {author} {\bibfnamefont {P.}~\bibnamefont
  {Huang}}\ and\ \bibinfo {author} {\bibfnamefont {C.~E.~M.}\ \bibnamefont
  {Wagner}},\ }\href {\doibase 10.1103/PhysRevD.90.015018} {\bibfield
  {journal} {\bibinfo  {journal} {Phys. Rev.}\ }\textbf {\bibinfo {volume}
  {D90}},\ \bibinfo {pages} {015018} (\bibinfo {year} {2014})},\ \Eprint
  {http://arxiv.org/abs/1404.0392} {arXiv:1404.0392 [hep-ph]} \BibitemShut
  {NoStop}%
\bibitem [{\citenamefont {Berlin}\ \emph {et~al.}(2015)\citenamefont {Berlin},
  \citenamefont {Gori}, \citenamefont {Lin},\ and\ \citenamefont
  {Wang}}]{Berlin:2015wwa}%
  \BibitemOpen
  \bibfield  {author} {\bibinfo {author} {\bibfnamefont {A.}~\bibnamefont
  {Berlin}}, \bibinfo {author} {\bibfnamefont {S.}~\bibnamefont {Gori}},
  \bibinfo {author} {\bibfnamefont {T.}~\bibnamefont {Lin}}, \ and\ \bibinfo
  {author} {\bibfnamefont {L.-T.}\ \bibnamefont {Wang}},\ }\href {\doibase
  10.1103/PhysRevD.92.015005} {\bibfield  {journal} {\bibinfo  {journal} {Phys.
  Rev.}\ }\textbf {\bibinfo {volume} {D92}},\ \bibinfo {pages} {015005}
  (\bibinfo {year} {2015})},\ \Eprint {http://arxiv.org/abs/1502.06000}
  {arXiv:1502.06000 [hep-ph]} \BibitemShut {NoStop}%
\bibitem [{\citenamefont {Cabrera}\ \emph {et~al.}(2019)\citenamefont
  {Cabrera}, \citenamefont {Casas}, \citenamefont {Delgado},\ and\
  \citenamefont {Robles}}]{Cabrera:2019gaq}%
  \BibitemOpen
  \bibfield  {author} {\bibinfo {author} {\bibfnamefont {M.~E.}\ \bibnamefont
  {Cabrera}}, \bibinfo {author} {\bibfnamefont {J.~A.}\ \bibnamefont {Casas}},
  \bibinfo {author} {\bibfnamefont {A.}~\bibnamefont {Delgado}}, \ and\
  \bibinfo {author} {\bibfnamefont {S.}~\bibnamefont {Robles}},\ }\href@noop {}
  {\  (\bibinfo {year} {2019})},\ \Eprint {http://arxiv.org/abs/1912.01758}
  {arXiv:1912.01758 [hep-ph]} \BibitemShut {NoStop}%
\bibitem [{\citenamefont {Sakharov}(1967)}]{Sakharov:1967dj}%
  \BibitemOpen
  \bibfield  {author} {\bibinfo {author} {\bibfnamefont {A.~D.}\ \bibnamefont
  {Sakharov}},\ }\href {\doibase 10.1070/PU1991v034n05ABEH002497} {\bibfield
  {journal} {\bibinfo  {journal} {Pisma Zh. Eksp. Teor. Fiz.}\ }\textbf
  {\bibinfo {volume} {5}},\ \bibinfo {pages} {32} (\bibinfo {year} {1967})},\
  \bibinfo {note} {[Usp. Fiz. Nauk161,no.5,61(1991)]}\BibitemShut {NoStop}%
\bibitem [{\citenamefont {Gavela}\ \emph
  {et~al.}(1994{\natexlab{a}})\citenamefont {Gavela}, \citenamefont
  {Hernandez}, \citenamefont {Orloff},\ and\ \citenamefont
  {Pene}}]{Gavela:1993ts}%
  \BibitemOpen
  \bibfield  {author} {\bibinfo {author} {\bibfnamefont {M.~B.}\ \bibnamefont
  {Gavela}}, \bibinfo {author} {\bibfnamefont {P.}~\bibnamefont {Hernandez}},
  \bibinfo {author} {\bibfnamefont {J.}~\bibnamefont {Orloff}}, \ and\ \bibinfo
  {author} {\bibfnamefont {O.}~\bibnamefont {Pene}},\ }\href {\doibase
  10.1142/S0217732394000629} {\bibfield  {journal} {\bibinfo  {journal} {Mod.
  Phys. Lett.}\ }\textbf {\bibinfo {volume} {A9}},\ \bibinfo {pages} {795}
  (\bibinfo {year} {1994}{\natexlab{a}})},\ \Eprint
  {http://arxiv.org/abs/hep-ph/9312215} {arXiv:hep-ph/9312215 [hep-ph]}
  \BibitemShut {NoStop}%
\bibitem [{\citenamefont {Huet}\ and\ \citenamefont
  {Sather}(1995)}]{Huet:1994jb}%
  \BibitemOpen
  \bibfield  {author} {\bibinfo {author} {\bibfnamefont {P.}~\bibnamefont
  {Huet}}\ and\ \bibinfo {author} {\bibfnamefont {E.}~\bibnamefont {Sather}},\
  }\href {\doibase 10.1103/PhysRevD.51.379} {\bibfield  {journal} {\bibinfo
  {journal} {Phys. Rev.}\ }\textbf {\bibinfo {volume} {D51}},\ \bibinfo {pages}
  {379} (\bibinfo {year} {1995})},\ \Eprint
  {http://arxiv.org/abs/hep-ph/9404302} {arXiv:hep-ph/9404302 [hep-ph]}
  \BibitemShut {NoStop}%
\bibitem [{\citenamefont {Gavela}\ \emph
  {et~al.}(1994{\natexlab{b}})\citenamefont {Gavela}, \citenamefont
  {Hernandez}, \citenamefont {Orloff}, \citenamefont {Pene},\ and\
  \citenamefont {Quimbay}}]{Gavela:1994dt}%
  \BibitemOpen
  \bibfield  {author} {\bibinfo {author} {\bibfnamefont {M.~B.}\ \bibnamefont
  {Gavela}}, \bibinfo {author} {\bibfnamefont {P.}~\bibnamefont {Hernandez}},
  \bibinfo {author} {\bibfnamefont {J.}~\bibnamefont {Orloff}}, \bibinfo
  {author} {\bibfnamefont {O.}~\bibnamefont {Pene}}, \ and\ \bibinfo {author}
  {\bibfnamefont {C.}~\bibnamefont {Quimbay}},\ }\href {\doibase
  10.1016/0550-3213(94)00410-2} {\bibfield  {journal} {\bibinfo  {journal}
  {Nucl. Phys.}\ }\textbf {\bibinfo {volume} {B430}},\ \bibinfo {pages} {382}
  (\bibinfo {year} {1994}{\natexlab{b}})},\ \Eprint
  {http://arxiv.org/abs/hep-ph/9406289} {arXiv:hep-ph/9406289 [hep-ph]}
  \BibitemShut {NoStop}%
\bibitem [{\citenamefont {Morrissey}\ and\ \citenamefont
  {Ramsey-Musolf}(2012)}]{Morrissey:2012db}%
  \BibitemOpen
  \bibfield  {author} {\bibinfo {author} {\bibfnamefont {D.~E.}\ \bibnamefont
  {Morrissey}}\ and\ \bibinfo {author} {\bibfnamefont {M.~J.}\ \bibnamefont
  {Ramsey-Musolf}},\ }\href {\doibase 10.1088/1367-2630/14/12/125003}
  {\bibfield  {journal} {\bibinfo  {journal} {New J. Phys.}\ }\textbf {\bibinfo
  {volume} {14}},\ \bibinfo {pages} {125003} (\bibinfo {year} {2012})},\
  \Eprint {http://arxiv.org/abs/1206.2942} {arXiv:1206.2942 [hep-ph]}
  \BibitemShut {NoStop}%
\bibitem [{\citenamefont {Haber}\ \emph {et~al.}(1979)\citenamefont {Haber},
  \citenamefont {Kane},\ and\ \citenamefont {Sterling}}]{HABER1979493}%
  \BibitemOpen
  \bibfield  {author} {\bibinfo {author} {\bibfnamefont {H.}~\bibnamefont
  {Haber}}, \bibinfo {author} {\bibfnamefont {G.}~\bibnamefont {Kane}}, \ and\
  \bibinfo {author} {\bibfnamefont {T.}~\bibnamefont {Sterling}},\ }\href
  {\doibase https://doi.org/10.1016/0550-3213(79)90225-6} {\bibfield  {journal}
  {\bibinfo  {journal} {Nuclear Physics B}\ }\textbf {\bibinfo {volume}
  {161}},\ \bibinfo {pages} {493 } (\bibinfo {year} {1979})}\BibitemShut
  {NoStop}%
\bibitem [{\citenamefont {Hall}\ and\ \citenamefont
  {Wise}(1981)}]{HALL1981397}%
  \BibitemOpen
  \bibfield  {author} {\bibinfo {author} {\bibfnamefont {L.~J.}\ \bibnamefont
  {Hall}}\ and\ \bibinfo {author} {\bibfnamefont {M.~B.}\ \bibnamefont
  {Wise}},\ }\href {\doibase https://doi.org/10.1016/0550-3213(81)90469-7}
  {\bibfield  {journal} {\bibinfo  {journal} {Nuclear Physics B}\ }\textbf
  {\bibinfo {volume} {187}},\ \bibinfo {pages} {397 } (\bibinfo {year}
  {1981})}\BibitemShut {NoStop}%
\bibitem [{\citenamefont {Donoghue}\ and\ \citenamefont
  {Li}(1979)}]{PhysRevD.19.945}%
  \BibitemOpen
  \bibfield  {author} {\bibinfo {author} {\bibfnamefont {J.~F.}\ \bibnamefont
  {Donoghue}}\ and\ \bibinfo {author} {\bibfnamefont {L.-F.}\ \bibnamefont
  {Li}},\ }\href {\doibase 10.1103/PhysRevD.19.945} {\bibfield  {journal}
  {\bibinfo  {journal} {Phys. Rev. D}\ }\textbf {\bibinfo {volume} {19}},\
  \bibinfo {pages} {945} (\bibinfo {year} {1979})}\BibitemShut {NoStop}%
\bibitem [{\citenamefont {Barger}\ \emph {et~al.}(1990)\citenamefont {Barger},
  \citenamefont {Hewett},\ and\ \citenamefont {Phillips}}]{PhysRevD.41.3421}%
  \BibitemOpen
  \bibfield  {author} {\bibinfo {author} {\bibfnamefont {V.}~\bibnamefont
  {Barger}}, \bibinfo {author} {\bibfnamefont {J.~L.}\ \bibnamefont {Hewett}},
  \ and\ \bibinfo {author} {\bibfnamefont {R.~J.~N.}\ \bibnamefont
  {Phillips}},\ }\href {\doibase 10.1103/PhysRevD.41.3421} {\bibfield
  {journal} {\bibinfo  {journal} {Phys. Rev. D}\ }\textbf {\bibinfo {volume}
  {41}},\ \bibinfo {pages} {3421} (\bibinfo {year} {1990})}\BibitemShut
  {NoStop}%
\bibitem [{\citenamefont {Aoki}\ \emph {et~al.}(2009)\citenamefont {Aoki},
  \citenamefont {Kanemura}, \citenamefont {Tsumura},\ and\ \citenamefont
  {Yagyu}}]{Aoki:2009ha}%
  \BibitemOpen
  \bibfield  {author} {\bibinfo {author} {\bibfnamefont {M.}~\bibnamefont
  {Aoki}}, \bibinfo {author} {\bibfnamefont {S.}~\bibnamefont {Kanemura}},
  \bibinfo {author} {\bibfnamefont {K.}~\bibnamefont {Tsumura}}, \ and\
  \bibinfo {author} {\bibfnamefont {K.}~\bibnamefont {Yagyu}},\ }\href
  {\doibase 10.1103/PhysRevD.80.015017} {\bibfield  {journal} {\bibinfo
  {journal} {Phys. Rev.}\ }\textbf {\bibinfo {volume} {D80}},\ \bibinfo {pages}
  {015017} (\bibinfo {year} {2009})},\ \Eprint {http://arxiv.org/abs/0902.4665}
  {arXiv:0902.4665 [hep-ph]} \BibitemShut {NoStop}%
\bibitem [{\citenamefont {Dorsch}\ \emph {et~al.}(2013)\citenamefont {Dorsch},
  \citenamefont {Huber},\ and\ \citenamefont {No}}]{Dorsch:2013wja}%
  \BibitemOpen
  \bibfield  {author} {\bibinfo {author} {\bibfnamefont {G.~C.}\ \bibnamefont
  {Dorsch}}, \bibinfo {author} {\bibfnamefont {S.~J.}\ \bibnamefont {Huber}}, \
  and\ \bibinfo {author} {\bibfnamefont {J.~M.}\ \bibnamefont {No}},\ }\href
  {\doibase 10.1007/JHEP10(2013)029} {\bibfield  {journal} {\bibinfo  {journal}
  {JHEP}\ }\textbf {\bibinfo {volume} {10}},\ \bibinfo {pages} {029} (\bibinfo
  {year} {2013})},\ \Eprint {http://arxiv.org/abs/1305.6610} {arXiv:1305.6610
  [hep-ph]} \BibitemShut {NoStop}%
\bibitem [{\citenamefont {Dorsch}\ \emph {et~al.}(2014)\citenamefont {Dorsch},
  \citenamefont {Huber}, \citenamefont {Mimasu},\ and\ \citenamefont
  {No}}]{Dorsch:2014qja}%
  \BibitemOpen
  \bibfield  {author} {\bibinfo {author} {\bibfnamefont {G.~C.}\ \bibnamefont
  {Dorsch}}, \bibinfo {author} {\bibfnamefont {S.~J.}\ \bibnamefont {Huber}},
  \bibinfo {author} {\bibfnamefont {K.}~\bibnamefont {Mimasu}}, \ and\ \bibinfo
  {author} {\bibfnamefont {J.~M.}\ \bibnamefont {No}},\ }\href {\doibase
  10.1103/PhysRevLett.113.211802} {\bibfield  {journal} {\bibinfo  {journal}
  {Phys. Rev. Lett.}\ }\textbf {\bibinfo {volume} {113}},\ \bibinfo {pages}
  {211802} (\bibinfo {year} {2014})},\ \Eprint {http://arxiv.org/abs/1405.5537}
  {arXiv:1405.5537 [hep-ph]} \BibitemShut {NoStop}%
\bibitem [{\citenamefont {Basler}\ \emph {et~al.}(2017)\citenamefont {Basler},
  \citenamefont {Krause}, \citenamefont {Muhlleitner}, \citenamefont
  {Wittbrodt},\ and\ \citenamefont {Wlotzka}}]{Basler:2016obg}%
  \BibitemOpen
  \bibfield  {author} {\bibinfo {author} {\bibfnamefont {P.}~\bibnamefont
  {Basler}}, \bibinfo {author} {\bibfnamefont {M.}~\bibnamefont {Krause}},
  \bibinfo {author} {\bibfnamefont {M.}~\bibnamefont {Muhlleitner}}, \bibinfo
  {author} {\bibfnamefont {J.}~\bibnamefont {Wittbrodt}}, \ and\ \bibinfo
  {author} {\bibfnamefont {A.}~\bibnamefont {Wlotzka}},\ }\href {\doibase
  10.1007/JHEP02(2017)121} {\bibfield  {journal} {\bibinfo  {journal} {JHEP}\
  }\textbf {\bibinfo {volume} {02}},\ \bibinfo {pages} {121} (\bibinfo {year}
  {2017})},\ \Eprint {http://arxiv.org/abs/1612.04086} {arXiv:1612.04086
  [hep-ph]} \BibitemShut {NoStop}%
\bibitem [{\citenamefont {Cline}\ and\ \citenamefont
  {Lemieux}(1997)}]{Cline:1996mga}%
  \BibitemOpen
  \bibfield  {author} {\bibinfo {author} {\bibfnamefont {J.~M.}\ \bibnamefont
  {Cline}}\ and\ \bibinfo {author} {\bibfnamefont {P.-A.}\ \bibnamefont
  {Lemieux}},\ }\href {\doibase 10.1103/PhysRevD.55.3873} {\bibfield  {journal}
  {\bibinfo  {journal} {Phys. Rev.}\ }\textbf {\bibinfo {volume} {D55}},\
  \bibinfo {pages} {3873} (\bibinfo {year} {1997})},\ \Eprint
  {http://arxiv.org/abs/hep-ph/9609240} {arXiv:hep-ph/9609240 [hep-ph]}
  \BibitemShut {NoStop}%
\bibitem [{\citenamefont {Fromme}\ \emph {et~al.}(2006)\citenamefont {Fromme},
  \citenamefont {Huber},\ and\ \citenamefont {Seniuch}}]{Fromme:2006cm}%
  \BibitemOpen
  \bibfield  {author} {\bibinfo {author} {\bibfnamefont {L.}~\bibnamefont
  {Fromme}}, \bibinfo {author} {\bibfnamefont {S.~J.}\ \bibnamefont {Huber}}, \
  and\ \bibinfo {author} {\bibfnamefont {M.}~\bibnamefont {Seniuch}},\ }\href
  {\doibase 10.1088/1126-6708/2006/11/038} {\bibfield  {journal} {\bibinfo
  {journal} {JHEP}\ }\textbf {\bibinfo {volume} {11}},\ \bibinfo {pages} {038}
  (\bibinfo {year} {2006})},\ \Eprint {http://arxiv.org/abs/hep-ph/0605242}
  {arXiv:hep-ph/0605242 [hep-ph]} \BibitemShut {NoStop}%
\bibitem [{\citenamefont {Haarr}\ \emph {et~al.}(2016)\citenamefont {Haarr},
  \citenamefont {Kvellestad},\ and\ \citenamefont {Petersen}}]{Haarr:2016qzq}%
  \BibitemOpen
  \bibfield  {author} {\bibinfo {author} {\bibfnamefont {A.}~\bibnamefont
  {Haarr}}, \bibinfo {author} {\bibfnamefont {A.}~\bibnamefont {Kvellestad}}, \
  and\ \bibinfo {author} {\bibfnamefont {T.~C.}\ \bibnamefont {Petersen}},\
  }\href@noop {} {\  (\bibinfo {year} {2016})},\ \Eprint
  {http://arxiv.org/abs/1611.05757} {arXiv:1611.05757 [hep-ph]} \BibitemShut
  {NoStop}%
\bibitem [{\citenamefont {Basler}\ \emph
  {et~al.}(2018{\natexlab{b}})\citenamefont {Basler}, \citenamefont
  {Mühlleitner},\ and\ \citenamefont {Wittbrodt}}]{Basler:2017uxn}%
  \BibitemOpen
  \bibfield  {author} {\bibinfo {author} {\bibfnamefont {P.}~\bibnamefont
  {Basler}}, \bibinfo {author} {\bibfnamefont {M.}~\bibnamefont
  {Mühlleitner}}, \ and\ \bibinfo {author} {\bibfnamefont {J.}~\bibnamefont
  {Wittbrodt}},\ }\href {\doibase 10.1007/JHEP03(2018)061} {\bibfield
  {journal} {\bibinfo  {journal} {JHEP}\ }\textbf {\bibinfo {volume} {03}},\
  \bibinfo {pages} {061} (\bibinfo {year} {2018}{\natexlab{b}})},\ \Eprint
  {http://arxiv.org/abs/1711.04097} {arXiv:1711.04097 [hep-ph]} \BibitemShut
  {NoStop}%
\bibitem [{\citenamefont {Bernon}\ \emph {et~al.}(2018)\citenamefont {Bernon},
  \citenamefont {Bian},\ and\ \citenamefont {Jiang}}]{Bernon:2017jgv}%
  \BibitemOpen
  \bibfield  {author} {\bibinfo {author} {\bibfnamefont {J.}~\bibnamefont
  {Bernon}}, \bibinfo {author} {\bibfnamefont {L.}~\bibnamefont {Bian}}, \ and\
  \bibinfo {author} {\bibfnamefont {Y.}~\bibnamefont {Jiang}},\ }\href
  {\doibase 10.1007/JHEP05(2018)151} {\bibfield  {journal} {\bibinfo  {journal}
  {JHEP}\ }\textbf {\bibinfo {volume} {05}},\ \bibinfo {pages} {151} (\bibinfo
  {year} {2018})},\ \Eprint {http://arxiv.org/abs/1712.08430} {arXiv:1712.08430
  [hep-ph]} \BibitemShut {NoStop}%
\bibitem [{\citenamefont {Wang}\ \emph {et~al.}(2019)\citenamefont {Wang},
  \citenamefont {Huang},\ and\ \citenamefont {Zhang}}]{Wang:2019pet}%
  \BibitemOpen
  \bibfield  {author} {\bibinfo {author} {\bibfnamefont {X.}~\bibnamefont
  {Wang}}, \bibinfo {author} {\bibfnamefont {F.~P.}\ \bibnamefont {Huang}}, \
  and\ \bibinfo {author} {\bibfnamefont {X.}~\bibnamefont {Zhang}},\
  }\href@noop {} {\  (\bibinfo {year} {2019})},\ \Eprint
  {http://arxiv.org/abs/1909.02978} {arXiv:1909.02978 [hep-ph]} \BibitemShut
  {NoStop}%
\bibitem [{\citenamefont {Kosowsky}\ \emph
  {et~al.}(1992{\natexlab{a}})\citenamefont {Kosowsky}, \citenamefont
  {Turner},\ and\ \citenamefont {Watkins}}]{Kosowsky:1991ua}%
  \BibitemOpen
  \bibfield  {author} {\bibinfo {author} {\bibfnamefont {A.}~\bibnamefont
  {Kosowsky}}, \bibinfo {author} {\bibfnamefont {M.~S.}\ \bibnamefont
  {Turner}}, \ and\ \bibinfo {author} {\bibfnamefont {R.}~\bibnamefont
  {Watkins}},\ }\href {\doibase 10.1103/PhysRevD.45.4514} {\bibfield  {journal}
  {\bibinfo  {journal} {Phys. Rev.}\ }\textbf {\bibinfo {volume} {D45}},\
  \bibinfo {pages} {4514} (\bibinfo {year} {1992}{\natexlab{a}})}\BibitemShut
  {NoStop}%
\bibitem [{\citenamefont {Paul}\ \emph {et~al.}(2019)\citenamefont {Paul},
  \citenamefont {Banerjee},\ and\ \citenamefont {Majumdar}}]{Paul:2019pgt}%
  \BibitemOpen
  \bibfield  {author} {\bibinfo {author} {\bibfnamefont {A.}~\bibnamefont
  {Paul}}, \bibinfo {author} {\bibfnamefont {B.}~\bibnamefont {Banerjee}}, \
  and\ \bibinfo {author} {\bibfnamefont {D.}~\bibnamefont {Majumdar}},\ }\href
  {\doibase 10.1088/1475-7516/2019/10/062} {\bibfield  {journal} {\bibinfo
  {journal} {JCAP}\ }\textbf {\bibinfo {volume} {1910}},\ \bibinfo {pages}
  {062} (\bibinfo {year} {2019})},\ \Eprint {http://arxiv.org/abs/1908.00829}
  {arXiv:1908.00829 [hep-ph]} \BibitemShut {NoStop}%
\bibitem [{\citenamefont {Kosowsky}\ and\ \citenamefont
  {Turner}(1993)}]{Kosowsky:1992vn}%
  \BibitemOpen
  \bibfield  {author} {\bibinfo {author} {\bibfnamefont {A.}~\bibnamefont
  {Kosowsky}}\ and\ \bibinfo {author} {\bibfnamefont {M.~S.}\ \bibnamefont
  {Turner}},\ }\href {\doibase 10.1103/PhysRevD.47.4372} {\bibfield  {journal}
  {\bibinfo  {journal} {Phys. Rev.}\ }\textbf {\bibinfo {volume} {D47}},\
  \bibinfo {pages} {4372} (\bibinfo {year} {1993})},\ \Eprint
  {http://arxiv.org/abs/astro-ph/9211004} {arXiv:astro-ph/9211004 [astro-ph]}
  \BibitemShut {NoStop}%
\bibitem [{\citenamefont {Huber}\ and\ \citenamefont
  {Konstandin}(2008)}]{Huber:2008hg}%
  \BibitemOpen
  \bibfield  {author} {\bibinfo {author} {\bibfnamefont {S.~J.}\ \bibnamefont
  {Huber}}\ and\ \bibinfo {author} {\bibfnamefont {T.}~\bibnamefont
  {Konstandin}},\ }\href {\doibase 10.1088/1475-7516/2008/09/022} {\bibfield
  {journal} {\bibinfo  {journal} {JCAP}\ }\textbf {\bibinfo {volume} {0809}},\
  \bibinfo {pages} {022} (\bibinfo {year} {2008})},\ \Eprint
  {http://arxiv.org/abs/0806.1828} {arXiv:0806.1828 [hep-ph]} \BibitemShut
  {NoStop}%
\bibitem [{\citenamefont {Kosowsky}\ \emph
  {et~al.}(1992{\natexlab{b}})\citenamefont {Kosowsky}, \citenamefont
  {Turner},\ and\ \citenamefont {Watkins}}]{Kosowsky:1992rz}%
  \BibitemOpen
  \bibfield  {author} {\bibinfo {author} {\bibfnamefont {A.}~\bibnamefont
  {Kosowsky}}, \bibinfo {author} {\bibfnamefont {M.~S.}\ \bibnamefont
  {Turner}}, \ and\ \bibinfo {author} {\bibfnamefont {R.}~\bibnamefont
  {Watkins}},\ }\href {\doibase 10.1103/PhysRevLett.69.2026} {\bibfield
  {journal} {\bibinfo  {journal} {Phys. Rev. Lett.}\ }\textbf {\bibinfo
  {volume} {69}},\ \bibinfo {pages} {2026} (\bibinfo {year}
  {1992}{\natexlab{b}})}\BibitemShut {NoStop}%
\bibitem [{\citenamefont {Kamionkowski}\ \emph {et~al.}(1994)\citenamefont
  {Kamionkowski}, \citenamefont {Kosowsky},\ and\ \citenamefont
  {Turner}}]{Kamionkowski:1993fg}%
  \BibitemOpen
  \bibfield  {author} {\bibinfo {author} {\bibfnamefont {M.}~\bibnamefont
  {Kamionkowski}}, \bibinfo {author} {\bibfnamefont {A.}~\bibnamefont
  {Kosowsky}}, \ and\ \bibinfo {author} {\bibfnamefont {M.~S.}\ \bibnamefont
  {Turner}},\ }\href {\doibase 10.1103/PhysRevD.49.2837} {\bibfield  {journal}
  {\bibinfo  {journal} {Phys. Rev.}\ }\textbf {\bibinfo {volume} {D49}},\
  \bibinfo {pages} {2837} (\bibinfo {year} {1994})},\ \Eprint
  {http://arxiv.org/abs/astro-ph/9310044} {arXiv:astro-ph/9310044 [astro-ph]}
  \BibitemShut {NoStop}%
\bibitem [{\citenamefont {Caprini}\ \emph {et~al.}(2008)\citenamefont
  {Caprini}, \citenamefont {Durrer},\ and\ \citenamefont
  {Servant}}]{Caprini:2007xq}%
  \BibitemOpen
  \bibfield  {author} {\bibinfo {author} {\bibfnamefont {C.}~\bibnamefont
  {Caprini}}, \bibinfo {author} {\bibfnamefont {R.}~\bibnamefont {Durrer}}, \
  and\ \bibinfo {author} {\bibfnamefont {G.}~\bibnamefont {Servant}},\ }\href
  {\doibase 10.1103/PhysRevD.77.124015} {\bibfield  {journal} {\bibinfo
  {journal} {Phys. Rev.}\ }\textbf {\bibinfo {volume} {D77}},\ \bibinfo {pages}
  {124015} (\bibinfo {year} {2008})},\ \Eprint {http://arxiv.org/abs/0711.2593}
  {arXiv:0711.2593 [astro-ph]} \BibitemShut {NoStop}%
\bibitem [{\citenamefont {Hindmarsh}\ \emph {et~al.}(2014)\citenamefont
  {Hindmarsh}, \citenamefont {Huber}, \citenamefont {Rummukainen},\ and\
  \citenamefont {Weir}}]{Hindmarsh:2013xza}%
  \BibitemOpen
  \bibfield  {author} {\bibinfo {author} {\bibfnamefont {M.}~\bibnamefont
  {Hindmarsh}}, \bibinfo {author} {\bibfnamefont {S.~J.}\ \bibnamefont
  {Huber}}, \bibinfo {author} {\bibfnamefont {K.}~\bibnamefont {Rummukainen}},
  \ and\ \bibinfo {author} {\bibfnamefont {D.~J.}\ \bibnamefont {Weir}},\
  }\href {\doibase 10.1103/PhysRevLett.112.041301} {\bibfield  {journal}
  {\bibinfo  {journal} {Phys. Rev. Lett.}\ }\textbf {\bibinfo {volume} {112}},\
  \bibinfo {pages} {041301} (\bibinfo {year} {2014})},\ \Eprint
  {http://arxiv.org/abs/1304.2433} {arXiv:1304.2433 [hep-ph]} \BibitemShut
  {NoStop}%
\bibitem [{\citenamefont {Giblin}\ and\ \citenamefont
  {Mertens}(2013)}]{Giblin:2013kea}%
  \BibitemOpen
  \bibfield  {author} {\bibinfo {author} {\bibfnamefont {J.~T.}\ \bibnamefont
  {Giblin}, \bibfnamefont {Jr.}}\ and\ \bibinfo {author} {\bibfnamefont
  {J.~B.}\ \bibnamefont {Mertens}},\ }\href {\doibase 10.1007/JHEP12(2013)042}
  {\bibfield  {journal} {\bibinfo  {journal} {JHEP}\ }\textbf {\bibinfo
  {volume} {12}},\ \bibinfo {pages} {042} (\bibinfo {year} {2013})},\ \Eprint
  {http://arxiv.org/abs/1310.2948} {arXiv:1310.2948 [hep-th]} \BibitemShut
  {NoStop}%
\bibitem [{\citenamefont {Giblin}\ and\ \citenamefont
  {Mertens}(2014)}]{Giblin:2014qia}%
  \BibitemOpen
  \bibfield  {author} {\bibinfo {author} {\bibfnamefont {J.~T.}\ \bibnamefont
  {Giblin}}\ and\ \bibinfo {author} {\bibfnamefont {J.~B.}\ \bibnamefont
  {Mertens}},\ }\href {\doibase 10.1103/PhysRevD.90.023532} {\bibfield
  {journal} {\bibinfo  {journal} {Phys. Rev.}\ }\textbf {\bibinfo {volume}
  {D90}},\ \bibinfo {pages} {023532} (\bibinfo {year} {2014})},\ \Eprint
  {http://arxiv.org/abs/1405.4005} {arXiv:1405.4005 [astro-ph.CO]} \BibitemShut
  {NoStop}%
\bibitem [{\citenamefont {Hindmarsh}\ \emph {et~al.}(2015)\citenamefont
  {Hindmarsh}, \citenamefont {Huber}, \citenamefont {Rummukainen},\ and\
  \citenamefont {Weir}}]{Hindmarsh:2015qta}%
  \BibitemOpen
  \bibfield  {author} {\bibinfo {author} {\bibfnamefont {M.}~\bibnamefont
  {Hindmarsh}}, \bibinfo {author} {\bibfnamefont {S.~J.}\ \bibnamefont
  {Huber}}, \bibinfo {author} {\bibfnamefont {K.}~\bibnamefont {Rummukainen}},
  \ and\ \bibinfo {author} {\bibfnamefont {D.~J.}\ \bibnamefont {Weir}},\
  }\href {\doibase 10.1103/PhysRevD.92.123009} {\bibfield  {journal} {\bibinfo
  {journal} {Phys. Rev.}\ }\textbf {\bibinfo {volume} {D92}},\ \bibinfo {pages}
  {123009} (\bibinfo {year} {2015})},\ \Eprint
  {http://arxiv.org/abs/1504.03291} {arXiv:1504.03291 [astro-ph.CO]}
  \BibitemShut {NoStop}%
\bibitem [{\citenamefont {Caprini}\ and\ \citenamefont
  {Durrer}(2006)}]{Caprini:2006jb}%
  \BibitemOpen
  \bibfield  {author} {\bibinfo {author} {\bibfnamefont {C.}~\bibnamefont
  {Caprini}}\ and\ \bibinfo {author} {\bibfnamefont {R.}~\bibnamefont
  {Durrer}},\ }\href {\doibase 10.1103/PhysRevD.74.063521} {\bibfield
  {journal} {\bibinfo  {journal} {Phys. Rev.}\ }\textbf {\bibinfo {volume}
  {D74}},\ \bibinfo {pages} {063521} (\bibinfo {year} {2006})},\ \Eprint
  {http://arxiv.org/abs/astro-ph/0603476} {arXiv:astro-ph/0603476 [astro-ph]}
  \BibitemShut {NoStop}%
\bibitem [{\citenamefont {Kahniashvili}\ \emph
  {et~al.}(2008{\natexlab{a}})\citenamefont {Kahniashvili}, \citenamefont
  {Kosowsky}, \citenamefont {Gogoberidze},\ and\ \citenamefont
  {Maravin}}]{Kahniashvili:2008pf}%
  \BibitemOpen
  \bibfield  {author} {\bibinfo {author} {\bibfnamefont {T.}~\bibnamefont
  {Kahniashvili}}, \bibinfo {author} {\bibfnamefont {A.}~\bibnamefont
  {Kosowsky}}, \bibinfo {author} {\bibfnamefont {G.}~\bibnamefont
  {Gogoberidze}}, \ and\ \bibinfo {author} {\bibfnamefont {Y.}~\bibnamefont
  {Maravin}},\ }\href {\doibase 10.1103/PhysRevD.78.043003} {\bibfield
  {journal} {\bibinfo  {journal} {Phys. Rev.}\ }\textbf {\bibinfo {volume}
  {D78}},\ \bibinfo {pages} {043003} (\bibinfo {year} {2008}{\natexlab{a}})},\
  \Eprint {http://arxiv.org/abs/0806.0293} {arXiv:0806.0293 [astro-ph]}
  \BibitemShut {NoStop}%
\bibitem [{\citenamefont {Kahniashvili}\ \emph
  {et~al.}(2008{\natexlab{b}})\citenamefont {Kahniashvili}, \citenamefont
  {Campanelli}, \citenamefont {Gogoberidze}, \citenamefont {Maravin},\ and\
  \citenamefont {Ratra}}]{Kahniashvili:2008pe}%
  \BibitemOpen
  \bibfield  {author} {\bibinfo {author} {\bibfnamefont {T.}~\bibnamefont
  {Kahniashvili}}, \bibinfo {author} {\bibfnamefont {L.}~\bibnamefont
  {Campanelli}}, \bibinfo {author} {\bibfnamefont {G.}~\bibnamefont
  {Gogoberidze}}, \bibinfo {author} {\bibfnamefont {Y.}~\bibnamefont
  {Maravin}}, \ and\ \bibinfo {author} {\bibfnamefont {B.}~\bibnamefont
  {Ratra}},\ }\href {\doibase 10.1103/PhysRevD.78.123006,
  10.1103/PhysRevD.79.109901} {\bibfield  {journal} {\bibinfo  {journal} {Phys.
  Rev.}\ }\textbf {\bibinfo {volume} {D78}},\ \bibinfo {pages} {123006}
  (\bibinfo {year} {2008}{\natexlab{b}})},\ \bibinfo {note} {[Erratum: Phys.
  Rev.D79,109901(2009)]},\ \Eprint {http://arxiv.org/abs/0809.1899}
  {arXiv:0809.1899 [astro-ph]} \BibitemShut {NoStop}%
\bibitem [{\citenamefont {Kahniashvili}\ \emph {et~al.}(2010)\citenamefont
  {Kahniashvili}, \citenamefont {Kisslinger},\ and\ \citenamefont
  {Stevens}}]{Kahniashvili:2009mf}%
  \BibitemOpen
  \bibfield  {author} {\bibinfo {author} {\bibfnamefont {T.}~\bibnamefont
  {Kahniashvili}}, \bibinfo {author} {\bibfnamefont {L.}~\bibnamefont
  {Kisslinger}}, \ and\ \bibinfo {author} {\bibfnamefont {T.}~\bibnamefont
  {Stevens}},\ }\href {\doibase 10.1103/PhysRevD.81.023004} {\bibfield
  {journal} {\bibinfo  {journal} {Phys. Rev.}\ }\textbf {\bibinfo {volume}
  {D81}},\ \bibinfo {pages} {023004} (\bibinfo {year} {2010})},\ \Eprint
  {http://arxiv.org/abs/0905.0643} {arXiv:0905.0643 [astro-ph.CO]} \BibitemShut
  {NoStop}%
\bibitem [{\citenamefont {Caprini}\ \emph {et~al.}(2009)\citenamefont
  {Caprini}, \citenamefont {Durrer},\ and\ \citenamefont
  {Servant}}]{Caprini:2009yp}%
  \BibitemOpen
  \bibfield  {author} {\bibinfo {author} {\bibfnamefont {C.}~\bibnamefont
  {Caprini}}, \bibinfo {author} {\bibfnamefont {R.}~\bibnamefont {Durrer}}, \
  and\ \bibinfo {author} {\bibfnamefont {G.}~\bibnamefont {Servant}},\ }\href
  {\doibase 10.1088/1475-7516/2009/12/024} {\bibfield  {journal} {\bibinfo
  {journal} {JCAP}\ }\textbf {\bibinfo {volume} {0912}},\ \bibinfo {pages}
  {024} (\bibinfo {year} {2009})},\ \Eprint {http://arxiv.org/abs/0909.0622}
  {arXiv:0909.0622 [astro-ph.CO]} \BibitemShut {NoStop}%
\bibitem [{\citenamefont {Gong}\ \emph {et~al.}(2015)\citenamefont {Gong} \emph
  {et~al.}}]{Gong:2014mca}%
  \BibitemOpen
  \bibfield  {author} {\bibinfo {author} {\bibfnamefont {X.}~\bibnamefont
  {Gong}} \emph {et~al.},\ }\bibfield  {booktitle} {\emph {\bibinfo {booktitle}
  {{Proceedings, 10th International LISA Symposium: Gainesville, Florida, USA,
  May 18-23, 2014}}},\ }\href {\doibase 10.1088/1742-6596/610/1/012011}
  {\bibfield  {journal} {\bibinfo  {journal} {J. Phys. Conf. Ser.}\ }\textbf
  {\bibinfo {volume} {610}},\ \bibinfo {pages} {012011} (\bibinfo {year}
  {2015})},\ \Eprint {http://arxiv.org/abs/1410.7296} {arXiv:1410.7296 [gr-qc]}
  \BibitemShut {NoStop}%
\bibitem [{\citenamefont {Harry}\ \emph {et~al.}(2006)\citenamefont {Harry},
  \citenamefont {Fritschel}, \citenamefont {Shaddock}, \citenamefont
  {Folkner},\ and\ \citenamefont {Phinney}}]{Harry:2006fi}%
  \BibitemOpen
  \bibfield  {author} {\bibinfo {author} {\bibfnamefont {G.~M.}\ \bibnamefont
  {Harry}}, \bibinfo {author} {\bibfnamefont {P.}~\bibnamefont {Fritschel}},
  \bibinfo {author} {\bibfnamefont {D.~A.}\ \bibnamefont {Shaddock}}, \bibinfo
  {author} {\bibfnamefont {W.}~\bibnamefont {Folkner}}, \ and\ \bibinfo
  {author} {\bibfnamefont {E.~S.}\ \bibnamefont {Phinney}},\ }\href {\doibase
  10.1088/0264-9381/23/24/C01, 10.1088/0264-9381/23/15/008} {\bibfield
  {journal} {\bibinfo  {journal} {Class. Quant. Grav.}\ }\textbf {\bibinfo
  {volume} {23}},\ \bibinfo {pages} {4887} (\bibinfo {year} {2006})},\ \bibinfo
  {note} {[Erratum: Class. Quant. Grav.23,7361(2006)]}\BibitemShut {NoStop}%
\bibitem [{\citenamefont {Seto}\ \emph {et~al.}(2001)\citenamefont {Seto},
  \citenamefont {Kawamura},\ and\ \citenamefont {Nakamura}}]{Seto:2001qf}%
  \BibitemOpen
  \bibfield  {author} {\bibinfo {author} {\bibfnamefont {N.}~\bibnamefont
  {Seto}}, \bibinfo {author} {\bibfnamefont {S.}~\bibnamefont {Kawamura}}, \
  and\ \bibinfo {author} {\bibfnamefont {T.}~\bibnamefont {Nakamura}},\ }\href
  {\doibase 10.1103/PhysRevLett.87.221103} {\bibfield  {journal} {\bibinfo
  {journal} {Phys. Rev. Lett.}\ }\textbf {\bibinfo {volume} {87}},\ \bibinfo
  {pages} {221103} (\bibinfo {year} {2001})},\ \Eprint
  {http://arxiv.org/abs/astro-ph/0108011} {arXiv:astro-ph/0108011 [astro-ph]}
  \BibitemShut {NoStop}%
\bibitem [{\citenamefont {Caprini}\ \emph {et~al.}(2016)\citenamefont {Caprini}
  \emph {et~al.}}]{Caprini:2015zlo}%
  \BibitemOpen
  \bibfield  {author} {\bibinfo {author} {\bibfnamefont {C.}~\bibnamefont
  {Caprini}} \emph {et~al.},\ }\href {\doibase 10.1088/1475-7516/2016/04/001}
  {\bibfield  {journal} {\bibinfo  {journal} {JCAP}\ }\textbf {\bibinfo
  {volume} {1604}},\ \bibinfo {pages} {001} (\bibinfo {year} {2016})},\ \Eprint
  {http://arxiv.org/abs/1512.06239} {arXiv:1512.06239 [astro-ph.CO]}
  \BibitemShut {NoStop}%
\bibitem [{\citenamefont {and}(2010)}]{Harry_2010}%
  \BibitemOpen
  \bibfield  {author} {\bibinfo {author} {\bibfnamefont {G.~M.~H.}\
  \bibnamefont {and}},\ }\href {\doibase 10.1088/0264-9381/27/8/084006}
  {\bibfield  {journal} {\bibinfo  {journal} {Classical and Quantum Gravity}\
  }\textbf {\bibinfo {volume} {27}},\ \bibinfo {pages} {084006} (\bibinfo
  {year} {2010})}\BibitemShut {NoStop}%
\bibitem [{\citenamefont {Schwaller}(2015)}]{Schwaller:2015tja}%
  \BibitemOpen
  \bibfield  {author} {\bibinfo {author} {\bibfnamefont {P.}~\bibnamefont
  {Schwaller}},\ }\href {\doibase 10.1103/PhysRevLett.115.181101} {\bibfield
  {journal} {\bibinfo  {journal} {Phys. Rev. Lett.}\ }\textbf {\bibinfo
  {volume} {115}},\ \bibinfo {pages} {181101} (\bibinfo {year} {2015})},\
  \Eprint {http://arxiv.org/abs/1504.07263} {arXiv:1504.07263 [hep-ph]}
  \BibitemShut {NoStop}%
\bibitem [{\citenamefont {Beniwal}\ \emph {et~al.}(2017)\citenamefont
  {Beniwal}, \citenamefont {Lewicki}, \citenamefont {Wells}, \citenamefont
  {White},\ and\ \citenamefont {Williams}}]{Beniwal:2017eik}%
  \BibitemOpen
  \bibfield  {author} {\bibinfo {author} {\bibfnamefont {A.}~\bibnamefont
  {Beniwal}}, \bibinfo {author} {\bibfnamefont {M.}~\bibnamefont {Lewicki}},
  \bibinfo {author} {\bibfnamefont {J.~D.}\ \bibnamefont {Wells}}, \bibinfo
  {author} {\bibfnamefont {M.}~\bibnamefont {White}}, \ and\ \bibinfo {author}
  {\bibfnamefont {A.~G.}\ \bibnamefont {Williams}},\ }\href {\doibase
  10.1007/JHEP08(2017)108} {\bibfield  {journal} {\bibinfo  {journal} {JHEP}\
  }\textbf {\bibinfo {volume} {08}},\ \bibinfo {pages} {108} (\bibinfo {year}
  {2017})},\ \Eprint {http://arxiv.org/abs/1702.06124} {arXiv:1702.06124
  [hep-ph]} \BibitemShut {NoStop}%
\bibitem [{\citenamefont {Cai}\ \emph {et~al.}(2017)\citenamefont {Cai},
  \citenamefont {Cao}, \citenamefont {Guo}, \citenamefont {Wang},\ and\
  \citenamefont {Yang}}]{Cai:2017cbj}%
  \BibitemOpen
  \bibfield  {author} {\bibinfo {author} {\bibfnamefont {R.-G.}\ \bibnamefont
  {Cai}}, \bibinfo {author} {\bibfnamefont {Z.}~\bibnamefont {Cao}}, \bibinfo
  {author} {\bibfnamefont {Z.-K.}\ \bibnamefont {Guo}}, \bibinfo {author}
  {\bibfnamefont {S.-J.}\ \bibnamefont {Wang}}, \ and\ \bibinfo {author}
  {\bibfnamefont {T.}~\bibnamefont {Yang}},\ }\href {\doibase
  10.1093/nsr/nwx029} {\bibfield  {journal} {\bibinfo  {journal} {Natl. Sci.
  Rev.}\ }\textbf {\bibinfo {volume} {4}},\ \bibinfo {pages} {687} (\bibinfo
  {year} {2017})},\ \Eprint {http://arxiv.org/abs/1703.00187} {arXiv:1703.00187
  [gr-qc]} \BibitemShut {NoStop}%
\bibitem [{\citenamefont {Buckley}\ and\ \citenamefont
  {Peter}(2018)}]{Buckley:2017ijx}%
  \BibitemOpen
  \bibfield  {author} {\bibinfo {author} {\bibfnamefont {M.~R.}\ \bibnamefont
  {Buckley}}\ and\ \bibinfo {author} {\bibfnamefont {A.~H.~G.}\ \bibnamefont
  {Peter}},\ }\href {\doibase 10.1016/j.physrep.2018.07.003} {\bibfield
  {journal} {\bibinfo  {journal} {Phys. Rept.}\ }\textbf {\bibinfo {volume}
  {761}},\ \bibinfo {pages} {1} (\bibinfo {year} {2018})},\ \Eprint
  {http://arxiv.org/abs/1712.06615} {arXiv:1712.06615 [astro-ph.CO]}
  \BibitemShut {NoStop}%
\bibitem [{\citenamefont {Alves}\ \emph {et~al.}(2019)\citenamefont {Alves},
  \citenamefont {Ghosh}, \citenamefont {Guo}, \citenamefont {Sinha},\ and\
  \citenamefont {Vagie}}]{Alves:2018jsw}%
  \BibitemOpen
  \bibfield  {author} {\bibinfo {author} {\bibfnamefont {A.}~\bibnamefont
  {Alves}}, \bibinfo {author} {\bibfnamefont {T.}~\bibnamefont {Ghosh}},
  \bibinfo {author} {\bibfnamefont {H.-K.}\ \bibnamefont {Guo}}, \bibinfo
  {author} {\bibfnamefont {K.}~\bibnamefont {Sinha}}, \ and\ \bibinfo {author}
  {\bibfnamefont {D.}~\bibnamefont {Vagie}},\ }\href {\doibase
  10.1007/JHEP04(2019)052} {\bibfield  {journal} {\bibinfo  {journal} {JHEP}\
  }\textbf {\bibinfo {volume} {04}},\ \bibinfo {pages} {052} (\bibinfo {year}
  {2019})},\ \Eprint {http://arxiv.org/abs/1812.09333} {arXiv:1812.09333
  [hep-ph]} \BibitemShut {NoStop}%
\bibitem [{\citenamefont {Shajiee}\ and\ \citenamefont
  {Tofighi}(2019)}]{Shajiee:2018jdq}%
  \BibitemOpen
  \bibfield  {author} {\bibinfo {author} {\bibfnamefont {V.~R.}\ \bibnamefont
  {Shajiee}}\ and\ \bibinfo {author} {\bibfnamefont {A.}~\bibnamefont
  {Tofighi}},\ }\href {\doibase 10.1140/epjc/s10052-019-6881-6} {\bibfield
  {journal} {\bibinfo  {journal} {Eur. Phys. J.}\ }\textbf {\bibinfo {volume}
  {C79}},\ \bibinfo {pages} {360} (\bibinfo {year} {2019})},\ \Eprint
  {http://arxiv.org/abs/1811.09807} {arXiv:1811.09807 [hep-ph]} \BibitemShut
  {NoStop}%
\bibitem [{\citenamefont {Bian}\ and\ \citenamefont
  {Tang}(2018)}]{Bian:2018mkl}%
  \BibitemOpen
  \bibfield  {author} {\bibinfo {author} {\bibfnamefont {L.}~\bibnamefont
  {Bian}}\ and\ \bibinfo {author} {\bibfnamefont {Y.-L.}\ \bibnamefont
  {Tang}},\ }\href {\doibase 10.1007/JHEP12(2018)006} {\bibfield  {journal}
  {\bibinfo  {journal} {JHEP}\ }\textbf {\bibinfo {volume} {12}},\ \bibinfo
  {pages} {006} (\bibinfo {year} {2018})},\ \Eprint
  {http://arxiv.org/abs/1810.03172} {arXiv:1810.03172 [hep-ph]} \BibitemShut
  {NoStop}%
\bibitem [{\citenamefont {Mohamadnejad}(2019)}]{Mohamadnejad:2019vzg}%
  \BibitemOpen
  \bibfield  {author} {\bibinfo {author} {\bibfnamefont {A.}~\bibnamefont
  {Mohamadnejad}},\ }\href@noop {} {\  (\bibinfo {year} {2019})},\ \Eprint
  {http://arxiv.org/abs/1907.08899} {arXiv:1907.08899 [hep-ph]} \BibitemShut
  {NoStop}%
\bibitem [{\citenamefont {Bertone}\ \emph {et~al.}(2019)\citenamefont {Bertone}
  \emph {et~al.}}]{Bertone:2019irm}%
  \BibitemOpen
  \bibfield  {author} {\bibinfo {author} {\bibfnamefont {G.}~\bibnamefont
  {Bertone}} \emph {et~al.},\ }\href@noop {} {\  (\bibinfo {year} {2019})},\
  \Eprint {http://arxiv.org/abs/1907.10610} {arXiv:1907.10610 [astro-ph.CO]}
  \BibitemShut {NoStop}%
\bibitem [{\citenamefont {Tanabashi}\ and\ \citenamefont
  {Hagiwara}(2018)}]{PhysRevD.98.030001}%
  \BibitemOpen
  \bibfield  {author} {\bibinfo {author} {\bibfnamefont {M.}~\bibnamefont
  {Tanabashi}}\ and\ \bibinfo {author} {\bibfnamefont {e.~a.}\ \bibnamefont
  {Hagiwara}} (\bibinfo {collaboration} {Particle Data Group}),\ }\href
  {\doibase 10.1103/PhysRevD.98.030001} {\bibfield  {journal} {\bibinfo
  {journal} {Phys. Rev. D}\ }\textbf {\bibinfo {volume} {98}},\ \bibinfo
  {pages} {030001} (\bibinfo {year} {2018})}\BibitemShut {NoStop}%
\bibitem [{\citenamefont {Chakrabarty}\ and\ \citenamefont
  {Mukhopadhyaya}(2017)}]{Chakrabarty:2016smc}%
  \BibitemOpen
  \bibfield  {author} {\bibinfo {author} {\bibfnamefont {N.}~\bibnamefont
  {Chakrabarty}}\ and\ \bibinfo {author} {\bibfnamefont {B.}~\bibnamefont
  {Mukhopadhyaya}},\ }\href {\doibase 10.1140/epjc/s10052-017-4705-0}
  {\bibfield  {journal} {\bibinfo  {journal} {Eur. Phys. J.}\ }\textbf
  {\bibinfo {volume} {C77}},\ \bibinfo {pages} {153} (\bibinfo {year}
  {2017})},\ \Eprint {http://arxiv.org/abs/1603.05883} {arXiv:1603.05883
  [hep-ph]} \BibitemShut {NoStop}%
\bibitem [{\citenamefont {Ivanov}(2007)}]{Ivanov:2006yq}%
  \BibitemOpen
  \bibfield  {author} {\bibinfo {author} {\bibfnamefont {I.~P.}\ \bibnamefont
  {Ivanov}},\ }\href {\doibase 10.1103/PhysRevD.76.039902,
  10.1103/PhysRevD.75.035001} {\bibfield  {journal} {\bibinfo  {journal} {Phys.
  Rev.}\ }\textbf {\bibinfo {volume} {D75}},\ \bibinfo {pages} {035001}
  (\bibinfo {year} {2007})},\ \bibinfo {note} {[Erratum: Phys.
  Rev.D76,039902(2007)]},\ \Eprint {http://arxiv.org/abs/hep-ph/0609018}
  {arXiv:hep-ph/0609018 [hep-ph]} \BibitemShut {NoStop}%
\bibitem [{\citenamefont {Maniatis}\ \emph {et~al.}(2006)\citenamefont
  {Maniatis}, \citenamefont {von Manteuffel}, \citenamefont {Nachtmann},\ and\
  \citenamefont {Nagel}}]{Maniatis:2006fs}%
  \BibitemOpen
  \bibfield  {author} {\bibinfo {author} {\bibfnamefont {M.}~\bibnamefont
  {Maniatis}}, \bibinfo {author} {\bibfnamefont {A.}~\bibnamefont {von
  Manteuffel}}, \bibinfo {author} {\bibfnamefont {O.}~\bibnamefont
  {Nachtmann}}, \ and\ \bibinfo {author} {\bibfnamefont {F.}~\bibnamefont
  {Nagel}},\ }\href {\doibase 10.1140/epjc/s10052-006-0016-6} {\bibfield
  {journal} {\bibinfo  {journal} {Eur. Phys. J.}\ }\textbf {\bibinfo {volume}
  {C48}},\ \bibinfo {pages} {805} (\bibinfo {year} {2006})},\ \Eprint
  {http://arxiv.org/abs/hep-ph/0605184} {arXiv:hep-ph/0605184 [hep-ph]}
  \BibitemShut {NoStop}%
\bibitem [{\citenamefont {Xu}(2017)}]{Xu:2017vpq}%
  \BibitemOpen
  \bibfield  {author} {\bibinfo {author} {\bibfnamefont {X.-J.}\ \bibnamefont
  {Xu}},\ }\href {\doibase 10.1103/PhysRevD.95.115019} {\bibfield  {journal}
  {\bibinfo  {journal} {Phys. Rev.}\ }\textbf {\bibinfo {volume} {D95}},\
  \bibinfo {pages} {115019} (\bibinfo {year} {2017})},\ \Eprint
  {http://arxiv.org/abs/1705.08965} {arXiv:1705.08965 [hep-ph]} \BibitemShut
  {NoStop}%
\bibitem [{\citenamefont {Arhrib}(2000)}]{Arhrib:2000is}%
  \BibitemOpen
  \bibfield  {author} {\bibinfo {author} {\bibfnamefont {A.}~\bibnamefont
  {Arhrib}},\ }in\ \href@noop {} {\emph {\bibinfo {booktitle} {{Workshop on
  Noncommutative Geometry, Superstrings and Particle Physics Rabat, Morocco,
  June 16-17, 2000}}}}\ (\bibinfo {year} {2000})\ \Eprint
  {http://arxiv.org/abs/hep-ph/0012353} {arXiv:hep-ph/0012353 [hep-ph]}
  \BibitemShut {NoStop}%
\bibitem [{\citenamefont {Ginzburg}\ and\ \citenamefont
  {Ivanov}(2003)}]{Ginzburg:2003fe}%
  \BibitemOpen
  \bibfield  {author} {\bibinfo {author} {\bibfnamefont {I.~F.}\ \bibnamefont
  {Ginzburg}}\ and\ \bibinfo {author} {\bibfnamefont {I.~P.}\ \bibnamefont
  {Ivanov}},\ }\href@noop {} {\  (\bibinfo {year} {2003})},\ \Eprint
  {http://arxiv.org/abs/hep-ph/0312374} {arXiv:hep-ph/0312374 [hep-ph]}
  \BibitemShut {NoStop}%
\bibitem [{\citenamefont {Ginzburg}\ and\ \citenamefont
  {Ivanov}(2005)}]{Ginzburg:2005dt}%
  \BibitemOpen
  \bibfield  {author} {\bibinfo {author} {\bibfnamefont {I.~F.}\ \bibnamefont
  {Ginzburg}}\ and\ \bibinfo {author} {\bibfnamefont {I.~P.}\ \bibnamefont
  {Ivanov}},\ }\href {\doibase 10.1103/PhysRevD.72.115010} {\bibfield
  {journal} {\bibinfo  {journal} {Phys. Rev.}\ }\textbf {\bibinfo {volume}
  {D72}},\ \bibinfo {pages} {115010} (\bibinfo {year} {2005})},\ \Eprint
  {http://arxiv.org/abs/hep-ph/0508020} {arXiv:hep-ph/0508020 [hep-ph]}
  \BibitemShut {NoStop}%
\bibitem [{\citenamefont {Dorsch}\ \emph {et~al.}(2017)\citenamefont {Dorsch},
  \citenamefont {Huber}, \citenamefont {Konstandin},\ and\ \citenamefont
  {No}}]{Dorsch:2016nrg}%
  \BibitemOpen
  \bibfield  {author} {\bibinfo {author} {\bibfnamefont {G.~C.}\ \bibnamefont
  {Dorsch}}, \bibinfo {author} {\bibfnamefont {S.~J.}\ \bibnamefont {Huber}},
  \bibinfo {author} {\bibfnamefont {T.}~\bibnamefont {Konstandin}}, \ and\
  \bibinfo {author} {\bibfnamefont {J.~M.}\ \bibnamefont {No}},\ }\href
  {\doibase 10.1088/1475-7516/2017/05/052} {\bibfield  {journal} {\bibinfo
  {journal} {JCAP}\ }\textbf {\bibinfo {volume} {1705}},\ \bibinfo {pages}
  {052} (\bibinfo {year} {2017})},\ \Eprint {http://arxiv.org/abs/1611.05874}
  {arXiv:1611.05874 [hep-ph]} \BibitemShut {NoStop}%
\bibitem [{\citenamefont {He}\ \emph {et~al.}(2001)\citenamefont {He},
  \citenamefont {Polonsky},\ and\ \citenamefont {Su}}]{He:2001tp}%
  \BibitemOpen
  \bibfield  {author} {\bibinfo {author} {\bibfnamefont {H.-J.}\ \bibnamefont
  {He}}, \bibinfo {author} {\bibfnamefont {N.}~\bibnamefont {Polonsky}}, \ and\
  \bibinfo {author} {\bibfnamefont {S.-f.}\ \bibnamefont {Su}},\ }\href
  {\doibase 10.1103/PhysRevD.64.053004} {\bibfield  {journal} {\bibinfo
  {journal} {Phys. Rev.}\ }\textbf {\bibinfo {volume} {D64}},\ \bibinfo {pages}
  {053004} (\bibinfo {year} {2001})},\ \Eprint
  {http://arxiv.org/abs/hep-ph/0102144} {arXiv:hep-ph/0102144 [hep-ph]}
  \BibitemShut {NoStop}%
\bibitem [{\citenamefont {Grimus}\ \emph
  {et~al.}(2008{\natexlab{a}})\citenamefont {Grimus}, \citenamefont {Lavoura},
  \citenamefont {Ogreid},\ and\ \citenamefont {Osland}}]{Grimus:2007if}%
  \BibitemOpen
  \bibfield  {author} {\bibinfo {author} {\bibfnamefont {W.}~\bibnamefont
  {Grimus}}, \bibinfo {author} {\bibfnamefont {L.}~\bibnamefont {Lavoura}},
  \bibinfo {author} {\bibfnamefont {O.~M.}\ \bibnamefont {Ogreid}}, \ and\
  \bibinfo {author} {\bibfnamefont {P.}~\bibnamefont {Osland}},\ }\href
  {\doibase 10.1088/0954-3899/35/7/075001} {\bibfield  {journal} {\bibinfo
  {journal} {J. Phys.}\ }\textbf {\bibinfo {volume} {G35}},\ \bibinfo {pages}
  {075001} (\bibinfo {year} {2008}{\natexlab{a}})},\ \Eprint
  {http://arxiv.org/abs/0711.4022} {arXiv:0711.4022 [hep-ph]} \BibitemShut
  {NoStop}%
\bibitem [{\citenamefont {Grimus}\ \emph
  {et~al.}(2008{\natexlab{b}})\citenamefont {Grimus}, \citenamefont {Lavoura},
  \citenamefont {Ogreid},\ and\ \citenamefont {Osland}}]{Grimus:2008nb}%
  \BibitemOpen
  \bibfield  {author} {\bibinfo {author} {\bibfnamefont {W.}~\bibnamefont
  {Grimus}}, \bibinfo {author} {\bibfnamefont {L.}~\bibnamefont {Lavoura}},
  \bibinfo {author} {\bibfnamefont {O.~M.}\ \bibnamefont {Ogreid}}, \ and\
  \bibinfo {author} {\bibfnamefont {P.}~\bibnamefont {Osland}},\ }\href
  {\doibase 10.1016/j.nuclphysb.2008.04.019} {\bibfield  {journal} {\bibinfo
  {journal} {Nucl. Phys.}\ }\textbf {\bibinfo {volume} {B801}},\ \bibinfo
  {pages} {81} (\bibinfo {year} {2008}{\natexlab{b}})},\ \Eprint
  {http://arxiv.org/abs/0802.4353} {arXiv:0802.4353 [hep-ph]} \BibitemShut
  {NoStop}%
\bibitem [{\citenamefont {Cynolter}\ and\ \citenamefont
  {Lendvai}(2008)}]{Cynolter:2008ea}%
  \BibitemOpen
  \bibfield  {author} {\bibinfo {author} {\bibfnamefont {G.}~\bibnamefont
  {Cynolter}}\ and\ \bibinfo {author} {\bibfnamefont {E.}~\bibnamefont
  {Lendvai}},\ }\href {\doibase 10.1140/epjc/s10052-008-0771-7} {\bibfield
  {journal} {\bibinfo  {journal} {Eur. Phys. J.}\ }\textbf {\bibinfo {volume}
  {C58}},\ \bibinfo {pages} {463} (\bibinfo {year} {2008})},\ \Eprint
  {http://arxiv.org/abs/0804.4080} {arXiv:0804.4080 [hep-ph]} \BibitemShut
  {NoStop}%
\bibitem [{\citenamefont {Barbieri}\ \emph {et~al.}(2004)\citenamefont
  {Barbieri}, \citenamefont {Pomarol}, \citenamefont {Rattazzi},\ and\
  \citenamefont {Strumia}}]{Barbieri:2004qk}%
  \BibitemOpen
  \bibfield  {author} {\bibinfo {author} {\bibfnamefont {R.}~\bibnamefont
  {Barbieri}}, \bibinfo {author} {\bibfnamefont {A.}~\bibnamefont {Pomarol}},
  \bibinfo {author} {\bibfnamefont {R.}~\bibnamefont {Rattazzi}}, \ and\
  \bibinfo {author} {\bibfnamefont {A.}~\bibnamefont {Strumia}},\ }\href
  {\doibase 10.1016/j.nuclphysb.2004.10.014} {\bibfield  {journal} {\bibinfo
  {journal} {Nucl. Phys.}\ }\textbf {\bibinfo {volume} {B703}},\ \bibinfo
  {pages} {127} (\bibinfo {year} {2004})},\ \Eprint
  {http://arxiv.org/abs/hep-ph/0405040} {arXiv:hep-ph/0405040 [hep-ph]}
  \BibitemShut {NoStop}%
\bibitem [{\citenamefont {Abbiendi}\ \emph {et~al.}(2013)\citenamefont
  {Abbiendi} \emph {et~al.}}]{Abbiendi:2013hk}%
  \BibitemOpen
  \bibfield  {author} {\bibinfo {author} {\bibfnamefont {G.}~\bibnamefont
  {Abbiendi}} \emph {et~al.} (\bibinfo {collaboration} {ALEPH, DELPHI, L3,
  OPAL, LEP}),\ }\href {\doibase 10.1140/epjc/s10052-013-2463-1} {\bibfield
  {journal} {\bibinfo  {journal} {Eur. Phys. J.}\ }\textbf {\bibinfo {volume}
  {C73}},\ \bibinfo {pages} {2463} (\bibinfo {year} {2013})},\ \Eprint
  {http://arxiv.org/abs/1301.6065} {arXiv:1301.6065 [hep-ex]} \BibitemShut
  {NoStop}%
\bibitem [{\citenamefont {Arbey}\ \emph {et~al.}(2018)\citenamefont {Arbey},
  \citenamefont {Mahmoudi}, \citenamefont {Stal},\ and\ \citenamefont
  {Stefaniak}}]{Arbey:2017gmh}%
  \BibitemOpen
  \bibfield  {author} {\bibinfo {author} {\bibfnamefont {A.}~\bibnamefont
  {Arbey}}, \bibinfo {author} {\bibfnamefont {F.}~\bibnamefont {Mahmoudi}},
  \bibinfo {author} {\bibfnamefont {O.}~\bibnamefont {Stal}}, \ and\ \bibinfo
  {author} {\bibfnamefont {T.}~\bibnamefont {Stefaniak}},\ }\href {\doibase
  10.1140/epjc/s10052-018-5651-1} {\bibfield  {journal} {\bibinfo  {journal}
  {Eur. Phys. J.}\ }\textbf {\bibinfo {volume} {C78}},\ \bibinfo {pages} {182}
  (\bibinfo {year} {2018})},\ \Eprint {http://arxiv.org/abs/1706.07414}
  {arXiv:1706.07414 [hep-ph]} \BibitemShut {NoStop}%
\bibitem [{\citenamefont {Akeroyd}(1999)}]{Akeroyd:1998dt}%
  \BibitemOpen
  \bibfield  {author} {\bibinfo {author} {\bibfnamefont {A.~G.}\ \bibnamefont
  {Akeroyd}},\ }\href {\doibase 10.1016/S0550-3213(98)00845-1} {\bibfield
  {journal} {\bibinfo  {journal} {Nucl. Phys.}\ }\textbf {\bibinfo {volume}
  {B544}},\ \bibinfo {pages} {557} (\bibinfo {year} {1999})},\ \Eprint
  {http://arxiv.org/abs/hep-ph/9806337} {arXiv:hep-ph/9806337 [hep-ph]}
  \BibitemShut {NoStop}%
\bibitem [{\citenamefont {Khachatryan}\ \emph
  {et~al.}(2015{\natexlab{a}})\citenamefont {Khachatryan} \emph
  {et~al.}}]{Khachatryan:2015qxa}%
  \BibitemOpen
  \bibfield  {author} {\bibinfo {author} {\bibfnamefont {V.}~\bibnamefont
  {Khachatryan}} \emph {et~al.} (\bibinfo {collaboration} {CMS}),\ }\href
  {\doibase 10.1007/JHEP11(2015)018} {\bibfield  {journal} {\bibinfo  {journal}
  {JHEP}\ }\textbf {\bibinfo {volume} {11}},\ \bibinfo {pages} {018} (\bibinfo
  {year} {2015}{\natexlab{a}})},\ \Eprint {http://arxiv.org/abs/1508.07774}
  {arXiv:1508.07774 [hep-ex]} \BibitemShut {NoStop}%
\bibitem [{\citenamefont {Aad}\ \emph {et~al.}(2013)\citenamefont {Aad} \emph
  {et~al.}}]{Aad:2013hla}%
  \BibitemOpen
  \bibfield  {author} {\bibinfo {author} {\bibfnamefont {G.}~\bibnamefont
  {Aad}} \emph {et~al.} (\bibinfo {collaboration} {ATLAS}),\ }\href {\doibase
  10.1140/epjc/s10052-013-2465-z} {\bibfield  {journal} {\bibinfo  {journal}
  {Eur. Phys. J.}\ }\textbf {\bibinfo {volume} {C73}},\ \bibinfo {pages} {2465}
  (\bibinfo {year} {2013})},\ \Eprint {http://arxiv.org/abs/1302.3694}
  {arXiv:1302.3694 [hep-ex]} \BibitemShut {NoStop}%
\bibitem [{\citenamefont {Aad}\ \emph {et~al.}(2015)\citenamefont {Aad} \emph
  {et~al.}}]{Aad:2014kga}%
  \BibitemOpen
  \bibfield  {author} {\bibinfo {author} {\bibfnamefont {G.}~\bibnamefont
  {Aad}} \emph {et~al.} (\bibinfo {collaboration} {ATLAS}),\ }\href {\doibase
  10.1007/JHEP03(2015)088} {\bibfield  {journal} {\bibinfo  {journal} {JHEP}\
  }\textbf {\bibinfo {volume} {03}},\ \bibinfo {pages} {088} (\bibinfo {year}
  {2015})},\ \Eprint {http://arxiv.org/abs/1412.6663} {arXiv:1412.6663
  [hep-ex]} \BibitemShut {NoStop}%
\bibitem [{\citenamefont {Khachatryan}\ \emph
  {et~al.}(2015{\natexlab{b}})\citenamefont {Khachatryan} \emph
  {et~al.}}]{Khachatryan:2015uua}%
  \BibitemOpen
  \bibfield  {author} {\bibinfo {author} {\bibfnamefont {V.}~\bibnamefont
  {Khachatryan}} \emph {et~al.} (\bibinfo {collaboration} {CMS}),\ }\href
  {\doibase 10.1007/JHEP12(2015)178} {\bibfield  {journal} {\bibinfo  {journal}
  {JHEP}\ }\textbf {\bibinfo {volume} {12}},\ \bibinfo {pages} {178} (\bibinfo
  {year} {2015}{\natexlab{b}})},\ \Eprint {http://arxiv.org/abs/1510.04252}
  {arXiv:1510.04252 [hep-ex]} \BibitemShut {NoStop}%
\bibitem [{\citenamefont {Aad}\ \emph {et~al.}(2016)\citenamefont {Aad} \emph
  {et~al.}}]{Aad:2015gba}%
  \BibitemOpen
  \bibfield  {author} {\bibinfo {author} {\bibfnamefont {G.}~\bibnamefont
  {Aad}} \emph {et~al.} (\bibinfo {collaboration} {ATLAS}),\ }\href {\doibase
  10.1140/epjc/s10052-015-3769-y} {\bibfield  {journal} {\bibinfo  {journal}
  {Eur. Phys. J.}\ }\textbf {\bibinfo {volume} {C76}},\ \bibinfo {pages} {6}
  (\bibinfo {year} {2016})},\ \Eprint {http://arxiv.org/abs/1507.04548}
  {arXiv:1507.04548 [hep-ex]} \BibitemShut {NoStop}%
\bibitem [{\citenamefont {Khachatryan}\ \emph
  {et~al.}(2015{\natexlab{c}})\citenamefont {Khachatryan} \emph
  {et~al.}}]{Khachatryan:2014jba}%
  \BibitemOpen
  \bibfield  {author} {\bibinfo {author} {\bibfnamefont {V.}~\bibnamefont
  {Khachatryan}} \emph {et~al.} (\bibinfo {collaboration} {CMS}),\ }\href
  {\doibase 10.1140/epjc/s10052-015-3351-7} {\bibfield  {journal} {\bibinfo
  {journal} {Eur. Phys. J.}\ }\textbf {\bibinfo {volume} {C75}},\ \bibinfo
  {pages} {212} (\bibinfo {year} {2015}{\natexlab{c}})},\ \Eprint
  {http://arxiv.org/abs/1412.8662} {arXiv:1412.8662 [hep-ex]} \BibitemShut
  {NoStop}%
\bibitem [{\citenamefont {Bauer}\ \emph {et~al.}(2018)\citenamefont {Bauer},
  \citenamefont {Klassen},\ and\ \citenamefont {Tenorth}}]{Bauer:2017fsw}%
  \BibitemOpen
  \bibfield  {author} {\bibinfo {author} {\bibfnamefont {M.}~\bibnamefont
  {Bauer}}, \bibinfo {author} {\bibfnamefont {M.}~\bibnamefont {Klassen}}, \
  and\ \bibinfo {author} {\bibfnamefont {V.}~\bibnamefont {Tenorth}},\ }\href
  {\doibase 10.1007/JHEP07(2018)107} {\bibfield  {journal} {\bibinfo  {journal}
  {JHEP}\ }\textbf {\bibinfo {volume} {07}},\ \bibinfo {pages} {107} (\bibinfo
  {year} {2018})},\ \Eprint {http://arxiv.org/abs/1712.06597} {arXiv:1712.06597
  [hep-ph]} \BibitemShut {NoStop}%
\bibitem [{\citenamefont {Amhis}\ \emph {et~al.}(2017)\citenamefont {Amhis}
  \emph {et~al.}}]{Amhis:2016xyh}%
  \BibitemOpen
  \bibfield  {author} {\bibinfo {author} {\bibfnamefont {Y.}~\bibnamefont
  {Amhis}} \emph {et~al.} (\bibinfo {collaboration} {HFLAV}),\ }\href {\doibase
  10.1140/epjc/s10052-017-5058-4} {\bibfield  {journal} {\bibinfo  {journal}
  {Eur. Phys. J.}\ }\textbf {\bibinfo {volume} {C77}},\ \bibinfo {pages} {895}
  (\bibinfo {year} {2017})},\ \Eprint {http://arxiv.org/abs/1612.07233}
  {arXiv:1612.07233 [hep-ex]} \BibitemShut {NoStop}%
\bibitem [{\citenamefont {Misiak}\ and\ \citenamefont
  {Steinhauser}(2017)}]{Misiak:2017bgg}%
  \BibitemOpen
  \bibfield  {author} {\bibinfo {author} {\bibfnamefont {M.}~\bibnamefont
  {Misiak}}\ and\ \bibinfo {author} {\bibfnamefont {M.}~\bibnamefont
  {Steinhauser}},\ }\href {\doibase 10.1140/epjc/s10052-017-4776-y} {\bibfield
  {journal} {\bibinfo  {journal} {Eur. Phys. J.}\ }\textbf {\bibinfo {volume}
  {C77}},\ \bibinfo {pages} {201} (\bibinfo {year} {2017})},\ \Eprint
  {http://arxiv.org/abs/1702.04571} {arXiv:1702.04571 [hep-ph]} \BibitemShut
  {NoStop}%
\bibitem [{\citenamefont {Karmakar}\ and\ \citenamefont
  {Rakshit}(2019)}]{Karmakar:2019vnq}%
  \BibitemOpen
  \bibfield  {author} {\bibinfo {author} {\bibfnamefont {S.}~\bibnamefont
  {Karmakar}}\ and\ \bibinfo {author} {\bibfnamefont {S.}~\bibnamefont
  {Rakshit}},\ }\href {\doibase 10.1103/PhysRevD.100.055016} {\bibfield
  {journal} {\bibinfo  {journal} {Phys. Rev.}\ }\textbf {\bibinfo {volume}
  {D100}},\ \bibinfo {pages} {055016} (\bibinfo {year} {2019})},\ \Eprint
  {http://arxiv.org/abs/1901.11361} {arXiv:1901.11361 [hep-ph]} \BibitemShut
  {NoStop}%
\bibitem [{\citenamefont {Achard}\ \emph {et~al.}(2001)\citenamefont {Achard}
  \emph {et~al.}}]{Achard:2001qw}%
  \BibitemOpen
  \bibfield  {author} {\bibinfo {author} {\bibfnamefont {P.}~\bibnamefont
  {Achard}} \emph {et~al.} (\bibinfo {collaboration} {L3}),\ }\href {\doibase
  10.1016/S0370-2693(01)01005-X} {\bibfield  {journal} {\bibinfo  {journal}
  {Phys. Lett.}\ }\textbf {\bibinfo {volume} {B517}},\ \bibinfo {pages} {75}
  (\bibinfo {year} {2001})},\ \Eprint {http://arxiv.org/abs/hep-ex/0107015}
  {arXiv:hep-ex/0107015 [hep-ex]} \BibitemShut {NoStop}%
\bibitem [{\citenamefont {Falkowski}\ \emph {et~al.}(2014)\citenamefont
  {Falkowski}, \citenamefont {Straub},\ and\ \citenamefont
  {Vicente}}]{Falkowski:2013jya}%
  \BibitemOpen
  \bibfield  {author} {\bibinfo {author} {\bibfnamefont {A.}~\bibnamefont
  {Falkowski}}, \bibinfo {author} {\bibfnamefont {D.~M.}\ \bibnamefont
  {Straub}}, \ and\ \bibinfo {author} {\bibfnamefont {A.}~\bibnamefont
  {Vicente}},\ }\href {\doibase 10.1007/JHEP05(2014)092} {\bibfield  {journal}
  {\bibinfo  {journal} {JHEP}\ }\textbf {\bibinfo {volume} {05}},\ \bibinfo
  {pages} {092} (\bibinfo {year} {2014})},\ \Eprint
  {http://arxiv.org/abs/1312.5329} {arXiv:1312.5329 [hep-ph]} \BibitemShut
  {NoStop}%
\bibitem [{\citenamefont {Gondolo}\ and\ \citenamefont
  {Gelmini}(1991)}]{Gondolo:1990dk}%
  \BibitemOpen
  \bibfield  {author} {\bibinfo {author} {\bibfnamefont {P.}~\bibnamefont
  {Gondolo}}\ and\ \bibinfo {author} {\bibfnamefont {G.}~\bibnamefont
  {Gelmini}},\ }\href {\doibase 10.1016/0550-3213(91)90438-4} {\bibfield
  {journal} {\bibinfo  {journal} {Nucl. Phys.}\ }\textbf {\bibinfo {volume}
  {B360}},\ \bibinfo {pages} {145} (\bibinfo {year} {1991})}\BibitemShut
  {NoStop}%
\bibitem [{\citenamefont {Semenov}(2009)}]{Semenov:2008jy}%
  \BibitemOpen
  \bibfield  {author} {\bibinfo {author} {\bibfnamefont {A.}~\bibnamefont
  {Semenov}},\ }\href {\doibase 10.1016/j.cpc.2008.10.012} {\bibfield
  {journal} {\bibinfo  {journal} {Comput. Phys. Commun.}\ }\textbf {\bibinfo
  {volume} {180}},\ \bibinfo {pages} {431} (\bibinfo {year} {2009})},\ \Eprint
  {http://arxiv.org/abs/0805.0555} {arXiv:0805.0555 [hep-ph]} \BibitemShut
  {NoStop}%
\bibitem [{\citenamefont {Belanger}\ \emph {et~al.}(2002)\citenamefont
  {Belanger}, \citenamefont {Boudjema}, \citenamefont {Pukhov},\ and\
  \citenamefont {Semenov}}]{Belanger:2001fz}%
  \BibitemOpen
  \bibfield  {author} {\bibinfo {author} {\bibfnamefont {G.}~\bibnamefont
  {Belanger}}, \bibinfo {author} {\bibfnamefont {F.}~\bibnamefont {Boudjema}},
  \bibinfo {author} {\bibfnamefont {A.}~\bibnamefont {Pukhov}}, \ and\ \bibinfo
  {author} {\bibfnamefont {A.}~\bibnamefont {Semenov}},\ }\href {\doibase
  10.1016/S0010-4655(02)00596-9} {\bibfield  {journal} {\bibinfo  {journal}
  {Comput. Phys. Commun.}\ }\textbf {\bibinfo {volume} {149}},\ \bibinfo
  {pages} {103} (\bibinfo {year} {2002})},\ \Eprint
  {http://arxiv.org/abs/hep-ph/0112278} {arXiv:hep-ph/0112278 [hep-ph]}
  \BibitemShut {NoStop}%
\bibitem [{\citenamefont {Billard}\ \emph {et~al.}(2014)\citenamefont
  {Billard}, \citenamefont {Strigari},\ and\ \citenamefont
  {Figueroa-Feliciano}}]{Billard:2013qya}%
  \BibitemOpen
  \bibfield  {author} {\bibinfo {author} {\bibfnamefont {J.}~\bibnamefont
  {Billard}}, \bibinfo {author} {\bibfnamefont {L.}~\bibnamefont {Strigari}}, \
  and\ \bibinfo {author} {\bibfnamefont {E.}~\bibnamefont
  {Figueroa-Feliciano}},\ }\href {\doibase 10.1103/PhysRevD.89.023524}
  {\bibfield  {journal} {\bibinfo  {journal} {Phys. Rev.}\ }\textbf {\bibinfo
  {volume} {D89}},\ \bibinfo {pages} {023524} (\bibinfo {year} {2014})},\
  \Eprint {http://arxiv.org/abs/1307.5458} {arXiv:1307.5458 [hep-ph]}
  \BibitemShut {NoStop}%
\bibitem [{\citenamefont {Belyaev}\ \emph {et~al.}(2013)\citenamefont
  {Belyaev}, \citenamefont {Christensen},\ and\ \citenamefont
  {Pukhov}}]{Belyaev:2012qa}%
  \BibitemOpen
  \bibfield  {author} {\bibinfo {author} {\bibfnamefont {A.}~\bibnamefont
  {Belyaev}}, \bibinfo {author} {\bibfnamefont {N.~D.}\ \bibnamefont
  {Christensen}}, \ and\ \bibinfo {author} {\bibfnamefont {A.}~\bibnamefont
  {Pukhov}},\ }\href {\doibase 10.1016/j.cpc.2013.01.014} {\bibfield  {journal}
  {\bibinfo  {journal} {Comput. Phys. Commun.}\ }\textbf {\bibinfo {volume}
  {184}},\ \bibinfo {pages} {1729} (\bibinfo {year} {2013})},\ \Eprint
  {http://arxiv.org/abs/1207.6082} {arXiv:1207.6082 [hep-ph]} \BibitemShut
  {NoStop}%
\bibitem [{\citenamefont {Sjostrand}\ \emph {et~al.}(2006)\citenamefont
  {Sjostrand}, \citenamefont {Mrenna},\ and\ \citenamefont
  {Skands}}]{Sjostrand:2006za}%
  \BibitemOpen
  \bibfield  {author} {\bibinfo {author} {\bibfnamefont {T.}~\bibnamefont
  {Sjostrand}}, \bibinfo {author} {\bibfnamefont {S.}~\bibnamefont {Mrenna}}, \
  and\ \bibinfo {author} {\bibfnamefont {P.~Z.}\ \bibnamefont {Skands}},\
  }\href {\doibase 10.1088/1126-6708/2006/05/026} {\bibfield  {journal}
  {\bibinfo  {journal} {JHEP}\ }\textbf {\bibinfo {volume} {05}},\ \bibinfo
  {pages} {026} (\bibinfo {year} {2006})},\ \Eprint
  {http://arxiv.org/abs/hep-ph/0603175} {arXiv:hep-ph/0603175 [hep-ph]}
  \BibitemShut {NoStop}%
\bibitem [{\citenamefont {Alwall}\ \emph {et~al.}(2011)\citenamefont {Alwall},
  \citenamefont {Herquet}, \citenamefont {Maltoni}, \citenamefont {Mattelaer},\
  and\ \citenamefont {Stelzer}}]{Alwall:2011uj}%
  \BibitemOpen
  \bibfield  {author} {\bibinfo {author} {\bibfnamefont {J.}~\bibnamefont
  {Alwall}}, \bibinfo {author} {\bibfnamefont {M.}~\bibnamefont {Herquet}},
  \bibinfo {author} {\bibfnamefont {F.}~\bibnamefont {Maltoni}}, \bibinfo
  {author} {\bibfnamefont {O.}~\bibnamefont {Mattelaer}}, \ and\ \bibinfo
  {author} {\bibfnamefont {T.}~\bibnamefont {Stelzer}},\ }\href {\doibase
  10.1007/JHEP06(2011)128} {\bibfield  {journal} {\bibinfo  {journal} {JHEP}\
  }\textbf {\bibinfo {volume} {06}},\ \bibinfo {pages} {128} (\bibinfo {year}
  {2011})},\ \Eprint {http://arxiv.org/abs/1106.0522} {arXiv:1106.0522
  [hep-ph]} \BibitemShut {NoStop}%
\bibitem [{\citenamefont {Alwall}\ \emph {et~al.}(2014)\citenamefont {Alwall},
  \citenamefont {Frederix}, \citenamefont {Frixione}, \citenamefont {Hirschi},
  \citenamefont {Maltoni}, \citenamefont {Mattelaer}, \citenamefont {Shao},
  \citenamefont {Stelzer}, \citenamefont {Torrielli},\ and\ \citenamefont
  {Zaro}}]{Alwall:2014hca}%
  \BibitemOpen
  \bibfield  {author} {\bibinfo {author} {\bibfnamefont {J.}~\bibnamefont
  {Alwall}}, \bibinfo {author} {\bibfnamefont {R.}~\bibnamefont {Frederix}},
  \bibinfo {author} {\bibfnamefont {S.}~\bibnamefont {Frixione}}, \bibinfo
  {author} {\bibfnamefont {V.}~\bibnamefont {Hirschi}}, \bibinfo {author}
  {\bibfnamefont {F.}~\bibnamefont {Maltoni}}, \bibinfo {author} {\bibfnamefont
  {O.}~\bibnamefont {Mattelaer}}, \bibinfo {author} {\bibfnamefont {H.~S.}\
  \bibnamefont {Shao}}, \bibinfo {author} {\bibfnamefont {T.}~\bibnamefont
  {Stelzer}}, \bibinfo {author} {\bibfnamefont {P.}~\bibnamefont {Torrielli}},
  \ and\ \bibinfo {author} {\bibfnamefont {M.}~\bibnamefont {Zaro}},\ }\href
  {\doibase 10.1007/JHEP07(2014)079} {\bibfield  {journal} {\bibinfo  {journal}
  {JHEP}\ }\textbf {\bibinfo {volume} {07}},\ \bibinfo {pages} {079} (\bibinfo
  {year} {2014})},\ \Eprint {http://arxiv.org/abs/1405.0301} {arXiv:1405.0301
  [hep-ph]} \BibitemShut {NoStop}%
\bibitem [{\citenamefont {Placakyte}(2011)}]{Placakyte:2011az}%
  \BibitemOpen
  \bibfield  {author} {\bibinfo {author} {\bibfnamefont {R.}~\bibnamefont
  {Placakyte}},\ }in\ \href@noop {} {\emph {\bibinfo {booktitle} {{Proceedings,
  31st International Conference on Physics in collisions (PIC 2011): Vancouver,
  Canada, August 28-September 1, 2011}}}}\ (\bibinfo {year} {2011})\ \Eprint
  {http://arxiv.org/abs/1111.5452} {arXiv:1111.5452 [hep-ph]} \BibitemShut
  {NoStop}%
\bibitem [{\citenamefont {Wainwright}(2012)}]{Wainwright:2011kj}%
  \BibitemOpen
  \bibfield  {author} {\bibinfo {author} {\bibfnamefont {C.~L.}\ \bibnamefont
  {Wainwright}},\ }\href {\doibase 10.1016/j.cpc.2012.04.004} {\bibfield
  {journal} {\bibinfo  {journal} {Comput. Phys. Commun.}\ }\textbf {\bibinfo
  {volume} {183}},\ \bibinfo {pages} {2006} (\bibinfo {year} {2012})},\ \Eprint
  {http://arxiv.org/abs/1109.4189} {arXiv:1109.4189 [hep-ph]} \BibitemShut
  {NoStop}%
\bibitem [{\citenamefont {Coleman}\ and\ \citenamefont
  {Weinberg}(1973)}]{Coleman:1973jx}%
  \BibitemOpen
  \bibfield  {author} {\bibinfo {author} {\bibfnamefont {S.~R.}\ \bibnamefont
  {Coleman}}\ and\ \bibinfo {author} {\bibfnamefont {E.~J.}\ \bibnamefont
  {Weinberg}},\ }\href {\doibase 10.1103/PhysRevD.7.1888} {\bibfield  {journal}
  {\bibinfo  {journal} {Phys. Rev.}\ }\textbf {\bibinfo {volume} {D7}},\
  \bibinfo {pages} {1888} (\bibinfo {year} {1973})}\BibitemShut {NoStop}%
\bibitem [{\citenamefont {Arnold}\ and\ \citenamefont
  {Espinosa}(1993)}]{Arnold:1992rz}%
  \BibitemOpen
  \bibfield  {author} {\bibinfo {author} {\bibfnamefont {P.~B.}\ \bibnamefont
  {Arnold}}\ and\ \bibinfo {author} {\bibfnamefont {O.}~\bibnamefont
  {Espinosa}},\ }\href {\doibase 10.1103/physrevd.50.6662.2,
  10.1103/PhysRevD.47.3546} {\bibfield  {journal} {\bibinfo  {journal} {Phys.
  Rev.}\ }\textbf {\bibinfo {volume} {D47}},\ \bibinfo {pages} {3546} (\bibinfo
  {year} {1993})},\ \bibinfo {note} {[Erratum: Phys. Rev.D50,6662(1994)]},\
  \Eprint {http://arxiv.org/abs/hep-ph/9212235} {arXiv:hep-ph/9212235 [hep-ph]}
  \BibitemShut {NoStop}%
\bibitem [{\citenamefont {Huang}\ and\ \citenamefont
  {Yu}(2018)}]{Huang:2017rzf}%
  \BibitemOpen
  \bibfield  {author} {\bibinfo {author} {\bibfnamefont {F.~P.}\ \bibnamefont
  {Huang}}\ and\ \bibinfo {author} {\bibfnamefont {J.-H.}\ \bibnamefont {Yu}},\
  }\href {\doibase 10.1103/PhysRevD.98.095022} {\bibfield  {journal} {\bibinfo
  {journal} {Phys. Rev.}\ }\textbf {\bibinfo {volume} {D98}},\ \bibinfo {pages}
  {095022} (\bibinfo {year} {2018})},\ \Eprint
  {http://arxiv.org/abs/1704.04201} {arXiv:1704.04201 [hep-ph]} \BibitemShut
  {NoStop}%
\bibitem [{\citenamefont {Blinov}\ \emph {et~al.}(2015)\citenamefont {Blinov},
  \citenamefont {Profumo},\ and\ \citenamefont {Stefaniak}}]{Blinov:2015vma}%
  \BibitemOpen
  \bibfield  {author} {\bibinfo {author} {\bibfnamefont {N.}~\bibnamefont
  {Blinov}}, \bibinfo {author} {\bibfnamefont {S.}~\bibnamefont {Profumo}}, \
  and\ \bibinfo {author} {\bibfnamefont {T.}~\bibnamefont {Stefaniak}},\ }\href
  {\doibase 10.1088/1475-7516/2015/07/028} {\bibfield  {journal} {\bibinfo
  {journal} {JCAP}\ }\textbf {\bibinfo {volume} {1507}},\ \bibinfo {pages}
  {028} (\bibinfo {year} {2015})},\ \Eprint {http://arxiv.org/abs/1504.05949}
  {arXiv:1504.05949 [hep-ph]} \BibitemShut {NoStop}%
\bibitem [{\citenamefont {Vieu}\ \emph {et~al.}(2018)\citenamefont {Vieu},
  \citenamefont {Morais},\ and\ \citenamefont {Pasechnik}}]{Vieu:2018nfq}%
  \BibitemOpen
  \bibfield  {author} {\bibinfo {author} {\bibfnamefont {T.}~\bibnamefont
  {Vieu}}, \bibinfo {author} {\bibfnamefont {A.~P.}\ \bibnamefont {Morais}}, \
  and\ \bibinfo {author} {\bibfnamefont {R.}~\bibnamefont {Pasechnik}},\ }\href
  {\doibase 10.1088/1475-7516/2018/07/014} {\bibfield  {journal} {\bibinfo
  {journal} {JCAP}\ }\textbf {\bibinfo {volume} {1807}},\ \bibinfo {pages}
  {014} (\bibinfo {year} {2018})},\ \Eprint {http://arxiv.org/abs/1801.02670}
  {arXiv:1801.02670 [hep-ph]} \BibitemShut {NoStop}%
\bibitem [{\citenamefont {Gil}\ \emph {et~al.}(2012)\citenamefont {Gil},
  \citenamefont {Chankowski},\ and\ \citenamefont {Krawczyk}}]{Gil:2012ya}%
  \BibitemOpen
  \bibfield  {author} {\bibinfo {author} {\bibfnamefont {G.}~\bibnamefont
  {Gil}}, \bibinfo {author} {\bibfnamefont {P.}~\bibnamefont {Chankowski}}, \
  and\ \bibinfo {author} {\bibfnamefont {M.}~\bibnamefont {Krawczyk}},\ }\href
  {\doibase 10.1016/j.physletb.2012.09.052} {\bibfield  {journal} {\bibinfo
  {journal} {Phys. Lett.}\ }\textbf {\bibinfo {volume} {B717}},\ \bibinfo
  {pages} {396} (\bibinfo {year} {2012})},\ \Eprint
  {http://arxiv.org/abs/1207.0084} {arXiv:1207.0084 [hep-ph]} \BibitemShut
  {NoStop}%
\bibitem [{\citenamefont {Linde}(1983)}]{Linde:1981zj}%
  \BibitemOpen
  \bibfield  {author} {\bibinfo {author} {\bibfnamefont {A.~D.}\ \bibnamefont
  {Linde}},\ }\href {\doibase 10.1016/0550-3213(83)90293-6,
  10.1016/0550-3213(83)90072-X} {\bibfield  {journal} {\bibinfo  {journal}
  {Nucl. Phys.}\ }\textbf {\bibinfo {volume} {B216}},\ \bibinfo {pages} {421}
  (\bibinfo {year} {1983})},\ \bibinfo {note} {[Erratum: Nucl.
  Phys.B223,544(1983)]}\BibitemShut {NoStop}%
\bibitem [{\citenamefont {Jinno}\ and\ \citenamefont
  {Takimoto}(2017)}]{Jinno:2016vai}%
  \BibitemOpen
  \bibfield  {author} {\bibinfo {author} {\bibfnamefont {R.}~\bibnamefont
  {Jinno}}\ and\ \bibinfo {author} {\bibfnamefont {M.}~\bibnamefont
  {Takimoto}},\ }\href {\doibase 10.1103/PhysRevD.95.024009} {\bibfield
  {journal} {\bibinfo  {journal} {Phys. Rev.}\ }\textbf {\bibinfo {volume}
  {D95}},\ \bibinfo {pages} {024009} (\bibinfo {year} {2017})},\ \Eprint
  {http://arxiv.org/abs/1605.01403} {arXiv:1605.01403 [astro-ph.CO]}
  \BibitemShut {NoStop}%
\bibitem [{\citenamefont {Jinno}\ and\ \citenamefont
  {Takimoto}(2019)}]{Jinno:2017fby}%
  \BibitemOpen
  \bibfield  {author} {\bibinfo {author} {\bibfnamefont {R.}~\bibnamefont
  {Jinno}}\ and\ \bibinfo {author} {\bibfnamefont {M.}~\bibnamefont
  {Takimoto}},\ }\href {\doibase 10.1088/1475-7516/2019/01/060} {\bibfield
  {journal} {\bibinfo  {journal} {JCAP}\ }\textbf {\bibinfo {volume} {1901}},\
  \bibinfo {pages} {060} (\bibinfo {year} {2019})},\ \Eprint
  {http://arxiv.org/abs/1707.03111} {arXiv:1707.03111 [hep-ph]} \BibitemShut
  {NoStop}%
\bibitem [{\citenamefont {Kozaczuk}(2015)}]{Kozaczuk:2015owa}%
  \BibitemOpen
  \bibfield  {author} {\bibinfo {author} {\bibfnamefont {J.}~\bibnamefont
  {Kozaczuk}},\ }\href {\doibase 10.1007/JHEP10(2015)135} {\bibfield  {journal}
  {\bibinfo  {journal} {JHEP}\ }\textbf {\bibinfo {volume} {10}},\ \bibinfo
  {pages} {135} (\bibinfo {year} {2015})},\ \Eprint
  {http://arxiv.org/abs/1506.04741} {arXiv:1506.04741 [hep-ph]} \BibitemShut
  {NoStop}%
\bibitem [{\citenamefont {Steinhardt}(1982)}]{Steinhardt:1981ct}%
  \BibitemOpen
  \bibfield  {author} {\bibinfo {author} {\bibfnamefont {P.~J.}\ \bibnamefont
  {Steinhardt}},\ }\href {\doibase 10.1103/PhysRevD.25.2074} {\bibfield
  {journal} {\bibinfo  {journal} {Phys. Rev.}\ }\textbf {\bibinfo {volume}
  {D25}},\ \bibinfo {pages} {2074} (\bibinfo {year} {1982})}\BibitemShut
  {NoStop}%
\bibitem [{\citenamefont {Ellis}\ \emph {et~al.}(2018)\citenamefont {Ellis},
  \citenamefont {Lewicki},\ and\ \citenamefont {No}}]{Ellis:2018mja}%
  \BibitemOpen
  \bibfield  {author} {\bibinfo {author} {\bibfnamefont {J.}~\bibnamefont
  {Ellis}}, \bibinfo {author} {\bibfnamefont {M.}~\bibnamefont {Lewicki}}, \
  and\ \bibinfo {author} {\bibfnamefont {J.~M.}\ \bibnamefont {No}},\ }\href
  {\doibase 10.1088/1475-7516/2019/04/003} {\  (\bibinfo {year} {2018}),\
  10.1088/1475-7516/2019/04/003},\ \bibinfo {note} {[JCAP1904,003(2019)]},\
  \Eprint {http://arxiv.org/abs/1809.08242} {arXiv:1809.08242 [hep-ph]}
  \BibitemShut {NoStop}%
\bibitem [{\citenamefont {Ellis}\ \emph {et~al.}(2019)\citenamefont {Ellis},
  \citenamefont {Lewicki}, \citenamefont {No},\ and\ \citenamefont
  {Vaskonen}}]{Ellis:2019oqb}%
  \BibitemOpen
  \bibfield  {author} {\bibinfo {author} {\bibfnamefont {J.}~\bibnamefont
  {Ellis}}, \bibinfo {author} {\bibfnamefont {M.}~\bibnamefont {Lewicki}},
  \bibinfo {author} {\bibfnamefont {J.~M.}\ \bibnamefont {No}}, \ and\ \bibinfo
  {author} {\bibfnamefont {V.}~\bibnamefont {Vaskonen}},\ }\href {\doibase
  10.1088/1475-7516/2019/06/024} {\bibfield  {journal} {\bibinfo  {journal}
  {JCAP}\ }\textbf {\bibinfo {volume} {1906}},\ \bibinfo {pages} {024}
  (\bibinfo {year} {2019})},\ \Eprint {http://arxiv.org/abs/1903.09642}
  {arXiv:1903.09642 [hep-ph]} \BibitemShut {NoStop}%
\bibitem [{\citenamefont {Alanne}\ \emph {et~al.}(2019)\citenamefont {Alanne},
  \citenamefont {Hugle}, \citenamefont {Platscher},\ and\ \citenamefont
  {Schmitz}}]{Alanne:2019bsm}%
  \BibitemOpen
  \bibfield  {author} {\bibinfo {author} {\bibfnamefont {T.}~\bibnamefont
  {Alanne}}, \bibinfo {author} {\bibfnamefont {T.}~\bibnamefont {Hugle}},
  \bibinfo {author} {\bibfnamefont {M.}~\bibnamefont {Platscher}}, \ and\
  \bibinfo {author} {\bibfnamefont {K.}~\bibnamefont {Schmitz}},\ }\href@noop
  {} {\  (\bibinfo {year} {2019})},\ \Eprint {http://arxiv.org/abs/1909.11356}
  {arXiv:1909.11356 [hep-ph]} \BibitemShut {NoStop}%
\bibitem [{\citenamefont {Thrane}\ and\ \citenamefont
  {Romano}(2013)}]{Thrane:2013oya}%
  \BibitemOpen
  \bibfield  {author} {\bibinfo {author} {\bibfnamefont {E.}~\bibnamefont
  {Thrane}}\ and\ \bibinfo {author} {\bibfnamefont {J.~D.}\ \bibnamefont
  {Romano}},\ }\href {\doibase 10.1103/PhysRevD.88.124032} {\bibfield
  {journal} {\bibinfo  {journal} {Phys. Rev.}\ }\textbf {\bibinfo {volume}
  {D88}},\ \bibinfo {pages} {124032} (\bibinfo {year} {2013})},\ \Eprint
  {http://arxiv.org/abs/1310.5300} {arXiv:1310.5300 [astro-ph.IM]} \BibitemShut
  {NoStop}%
\bibitem [{\citenamefont {Dev}\ \emph {et~al.}(2019)\citenamefont {Dev},
  \citenamefont {Ferrer}, \citenamefont {Zhang},\ and\ \citenamefont
  {Zhang}}]{Dev:2019njv}%
  \BibitemOpen
  \bibfield  {author} {\bibinfo {author} {\bibfnamefont {P.~S.~B.}\
  \bibnamefont {Dev}}, \bibinfo {author} {\bibfnamefont {F.}~\bibnamefont
  {Ferrer}}, \bibinfo {author} {\bibfnamefont {Y.}~\bibnamefont {Zhang}}, \
  and\ \bibinfo {author} {\bibfnamefont {Y.}~\bibnamefont {Zhang}},\ }\href
  {\doibase 10.1088/1475-7516/2019/11/006} {\bibfield  {journal} {\bibinfo
  {journal} {JCAP}\ }\textbf {\bibinfo {volume} {1911}},\ \bibinfo {pages}
  {006} (\bibinfo {year} {2019})},\ \Eprint {http://arxiv.org/abs/1905.00891}
  {arXiv:1905.00891 [hep-ph]} \BibitemShut {NoStop}%
\bibitem [{\citenamefont {Khachatryan}\ \emph {et~al.}(2017)\citenamefont
  {Khachatryan} \emph {et~al.}}]{Khachatryan:2016whc}%
  \BibitemOpen
  \bibfield  {author} {\bibinfo {author} {\bibfnamefont {V.}~\bibnamefont
  {Khachatryan}} \emph {et~al.} (\bibinfo {collaboration} {CMS}),\ }\href
  {\doibase 10.1007/JHEP02(2017)135} {\bibfield  {journal} {\bibinfo  {journal}
  {JHEP}\ }\textbf {\bibinfo {volume} {02}},\ \bibinfo {pages} {135} (\bibinfo
  {year} {2017})},\ \Eprint {http://arxiv.org/abs/1610.09218} {arXiv:1610.09218
  [hep-ex]} \BibitemShut {NoStop}%
\end{thebibliography}%
\end{document}